\PassOptionsToPackage{numbers,sort&compress}{natbib}
\documentclass[preprintnumbers,amsmath,amssymb,prd, notitlepage,nofootinbib, superscriptaddress]{revtex4}

\usepackage{amsfonts,amssymb,amsmath}
\usepackage{gensymb}
\usepackage{color}
\usepackage{graphicx}
\usepackage{multirow}
\usepackage[utf8]{inputenc}
\usepackage{aas_macros}
\usepackage{hyperref}

\usepackage[normalem]{ulem}

\newcommand{\greaterthanapprox}{\mathrel{\vcenter{
  \offinterlineskip\halign{\hfil$##$\cr
    >\cr\noalign{\kern2pt}\sim\cr\noalign{\kern-2pt}}}}}
    
    \newcommand{\lessthanapprox}{\mathrel{\vcenter{
  \offinterlineskip\halign{\hfil$##$\cr
    <\cr\noalign{\kern2pt}\sim\cr\noalign{\kern-2pt}}}}}
    
\newcommand{\lb}{\left(}
\newcommand{\rb}{\right)}

\newcommand{\be}{\begin{equation}}        
\newcommand{\ee}{\end{equation}}

\newcommand{\fiducialmeasurement}{$A=0.56\pm0.24$}
\newcommand{\fiducialmeasurementbest}{$A=0.82\pm0.21$}
\newcommand{\fiducialmeasurementP}{$\frac{P_0}{P_{\mathrm{fid}}}=0.56\pm0.24$}
\newcommand{\fiducialmeasurementbestP}{$\frac{P_0}{P_{\mathrm{fid}}}=0.82\pm0.21$}

\begin{document}

\title{Cross-correlation of the thermal Sunyaev--Zel'dovich and CMB lensing signals in \textit{Planck} PR4 data with robust CIB decontamination}

\author{Fiona McCarthy}
\email{fiona.mccarthy0@gmail.com}

\affiliation{Center for Computational Astrophysics, Flatiron Institute, New York, NY, USA 10010}

\affiliation{DAMTP, Centre for Mathematical Sciences, Wilberforce Road, Cambridge CB3 0WA, UK
}

\author{J.~Colin Hill}
\email{jch2200@columbia.edu}

\affiliation{Department of Physics, Columbia University, 538 West 120th Street, New York, NY, USA 10027}

\date{\today}

\begin{abstract}

We use the full-mission \textit{Planck} PR4 data to measure the CMB lensing convergence ($\kappa$) -- thermal Sunyaev-Zel'dovich (tSZ, $y$) cross-correlation signal, $C_\ell^{y\kappa}$. This is only the second measurement to date of this signal, following Hill \& Spergel~(2014)~\cite{2014JCAP...02..030H}. {We perform the measurement using foreground-cleaned tSZ maps built from the PR4 frequency maps via a tailored needlet internal linear combination (NILC) code in our companion paper~\cite{Paper1}, in combination with the \textit{Planck} PR4 $\kappa$ maps and various systematic-mitigated PR3 $\kappa$ maps.}   A serious systematic is the residual cosmic infrared background (CIB) signal in the tSZ map, as the high CIB -- $\kappa$ cross-correlation can significantly bias the inferred tSZ -- $\kappa$ cross-correlation.  We mitigate this contamination by deprojecting the CIB in our NILC algorithm, using a moment deprojection approach to avoid leakage due to incorrect modelling of the CIB frequency dependence. We validate this method on mm-wave sky simulations. {We fit a theoretical halo model to our measurement, finding a best-fit amplitude of \fiducialmeasurementbest~(for the highest signal-to-noise PR4 $\kappa$ map) or \fiducialmeasurement~(for a PR3 $\kappa$ map built from a tSZ-deprojected CMB map), indicating that the data are consistent with our fiducial model within $\approx 1$-$2\sigma$. Although our error bars are similar to those of the previous measurement~\cite{2014JCAP...02..030H}, our method is significantly more robust to CIB contamination.} {Our moment-deprojection approach lays the foundation for future measurements of this signal with higher signal-to-noise $\kappa$  and $y$ maps from ground-based telescopes,  which will precisely probe the astrophysics of the intracluster medium of galaxy groups and clusters in the intermediate-mass ($M\sim 10^{13} -10^{14} h^{-1} M_\odot$), high-$z$ ($z\lessthanapprox 1.5${, c.f. $z\lessthanapprox 0.8$ for the tSZ auto-power signal)} regime{, as well as CIB-decontaminated measurements of tSZ cross-correlations with other large-scale structure probes}. }
\end{abstract}
 
\maketitle

\section{Introduction}

The cosmic microwave background (CMB) photons have been travelling through the Universe almost since it began, having been released about 380,000 years after the Big Bang. They carry valuable information not only about the state of the Universe at that time, but also about the structures they have encountered along the way, including information about the growth of structure and astrophysical signals. These late-Universe effects, referred to as CMB ``secondary'' anisotropies, have long been appreciated as late-Universe large-scale structure (LSS) probes.  CMB lensing and the Sunyaev-Zel'dovich effect are particularly well-known examples.

CMB lensing --- the anisotropies induced by the weak gravitational lensing deflection of CMB photons as they pass near matter overdensities and underdensities --- is an extremely well-understood LSS probe and traces the growth of \textit{all of the post-recombination structure} in the Universe (along our past lightcone). First detected statistically in cross-correlation with galaxy surveys~\cite{2007PhRvD..76d3510S} using \textit{WMAP} data, it has been detected at high significance by the Atacama Cosmology Telescope (ACT), the South Pole Telescope (SPT), and \textit{Planck}~\cite{PhysRevLett.107.021301, 2012ApJ...756..142V, 2017ApJ...849..124O, 2020A&A...641A...8P,2023arXiv230405202Q}. CMB lensing probes matter at nearly all redshifts, but is particularly sensitive to structure in the range $z\sim0.5-5$. CMB lensing is usually detected by reconstructing a convergence ($\kappa$) map from the observed CMB temperature and polarization anisotropies using optimal quadratic estimators (see, e.g.,~\cite{2003PhRvD..67h3002O,2021PhRvD.103h3524M}), which exploit the well-understood non-Gaussian statistics induced in a Gaussian primary CMB by lensing.

The Sunyaev-Zel'dovich (SZ) effect~\cite{1969Ap&SS...4..301Z,1970Ap&SS...7....3S} is sourced by the Compton scattering of CMB photons from free electrons in the late Universe. These are electrons which fill the Universe after ``reionization'' at $z\sim7-10$, when the radiation from the first stars reionized the hydrogen atoms which had been neutral since recombination (the release of the CMB). In particular, the \textit{thermal} SZ (tSZ) effect is sourced when the electron in the interaction has a high temperature and up-scatters the CMB photon, changing its frequency and sourcing a spectral distortion in the CMB. The tSZ anisotropies are subdominant with respect to the primary CMB anisotropies, but the unique tSZ frequency dependence can be exploited to separate the primary CMB and the anisotropies induced by the tSZ effect using multi-frequency measurements.  While it was first detected by making targeted observations of clusters in millimeter wavelengths~\cite{1978Natur.275...40B,1981MNRAS.197..571B}, it is now used to discover clusters at high signal-to-noise~\cite{2016A&A...594A..27P,2020ApJS..247...25B,2021ApJS..253....3H} and can be studied at the map (or power spectrum) level by making an all-sky $y$-map~\cite{2016A&A...594A..22P,2022MNRAS.509..300T,2020PhRvD.102b3534M,2022ApJS..258...36B,Paper1,2023arXiv230701258C}.  As the electrons that source the tSZ effect must have very high temperature, it is mostly a very low-$z$ probe ($z\lessthanapprox1$), as the massive clusters with the highest temperatures (which dominate the signal) take nearly a Hubble time to form.

CMB lensing and the tSZ effect provide complementary probes of cosmology. As a very well-theoretically-understood, mostly linear signal, CMB lensing can be used to robustly infer cosmological parameters~\cite{2020A&A...641A...8P,2020ApJ...888..119B,2023arXiv230405203M} via the power spectrum of a lensing convergence map, which is a direct probe of the (redshift-integrated) matter power spectrum. In contrast, the tSZ effect is sensitive to cosmology primarily as a probe of the halo mass function (HMF), which quantifies the abundance of collapsed objects as a function of their mass.  In particular, as the tSZ effect probes the highest-mass  objects (and thus the \textit{rarest} density-field peaks) in the Universe, it probes directly the high-mass tail of the HMF, which is especially sensitive to cosmology. This information can be accessed either by making a cluster catalogue from high signal-to-noise ratio tSZ-selected objects or through measuring the power spectrum (or various other $N$-point statistics, including the bispectrum/skewness~\cite{2012PhRvD..86l2005W,2013PhRvD..87b3527H,2014ApJ...784..143C} or 1-point PDF~\cite{2014arXiv1411.8004H,2019PhRvD..99j3511T}) of an all- (or partial-)sky tSZ map. Simultaneously, the tSZ effect probes the astrophysics of the intracluster medium (ICM) in which the electrons live, probing the  pressure profile of galaxy clusters. Any cosmology constraint derived from the analysis of a single statistic of the tSZ field is thus degenerate with astrophysical uncertainties.

As the tSZ effect is primarily sourced by objects at lower redshifts than those that dominate the CMB lensing redshift kernel, the two signals contain significant independent information, with a correlation coefficient (on perfectly measured signals) predicted to be around 30-40\% (for current tSZ models)~\cite{2014JCAP...02..030H}. Their cross-correlation weights the higher-$z$ contribution to the tSZ signal higher than in the tSZ auto-power spectrum, and thus probes higher-$z$ ICM physics (and thus physics of clusters with lower mass) than the tSZ signal alone. In addition, because somewhat lower-mass  halos (which are much more common than the highest-mass halos) contribute most of the CMB lensing signal, the tSZ--lensing cross-correlation probes lower-mass halos than the tSZ auto-correlation does. Importantly, the degeneracy between ICM astrophysics and cosmological parameters is in a slightly different direction than for the tSZ auto-spectrum, and so the combination of these two signals (e.g., in a joint analysis of the auto- and cross-spectra) can break these degeneracies and separately constrain astrophysics and cosmology. The tSZ -- CMB lensing cross-correlation signal has been detected only once before, in Ref.~\cite{2014JCAP...02..030H} (hereafter HS14), using the nominal-mission \textit{Planck} data.  In this work, we measure the signal with the final PR4 (\texttt{NPIPE}) release of the \textit{Planck} data~\cite{2020A&A...643A..42P}.

To measure the signal, we construct all-sky Compton-$y$ maps (tSZ maps) from the single-frequency \texttt{NPIPE} maps, using a needlet internal linear combination (NILC) method~\cite{2009A&A...493..835D}. We describe our NILC pipeline and characterize our $y$-maps in detail in a companion paper~\cite{Paper1} {(hereafter ``Paper I'')}. The needlet ILC method allows us to build a linear combination of frequency maps with weights localized in both pixel and harmonic space.  We cross-correlate our $y$-maps with various publicly available $\kappa$ maps constructed from the \textit{Planck} data, each with complementary systematics, including maps built from both the 2018 (PR3) data~\cite{2020A&A...641A...8P} and the PR4 (\texttt{NPIPE}) data~\cite{2022JCAP...09..039C}. In particular, we use: (1) a {PR3} $\kappa$ map reconstructed with tSZ-deprojected temperature maps, in order to avoid spurious $\langle yyy \rangle$  three-point correlations in the measured signal; (2) a {PR3} $\kappa$ map with the signal at the location of tSZ clusters restored (it is standard to mask the brightest tSZ sources in the sky in the $\kappa$ analysis mask), to avoid a systematic low bias to our measurement; and (3) the standard minimum-variance \texttt{NPIPE} $\kappa$ map~\cite{2022JCAP...09..039C}, which has lower noise than that of the 2018 $\kappa$ maps, but no mitigation of any of the previously mentioned systematics. Our measurements with all the different $\kappa$ maps are consistent.

Any measurement of an all-sky mm-wave signal necessarily includes some contamination from the cosmic infrared background (CIB), which is an unresolved, diffuse (at the resolution of CMB experiments) emission sourced by dusty star-forming galaxies at high redshift. As the redshifts that are most efficient for CMB lensing overlap significantly with the redshifts at which the CIB is sourced, the CIB has a very high correlation coefficient with CMB lensing, reaching $80$-$90$\%~\cite{2003ApJ...590..664S,2014A&A...571A..18P} --- much larger than that of the tSZ signal. Thus, the bias induced by the correlation of residual CIB in a tSZ map with the CMB lensing convergence is a significant systematic in any measurement of the tSZ -- CMB lensing signal. We avoid this bias by deprojecting the CIB in our NILC tSZ map, i.e., requiring that the NILC weights have zero response to an assumed CIB spectral energy distribution (SED).  However, such a method is sensitive to the modelling of the CIB SED, which encodes its frequency dependence; additionally, it assumes that the CIB can be described as (e.g.) a perfect modified blackbody with no inter-frequency decorrelation. This is not a fully valid assumption, as the CIB does in fact display up to $\approx 10$\% decorrelation between frequency channels (or higher, if the channels are widely separated), because different frequency channels are sensitive to the emission at different redshift ranges that has been redshifted into the appropriate frequency (with high-frequency channels more sensitive to low-$z$ emission that has experienced less redshift)~\cite{2013ApJ...772...77V,2014A&A...571A..30P,2017MNRAS.466..286M,2019ApJ...883...75L}. We avoid these systematics by using a moment-based approach, as suggested in~\cite{2017MNRAS.472.1195C}. This approach desensitizes our measurement to CIB SED modelling systematics, as we verify using detailed simulations of the microwave sky that include both Galactic foregrounds from PySM3~\cite{2017MNRAS.469.2821T} and extragalactic signals from Websky~\cite{2020JCAP...10..012S}.  Using these simulations, we show that we can recover an unbiased measurement of the tSZ -- CMB lensing cross-correlation signal after appropriate moment-based CIB deprojection.

We constrain the astrophysics of the halos sourcing the tSZ -- CMB lensing cross-correlation by comparing our $C_\ell^{y\kappa}$ measurements to a theoretical model, in particular by fitting a scale-independent amplitude parameter $A$ to the data, where $A=1$ corresponds to the prediction of our fiducial model, based on the pressure profile of Ref.~\cite{2012ApJ...758...74B}. Our tightest constraint (from the \texttt{NPIPE} $\kappa$ cross-correlation) is \fiducialmeasurementbest; i.e., our measurement is consistent at $\approx 1\sigma$ with the fiducial model. For the other $\kappa$ maps, our cross-correlation measurements remain broadly consistent, although the measurement with the tSZ-deprojected $\kappa$ of~\fiducialmeasurement\ is roughly $2\sigma$ lower than the fiducial model. These results are in ${1-2\sigma}$ agreement with the measurement of HS14, who found $A=1.10\pm0.22$ for the same fiducial model. Taken at face value, our measurement has similar error bars to those of HS14, which is surprising given the  improved data quality in our work, but we emphasize that our signal-to-noise is significantly reduced by our use of the moment deprojection method to clean the CIB.  Without this penalty, our signal-to-noise would surpass that of HS14 significantly, but we find that such methods are needed in order to obtain a robust measurement of the signal.  We note that most previous such measurements, including, e.g., that of HS14 and~\cite{2014PhRvD..89b3508V,2017MNRAS.471.1565H,2022A&A...660A..27T}, were not validated on detailed, non-Gaussian sky simulations containing tSZ, CIB, lensing, and other fields with realistic correlations, whereas ours is.\footnote{Validation using Websky-based simulations, similar to ours, was performed for the tSZ -- galaxy weak lensing cross-correlation measurements in Refs.~\cite{2022PhRvD.105l3525G,2022PhRvD.105l3526P}.}  We thus emphasize that our $C_\ell^{y\kappa}$ measurement is much more robust to CIB contamination than that of HS14, which employed a simpler CIB-cleaning method that required an (unphysical) assumption that the intrinsic CIB-$y$ cross-correlation vanishes.  While our final astrophysical constraints are similar to those of HS14, their robustness and stability are validated with significantly more powerful and thorough methods.

This paper is structured as follows. In Section~\ref{sec:theory} we describe the theory of the tSZ signal and CMB lensing signal, and the halo model that we use to model their cross-correlation. In Section~\ref{sec:data} we describe the datasets we use to detect the signal.
{In Section~\ref{sec:ilc_ymap} we discuss the foreground mitigation techniques for the ISW and CIB biases, including describing the components we deproject from our NILC $y$-maps and discussing the frequency coverage in the $y$-map construction. } In Section~\ref{sec:clyk_meas} we describe our measurement of the tSZ -- CMB lensing cross-correlation signal $C_\ell^{y\kappa}$ and present the results. { Section~\ref{sec:CIB_systematics} in particular explicitly illustrates our use of different combinations of $y$-maps released in Paper I in order to remove CIB residuals in our final measurement.} In Section~\ref{sec:pipeline_validation} we present the results of our pipeline validation on simulations, demonstrating that our measurement is unbiased to CIB systematics, and comment on the consistency with the previous measurement of this signal. In Section~\ref{sec:analysis} we analyze the signal, quantifying the significance of our detection and constraining {separately} the ICM physics that controls the pressure-mass relation, {and cosmological parameters}.  We discuss our results and conclude in Section~\ref{sec:conclusion}.

We assume the cosmology of~\cite{2020A&A...641A...6P} throughout: $\{H_0=67.32 \,\, \mathrm{km/s/Mpc};\,\sigma_8 = 0.812;\, n_s = 0.96605;  \,\Omega_b h^2 = 0.022383;\, {\Omega_{{\mathrm{cdm}}}h^2}=0.12011\}$, where $H_0\equiv100 h$ is the Hubble parameter today; $\sigma_8$ is the amplitude of linear fluctuations on a scale of $8 h^{-1}$ Mpc today (i.e., the linear matter power spectrum integrated over all scales, smoothed with a top-hat window function of $8 h^{-1}$ Mpc); {$n_s$ is the spectral index of the primordial fluctuation power spectrum}; $\Omega_b h^2$ is the physical density of baryons today; and $\Omega_{{\mathrm{cdm}}} h^2$ is the physical density of dark matter today.

\section{Theory}\label{sec:theory}

We model the tSZ signal and the CMB lensing signal using the halo model. In this section, we present the theory and model we use to calculate their cross-power spectrum $C_\ell^{y \kappa}$.

We perform all theory calculations throughout with \texttt{class\_sz}\footnote{\url{https://github.com/borisbolliet/class_sz}}~\cite{2018MNRAS.477.4957B,2020MNRAS.497.1332B,2022arXiv220807847B}, which is an extension of the cosmological Boltzmann solver \texttt{class}\footnote{\url{https://lesgourg.github.io/class_public/class.html}}~\cite{2011JCAP...07..034B}. {Sample code showing how to use \texttt{class\_sz} to calculate the observables is available at \url{https://github.com/fmccarthy/tSZ_kappa_halomodel_classsz}.}

\subsection{The tSZ anisotropies}
The tSZ temperature anisotropy at sky position $\hat n$ observed at frequency $\nu$ is given by
\be
\frac{\Delta T^{\mathrm{tSZ}}(\hat n, \nu)}{T_{\mathrm{CMB}}} = g_\nu y (\hat n),
\ee
where $g_\nu$ is the spectral function of the tSZ effect~\cite{1970Ap&SS...7....3S}:
\be
g_\nu =x\coth\lb\frac{x}{2}\rb-4 \,, \label{gnu_tsz}
\ee
with $x\equiv\frac{h\nu}{k_B T_{CMB}}$. Here $h$ is Planck's constant; $k_B$ is Boltzmann's constant; $T_{\mathrm{CMB}} = 2.726$ K is the mean temperature of the CMB~\cite{1996ApJ...473..576F,1999ApJ...512..511M,2009ApJ...707..916F}; and $y(\hat n)$ is the dimensionless (and frequency-independent) Compton $y$-parameter that quantifies the integral of the electron pressure along the line of sight (LOS):
\be
y(\hat n) = \frac{\sigma_T}{m_e c^2}\int _{\mathrm{LOS}} d\chi \, a(\chi) P_e(\chi, \hat n) \,.
\ee
Here $\sigma_T$ is the Thomson scattering cross-section; $m_e$ is the electron mass; $c$ is the speed of light; $a(\chi)$ is the scale factor at comoving distance $\chi$; and $P_e(\chi, \hat n)$ is the electron pressure at $(\chi, \hat n)$.  The integral is done over comoving distance $\chi$ from today to the beginning of the epoch of reionization at redshift $z\sim7-10$. 

We model the distribution of $P_e(\chi, \hat n)$ within the halo model; the details are given below in Section~\ref{sec:tsz_halo}.

\subsection{The CMB lensing potential}

The CMB lensing convergence at sky position $\hat n$ is given by a line-of-sight integral over  $\delta(\chi, \hat n)$, which denotes the matter overdensity at $(\chi, \hat n)$:
\be
\kappa(\hat n) = \int _0 ^{\chi_S}d\chi W^\kappa(\chi) \delta(\chi, \hat n)
\ee
where the CMB lensing convergence kernel $W^\kappa(\chi)$ is given by
\be
W^\kappa(\chi) = \frac{3}{2}\left(\frac{H_0}{c}\right)^2 \frac{\Omega_m}{a}\chi\left(1-\frac{\chi}{\chi_S}\right),\label{W_k_chi}
\ee
with $\chi_S$ the comoving distance to the surface of last scattering, at which the CMB was released (corresponding to $z\simeq 1100$), {and $\Omega_m$ the matter density parameter}.  For a review of CMB lensing theory, see~\cite{2006PhR...429....1L}.

\subsection{The halo model}

The halo model (see, e.g.,~\cite{2002PhR...372....1C} for a review) has long been used to model the distribution of the large-scale structure of our Universe~\cite{1974ApJ...187..425P,2000MNRAS.318..203S}.  
Within the halo model, the continuous overdensity field is replaced by a discrete distribution of ``halos'', which cluster according to the underlying density field but with a bias that is scale-independent on large scales. 
These halos are modeled as having a spherically symmetric density profile $\rho(r)$, where $r$ is the distance from the center of the halo. In the simplest picture, all halo properties and observables are functions only of the halo mass $M$ and the redshift $z$. The halos are distributed according to the halo mass function (HMF) $\frac{dN}{dM}(M,z)$, which gives the comoving differential number density of halos of mass $M$ at redshift $z$. {Note that the halo model makes several unrealistic simplifying assumptions, such as assuming that all halos are spherical and that all properties of the halo and its constituent galaxy (or galaxies) are functions only of the host halo mass and redshift, neglecting real effects due to scatter and environment.}

Within the halo model, all correlation functions are split into various $N$-halo terms, where $N$ is the number of halos whose profiles appear in that term of the correlation function. For the two-point correlation function, this means that there is a $2$-halo term, which describes the correlations between two different halos (due to clustering), and a $1$-halo term, which describes the correlations within a single halo. The halo mass function is an important quantity for calculating these correlations, as is the \textit{halo bias}, which is important on scales where halo clustering is relevant (i.e., the two-halo term). The halo bias quantifies the bias of the halo distribution with respect to the underlying dark matter distribution: 
\be
\delta_h(M,z) = b_h(M,z)\delta
\ee
where $\delta$ is the underlying (continuous) dark matter overdensity; $\delta_h$ is the halo overdensity; and $b_h(M,z)$ is the halo bias. It is assumed that halos are spherically symmetric and that the dark matter within them follows a continuous density profile; in our case we take this to be a Navarro--Frenk--White (NFW) profile~\cite{1997ApJ...490..493N}
\be
\rho(r,M,z) =\rho_s \lb\frac{r}{r_s}\rb^{-1} \lb1+\frac{r}{r_s}\rb^{-2}\label{NFW_profile}
\ee
where $\rho_s$ is a characteristic density and $r_s$ is a characteristic NFW scale radius.  
In our halo model calculations, we use the halo mass function of~\cite{2008ApJ...688..709T} and the halo bias of~\cite{2010ApJ...724..878T}, as implemented in \texttt{class\_sz} (for implementation details, see the appendices of~\cite{2022arXiv220807847B}).

In general, to define quantities that depend on a halo mass, one requires a definition of halo mass. Several common definitions exist; for example, the spherical overdensity (SO) mass, which is the mass $M_{\Delta c,m}$ within the radius $R_{\Delta c,m}$ within which the mean density of the halo is $\Delta$ times either the critical density $\rho_{\mathrm{cr}}(z)$ (in the case of the subscript $c$) or the mean background matter density $\rho_m(z)$ (in the case of the subscript $m$).   
The halo mass function of~\cite{2008ApJ...688..709T} was defined for $M_{\Delta m}$, at several values of $\Delta$. As the pressure profiles we will use to describe the tSZ signal (see below in Section~\ref{sec:tsz_halo}) are defined for $M_{200c}$, we use a halo mass function that is defined for this mass,  by using $M_{\Delta m}$ with a $z$-dependent $\Delta=200/\Omega_m(z)$, which is appropriate as $\rho_m(z)=\Omega_m(z) \rho_\mathrm{cr}(z)$ (as implemented directly in \texttt{class\_sz}).

In all our integrals over mass, we use the integration limits $10^{10}  \, h^{-1}M_\odot< M < 10^{15.5} \, h^{-1} M_\odot$. In our integrals over $z$, we use the integration limits $0.005<z<10$. {Note that this neglects the contribution from halos below the lower mass integration limit, for which the HMF is not calibrated.}  
{We account for the contribution from these halos using counter-terms~\cite{2016PhRvD..93f3512S} (see Appendix B2 of~\cite{2022arXiv220807847B} for further details).} {Additionally, we extrapolate the HMF to higher $z$ than where it was calibrated; following the recommendation of Ref.~\cite{2008ApJ...688..709T}, we deal with this regime by removing the $z$-evolution of the halo multiplicity function $f(\sigma)$ and replacing it with its values at $z=2.5$ for all $z>2.5$. }

{Finally, we cut all of our radial profiles at $r=2 R_{200m}$. This is a \textit{choice}, and in practice observables depend on this cut-off as the NFW profile does not converge as $r\rightarrow\infty$ (although the profiles we use for the tSZ signal do converge). We choose this value as we find that it allows us to reproduce most accurately a measurement of the $C_\ell^{y\kappa}$ signal from the hydrodynamical simulations from which the tSZ profiles were extracted~\cite{2015ApJ...812..154B} {(see Fig.~\ref{fig:compare_sims_calculation} in Appendix~\ref{app:cutoff})}.  Choosing a cut-off for the NFW profile that is not equal to the radius we use to define the mass of our halos leads to some subtleties in the matter power spectrum calculation; we discuss this in Appendix~\ref{app:cutoff}.}

\subsection{tSZ power spectrum within the halo model}\label{sec:tsz_halo}

In~\cite{1999ApJ...526L...1K,2002MNRAS.336.1256K},  a halo model prescription for the calculation of the auto-power spectrum of the tSZ effect was presented. A pressure-mass relation $P_e(z,M)$ is assumed, and (as halos are assumed to be spherically symmetric), a three-dimensional pressure profile $P_{3D}(z,M,r)$ can be written down that describes the pressure at distance $r$ from the center of the halo. The power spectrum, which involves the Fourier transform of this pressure profile, can then be written as
\be
C_\ell^{yy}=C_\ell^{yy,2h}+C_\ell^{yy,1h}
\ee
where $C_\ell^{yy,2h}$ is the two-halo term that depends on the distribution and clustering of the halos, and $C_\ell^{yy,1h}$ is the one-halo term that probes the distribution of the pressure within each halo. As $C_\ell^{yy}$ is dominated (at $\ell\greaterthanapprox300$) by the one-halo term, Ref.~\cite{2002MNRAS.336.1256K} neglected the two-halo term; however, for an in-depth discussion and presentation of this (and the one-halo term), see~\cite{2013PhRvD..88f3526H} (see also~\cite{1999ApJ...526L...1K}). The final expressions (in the flat-sky and Limber approximations~\cite{1953ApJ...117..134L}) are
\begin{align}
C^{yy,}_\ell{}^{2h} =& \int d\chi \, \chi^2 \left(\int dM \frac{dN}{dM} b_h(M,z)\tilde y_\ell(M,z)\right)^2 P_{\mathrm{lin}}\left(k=\frac{\ell+1/2}{\chi},z\right)\label{yy_2h}\\
C^{yy,}_\ell{}^{1h} =&  \int  d\chi \, \chi^2\int dM \frac{dN}{dM}\left|\tilde y_\ell(M,z)\right|^2,\label{yy_1h}
\end{align}
where $P_{\mathrm{lin}}\left(k,z\right)$ is the linear matter power spectrum and both expressions depend on $\tilde{y}_\ell(M,z)$, the 2-dimensional Limber projection of the 3-dimensional Fourier transform of the pressure profile, which is given by
\be
\tilde y_\ell(M,z) = \frac{4\pi r_y}{\ell_y^2}\int dx x^2\frac{\sin\left({\left(\ell+1/2\right) x/\ell_y}\right)}{\left({\left(\ell+1/2\right) x/\ell_y}\right)} y_{3D}(x, M,z) \,. \label{ytilde_def}
\ee
Here $r_y$ is a characteristic scale radius of the three-dimensional pressure profile, for which we use $r_y=R_{200c}$; $\ell_y=\frac{a(\chi)\chi}{r_y}$ is the characteristic multipole moment of $r_y$; and $y_{3D}(x,M,z)$ is the 3-dimensional Compton-$y$ profile of a halo of mass $M$ at redshift $z$ for which a model is required (note the integral in Equation~\eqref{ytilde_def} is in terms of the dimensionless parameter $x=\frac{r}{r_y}$ while the model for $ y_{3D}(x, M,z)$ is usually expressed in terms of the dimensionful distance $r$ from the halo center). The Compton-$y$ profile is directly related to the \textit{pressure} profile $P_{3D}$ 
by
\be
y_{3D} = \frac{\sigma_T}{m_e c^2}P_{3D}.
\ee 

As our fiducial model for $P_{3D}$, we use the generalized NFW-based fitting functions of~\cite{2012ApJ...758...74B}. These are given by
\be
P_{3D}(x,M,z)= P_\Delta P_0 \lb x/x_c\rb^\gamma\lb1+\lb{x/x_c}\rb^\alpha\rb^{-\beta}\label{pressure_profile}
\ee
where $P_\Delta$, the self-similar amplitude for pressure, provides the physical dimensions of pressure:
\be
P_\Delta = \frac{G M_\Delta \Delta \rho_{\mathrm{cr}}(z) f_b}{2 R_{\Delta c}} \,.
\ee
Here $G$ is Newton's constant and $f_b\equiv\frac{\Omega_b}{\Omega_m}$ is the baryon fraction.  Recall that the mean density of the halo within $R_{\Delta c}$ is $\Delta\rho_{\mathrm{cr}}(z)$ (we are working with $\Delta=200$). In Equation~\eqref{pressure_profile}, the amplitude parameter $P_0$, the core-radius parameter $x_c$, and the power-law index $\beta$ are parameters that are fit to hydrodynamical simulations~\cite{2010ApJ...725...91B}, while $\alpha$ and $\gamma$ are fixed at $\alpha=1$ and $\gamma=-0.3$~\cite{2012ApJ...758...74B}. $P_0$, $x_c$, and $\beta$ are allowed to have mass and redshift dependence and each are of the form
\be
A=A_0\lb\frac{M_{200}}{10^{14}M_\odot}\rb^{\alpha_m}(1+z)^{\alpha_z},\label{parameter_scaling_def}
\ee
with $A_0$, $\alpha_m$, and $\alpha_z$ fit separately for each parameter. The best-fit values (to the simulations of~\cite{2010ApJ...725...91B}) are listed in Table 1 of~\cite{2012ApJ...758...74B}; for completeness, we repeat them in Table~\ref{tab:pressure_profile_parameters}. This model is implemented directly in \texttt{class\_sz}. 

\begin{table}
\begin{tabular}{|c||c|c|c|}\hline
Parameter & $A_0$ &$\alpha_m$ & $\alpha_z$\\\hline\hline
$P_0$ & 18.1 & 0.154 & -0.758\\\hline
$x_c$ & 0.497 & -0.00865 & 0.731 \\\hline
$\beta$ & 4.35 & 0.0393 & 0.415\\\hline
\end{tabular}\caption{The best-fit values from Ref.~\cite{2012ApJ...758...74B} for the three-dimensional pressure profile of Equation~\eqref{pressure_profile}. The parameters $A_0$, $\alpha_m$, and $\alpha_z$ are defined in Equation~\eqref{parameter_scaling_def}.}\label{tab:pressure_profile_parameters}
\end{table}

\subsection{CMB lensing power spectrum within the halo model}\label{sec:cmblensing_halo}

We can also model the matter power spectrum within the halo model, which can then be weighted appropriately with the CMB lensing kernel and integrated against redshift to calculate the CMB lensing power spectrum. Again, there are both one-halo and two-halo contributions. The final expressions are similar in form to Equations~\eqref{yy_2h} and~\eqref{yy_1h}, but with $\tilde y_\ell$ replaced by the CMB-lensing-kernel-weighted 2-dimensional mass profile $\tilde \kappa_\ell$: 
\begin{align}
C_\ell^{\kappa\kappa,2h} =& \int d\chi \chi^2 \left(\int dM \frac{dN}{dM}b_h (M,z) \tilde \kappa_\ell(M,z)\right)^2 P_{\mathrm{lin}}\left(k=\frac{\ell+1/2}{\chi},z)\right);\\
C_\ell^{\kappa\kappa,1h} =& \int d\chi \chi^2 \int dM \frac{dN}{dM}\left|\tilde \kappa_\ell(M,z)\right|^2,\\
\end{align}
with
\be
\tilde\kappa_\ell(M,z) = \frac{4\pi r_s a^2(z) W^\kappa(\chi(z))}{\bar{\rho}_m^0\ell_s^2}\int dx x^2\frac{\sin\lb \lb\ell+1/2\rb x/\ell_s\rb}{\lb\ell+1/2\rb x/\ell_s}\rho(x r_s,M,z)
\,,
\ee
where $r_s$ is the characteristic scale radius for $\kappa$, which is indeed the NFW scale radius of Equation~\eqref{NFW_profile}, and $\ell_s$ is the characteristic multipole $\ell_s=\frac{a(\chi)\chi}{r_s}$.

Care must be taken when the mass integral is evaluated in the calculation of the matter power spectrum (and thus the CMB lensing power spectrum). In particular, there is a significant contribution to the signal from low-mass halos below the range for which the halo mass function has been calibrated; also, the result depends critically on any low mass cut-off imposed in the integral. This is in contrast to the case of the tSZ power spectrum, where any contribution from low-mass halos is severely downweighted by their low gas pressure amplitude.

We account for the contribution from lower-mass halos as follows. We require that the halo mass function and bias obey the consistency condition
 \be
\int _0^\infty dM\frac{dN}{dM} b_h(M,z)\frac{M}{\bar{\rho}_m(z)} =1 \,.
\ee
This condition ensures that the dark matter is unbiased with respect to itself. Following Ref.~\cite{2016PhRvD..93f3512S}, we write this as
\be
\int _0^\infty dM\frac{dN}{dM} b_h(M,z)\frac{M}{\bar{\rho}_m(z)}  = 1 = \int _0^{M_{\mathrm{cut}}} dM\frac{dN}{dM} b_h(M,z)\frac{M}{\bar{\rho}_m(z)}  +\int _{M_\mathrm{cut}}^\infty dM\frac{dN}{dM}\frac{M}{\bar{\rho}_m(z)}  b_h(M,z) \,,
\ee
where $M_{\mathrm{cut}}$ is the low-mass cut-off of the integral we use in practice when calculating halo model quantities.  This allows us to define (and explicitly calculate) a counter-term explicitly as the integral over all the low-mass halos:
\be
b_{\mathrm{cut}} \equiv 1 - \int _{M_\mathrm{cut}}^\infty dM\frac{dN}{dM}\frac{M}{\bar{\rho}_m(z)} b_h(M,z) \,.
\ee
We can then replace $\int_0^\infty \frac{dN}{dM}b_h(M,z)$ with $\int_{M_{\mathrm{cut}-\epsilon}}^\infty\left(\frac{dN}{dM}b_h(M,z)+b_{\mathrm{cut}} \delta(M-M_{\mathrm{cut}})\right)$, where $\epsilon$ is an infinitesimal mass. This allows us to account for the contribution from the lower-mass halos in the two-halo power spectrum while preserving the consistency conditions and also, importantly, not needing to model any properties of the low-mass halos below the cut-off. Note that this issue is not as serious in the one-halo term, which is dominated by the massive halos.

\subsection{tSZ -- CMB lensing power spectrum within the halo model}\label{sec:tszcmb_halo}

We can use the halo model formalism to write down an expression for the tSZ -- CMB lensing cross-correlation $C_\ell^{y\kappa}$. As in HS14, we use one factor of $\tilde{y}_\ell(M,z)$ and one factor of $\tilde{\kappa}_\ell(M,z)$:
\begin{widetext}
\begin{align}
C^{y\kappa,}_\ell{}^{2h} =& \int d\chi \, \chi^2 \left( \frac{dN}{dM} b_h(M,z)\tilde y_\ell(M,z)\right)\left( \frac{dN}{dM} b_h(M,z)\tilde \kappa_\ell(M,z)\right) P_{\mathrm{lin}}\left(k=\frac{\ell+1/2}{\chi},z\right)\\
C^{y\kappa,}_\ell{}^{1h} =&  \int d\chi \, \chi^2\int \frac{dN}{dM}\tilde y_\ell(M,z)\tilde \kappa_\ell(M,z) \,.
\end{align}
\end{widetext}
Again, we must replace the contribution from the low-mass halos to $\tilde{\kappa}_\ell$ in the two-halo term with an appropriate counter-term as described above. {However, while the counter-terms for $y$ are also well-defined according to the formalism of~\cite{2016PhRvD..93f3512S}, we choose \textit{not} to add them. The counter-terms account for the contribution of the halos below the low-mass limit, which we take to be $M=10^{10} h^{-1}M_\odot $; this is already extrapolated far below the mass range in which the $y$ profiles were fit, and indeed we expect very little contribution from such low-mass objects. Thus, the counter-term addition in $C_\ell^{y\kappa}$ probes no gas in such low-mass halos, and only the clustering of the low-$z$ lensing halos with the gas in the larger-mass halos through the 2-halo term.}

\subsubsection*{Comparison to simulations}

{In Appendix~\ref{app:cutoff}, we explicitly compare our halo model prediction to a measurement of the tSZ -- CMB lensing cross-power spectrum in the hydrodynamical simulations from which the pressure profile model was calibrated~\cite{2015ApJ...812..154B}. While there is good agreement at high $\ell$, at $\ell\lessthanapprox2000$ (which is the regime where we measure the signal), the halo model under-predicts the total signal.}

{As the simulations were performed with different cosmological parameters than those we use for our theory model, we cannot use the measurement from simulations as a template with which to model the signal. However, we can use it to calibrate a ratio $\frac{C_\ell^{y\kappa, \mathrm{simulation}}}{C_\ell^{y\kappa, \mathrm{halo-model}}}$ (where $C_\ell^{y\kappa, \mathrm{halo-model}}$ is calculated at the same cosmological parameters as the simulation), which we can assume to be cosmology-independent at first order. We can then use this ratio to rescale our halo model calculation to account for this mismatch, which is likely due to unbound gas blown out of halos into the IGM.}

{While we choose to retain the un-rescaled halo model theory prediction in all plots and when we quote our fiducial constraints, we will also quote constraints on an amplitude of this rescaled halo model prediction when we constrain the model in Section~\ref{sec:simulation_correction_constraints}. Of course, the best-fit amplitude for this model will necessarily be lower than that of the unrescaled model, as the rescaled model predicts a higher signal by $\sim25\%$ over our multipole range of interest.

\subsection{Halo mass and redshift dependence of the signal}

To source the tSZ effect, the electrons involved in the Compton scattering process must have high temperature. They inherit this high energy from the gravitational collapse of massive structures, and thus the highest-temperature electrons are in the most massive clusters ($T_e \sim M^{2/3}$ in a self-similar ICM model~\cite{1986MNRAS.222..323K}). These are primarily located at low $z$, as the formation of such large structures requires nearly a Hubble time. Thus, the tSZ effect is primarily a low-$z$ probe, with most ($>90\%$ of the contribution to $C_\ell^{yy}$ in the range $\ell<2000$) of the signal sourced at $z<1$ (and $>95\%$ of the signal for $\ell<1000$). The cross-correlation with CMB lensing, while also being a low-$z$ probe (with closer to $\sim 70\%$ of the contribution to $C_\ell^{y\kappa}$ being sourced at $z<1$ for $\ell<2000$), is weighted more towards high redshifts than $C_\ell^{yy}$, as the objects with high efficiency for CMB lensing are located at higher $z$ due to the lensing kernel. We illustrate this in Figure~\ref{fig:redshift_dependence_signal}, where we plot $C_\ell^{yy}(z>z_{\mathrm{min}})$ and $C_\ell^{y\kappa}(z>z_{\mathrm{min}})$, i.e., the contributions to the respective signals from $z>z_{\mathrm{min}}$.

\begin{figure*}
\includegraphics[width=0.49\textwidth]{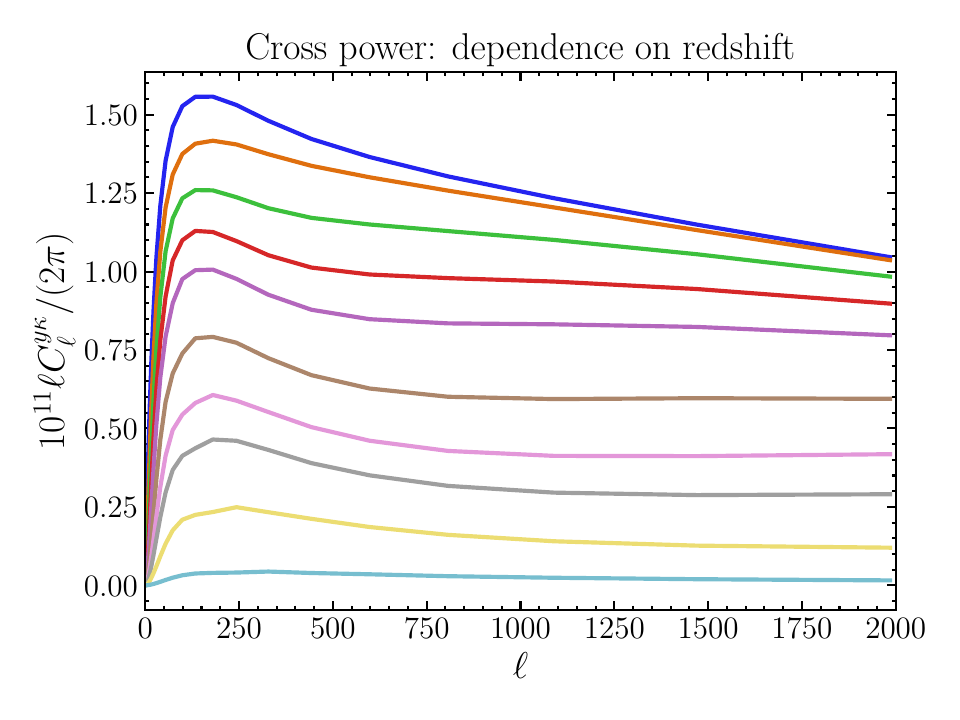}
\includegraphics[width=0.49\textwidth]{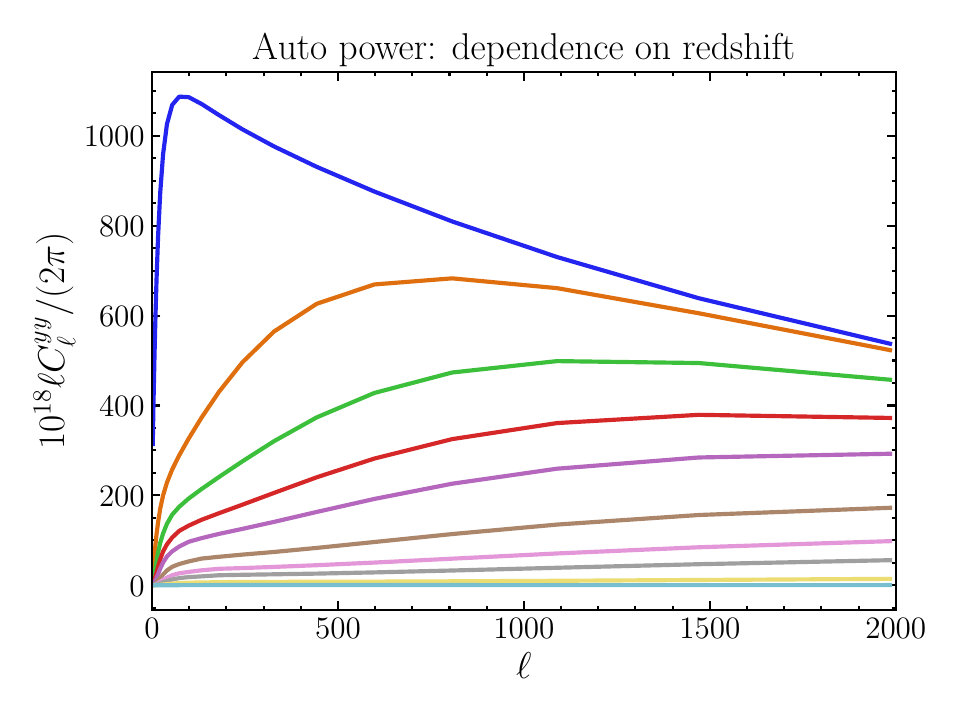}
\includegraphics[width=0.6\textwidth]{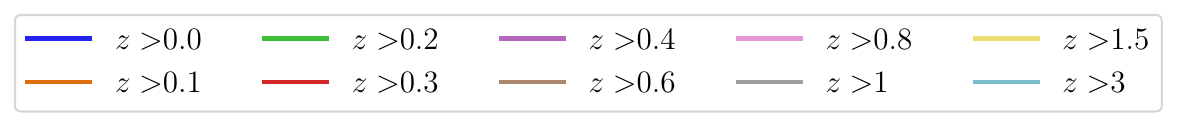}
\caption{The contribution to $C_\ell^{yy}$ and $C_\ell^{y\kappa}$ from different redshifts. $C_\ell^{yy}$ has a significant contribution even from $z<0.1$, while $C_\ell^{y\kappa}$ is more evenly distributed in the redshift range $0<z<1$, and also receives non-negligible contributions ($\approx 25\%$ of the total signal) from $z>1$. }\label{fig:redshift_dependence_signal}
\end{figure*}

In Figure~\ref{fig:mass_dependence_signal} we show the dependence of the signals on the highest mass included in the halo model integral. These plots illustrate the significantly lower mass range probed by $C_\ell^{y\kappa}$ compared to $C_\ell^{yy}$. Roughly $50\%$ of the $C_\ell^{y\kappa}$ signal comes from halos with $M>10^{14} \, h^{-1} M_\odot$; the equivalent number for the auto-power spectrum is $\approx 90\%$.  This is due to the combination of two reasons: first, the CMB lensing signal is sourced at higher $z$, where there are fewer high-mass halos, and so is preferentially probing a redshift regime with lower-mass halos than the tSZ signal. Second, at every $z$, the highest mass halos that source the largest tSZ signal are very rare, and contribute fractionally less of the \textit{total} mass at a given redshift than the lower-mass halos. Thus most of the CMB lensing signal (which is weighted by overall mass) is sourced by the lower-mass halos that host most of the mass. 
\begin{figure*}
\includegraphics[width=0.49\textwidth]{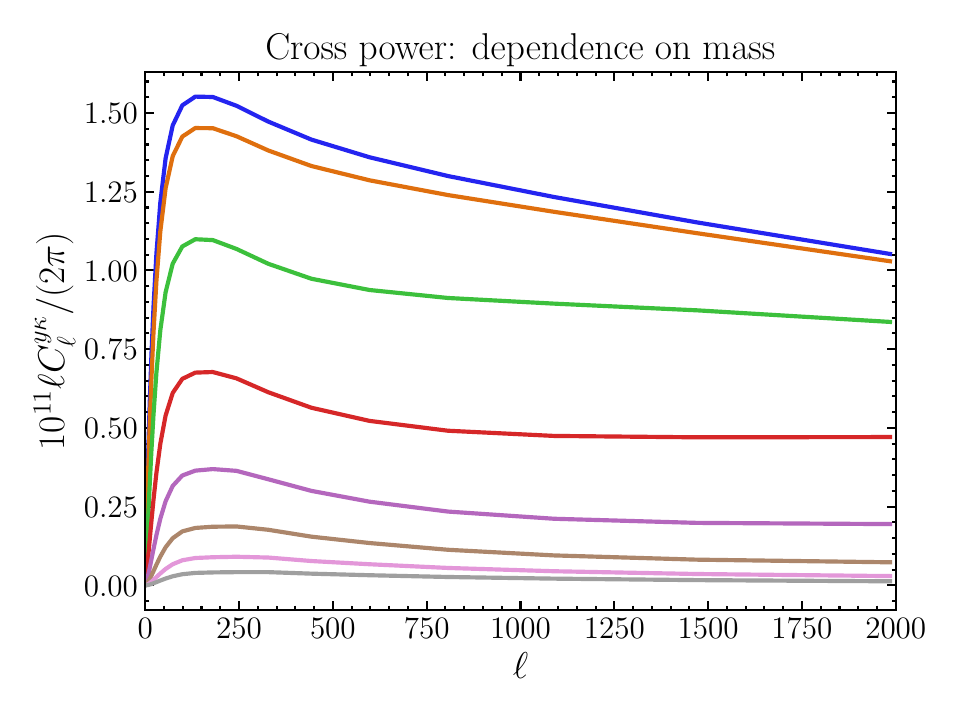}
\includegraphics[width=0.49\textwidth]{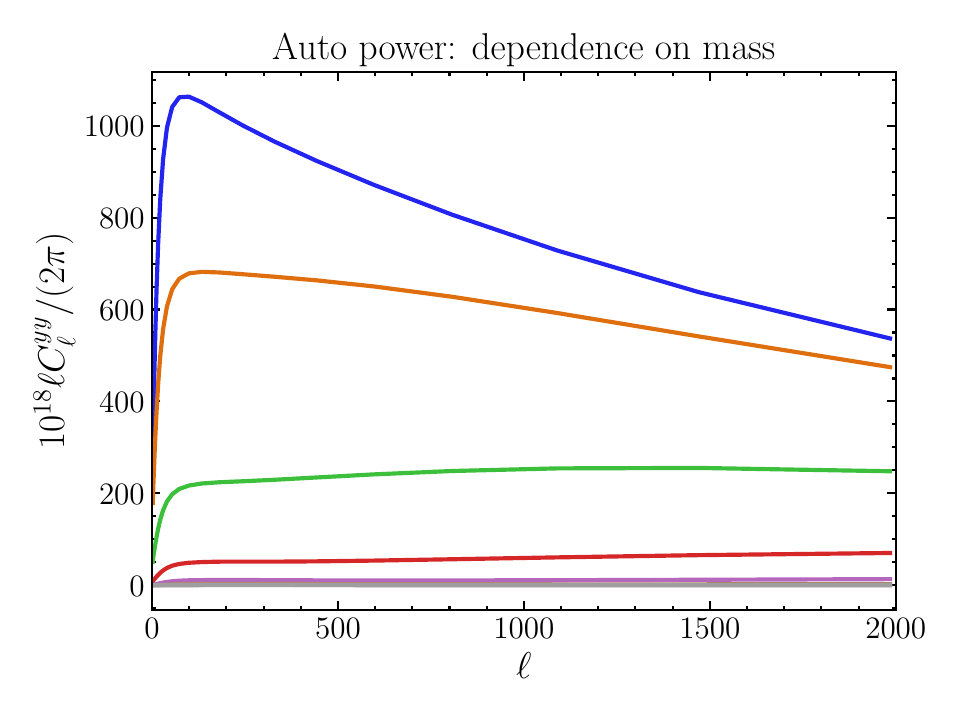}
\includegraphics[width=0.8\textwidth]{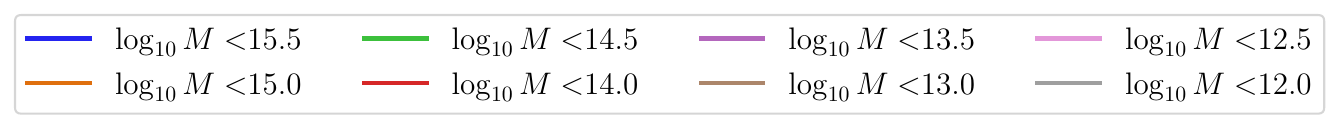}
\caption{ $C_\ell^{yy}$ and $C_\ell^{y\kappa}$ as calculated with different maximum masses in the halo model integral. The $C_\ell^{yy}$ signal is highly concentrated in the highest mass range with $M>10^{14.5} h^{-1}M_\odot$.  The $C_\ell^{y\kappa}$ signal probes lower-mass halos, with significant contributions down to $M>10^{13.5} h^{-1}M_\odot$. The units of $M$ in the legend are $h^{-1}M_\odot$.}\label{fig:mass_dependence_signal}
\end{figure*}

\section{Data}\label{sec:data}

\subsection{Intensity data}

We analyze Compton-$y$ maps constructed by applying a {NILC} pipeline to the single-frequency maps from the  \textit{Planck} \texttt{NPIPE} data release (PR4)~\cite{2020A&A...643A..42P}.  Our baseline data set includes all maps from 30 to 545 GHz.  We compare various choices of component deprojection in these maps in order to minimize residual CIB contamination. We discuss the $y$-maps and our needlet ILC pipeline in detail in Paper I.

\subsection{CMB lensing data}\label{sec:cmblensing_data}

Maps of the CMB lensing potential $\phi$ (or convergence $\kappa$) can be reconstructed from CMB temperature and polarization anisotropy maps by taking advantage of the well-theoretically-understood non-Gaussianities and statistical anisotropies induced in an underlying (assumed Gaussian and statistically isotropic) primary CMB map by the intervening gravitational potential. Optical quadratic estimators (QEs)~\cite{2003PhRvD..67h3002O, 2021PhRvD.103h3524M} have been derived for such an operation. 
The lowest-noise estimation of $\kappa$ from \textit{Planck} data used the globally optimal estimator  (``generalized minimum variance (GMV) QE'') of Ref.~\cite{2021PhRvD.103h3524M} applied to the PR4 \textit{Planck} \texttt{NPIPE} maps~\cite{2022JCAP...09..039C}; previous measurements used the minimum-variance (MV) QE of Ref.~\cite{2003PhRvD..67h3002O} (the Hu--Okamoto QE).

This lowest-noise \texttt{NPIPE} measurement is not necessarily the optimal one for our analysis, as the lensing analysis as part of the 2018 \textit{Planck} data release (PR3)~\cite{2020A&A...641A...8P} provides $\kappa$ estimations with various analysis choices that are more suitable for a robust measurement of  $C_\ell^{y\kappa}$. We consider the following $\kappa$ maps for our $C_\ell^{y\kappa}$ measurement:
\begin{enumerate}

\item The $\kappa$ reconstruction from the 2018 \textit{Planck} release which applied the Hu--Okamoto QE to tSZ-deprojected CMB maps to reconstruct a $\kappa$ map free of any bias due to residual tSZ signal\footnote{This map is available at \url{http://pla.esac.esa.int/pla/aio/product-action?COSMOLOGY.FILE_ID=COM_Lensing-Szdeproj_4096_R3.00.tgz
}.}. As any $C_{\ell}^{y \hat \kappa}$ measurement using a QE is in fact a $\langle T T T \rangle$ three-point function (with one $T$ leg coming from the $y$ estimate and two coming from the quadratic-in-$T$ $\hat \kappa$ estimate), the deprojection of $y$ in the temperature map used to estimate $\hat \kappa$ avoids any potential $\langle yyy \rangle$ contribution from the bispectrum of the Compton-$y$ field in the $C_{\ell}^{y \hat \kappa}$ measurement (or mixed bispectra of the form $\langle y y T_{\rm CIB} \rangle$ or $\langle y y T_{\rm radio} \rangle$).\footnote{Of course, $C_{\ell}^{y \hat \kappa}$ also includes contributions from polarization legs in the lensing reconstruction, e.g., $\langle T E B\rangle$, $\langle T T E \rangle$, etc.}

\item The $\kappa$ reconstruction from the 2018 \textit{Planck} release without the tSZ-selected galaxy cluster mask applied.\footnote{This map is available at \url{http://pla.esac.esa.int/pla/aio/product-action?COSMOLOGY.FILE_ID=COM_Lensing-Sz_4096_R3.00.tgz}.} It is standard to provide $\kappa$ maps with analysis masks that mask out these clusters, as the lensing reconstruction can become biased at these locations.  Measurement of the $C_\ell^{y\kappa}$ signal using such a mask, which is  highly correlated with the tSZ signal, would bias the signal low (although most of the tSZ -- CMB lensing cross-correlation comes from much lower halo masses than those found in such a catalog, as shown in Figure~\ref{fig:mass_dependence_signal}).

\item The lowest-noise estimation of $\kappa$ from the \texttt{NPIPE} \textit{Planck} data.\footnote{This map is available on NERSC at \texttt{\$CFS/cmb/data/planck2020/PR4\_lensing/PR4\_klm\_dat\_p.fits}.}
\end{enumerate}
We measure $C_\ell^{y\kappa}$ separately using all three of the above $\kappa$ reconstructions from \textit{Planck} (\texttt{NPIPE}; tSZ-deprojected; tSZ-clusters-unmasked) and compare the signal and the signal-to-noise ratio (SNR) in every case. We find broadly consistent signals with all maps, with the \texttt{NPIPE} measurement having the highest SNR. However, we find below that the tSZ-deprojected measurement is {slightly lower},  {by $\approx1\sigma$}.\footnote{{Ideally, we would additionally make the measurement with a polarization-only $\kappa$ map, which is immune to many of the extragalactic foregrounds that are present in the temperature-based lensing reconstruction. However, this reconstruction (for \textit{Planck}) is too noisy to extract any meaningful information in the tSZ cross-correlation studied here.}}

\section{Foreground Mitigation}\label{sec:ilc_ymap}

In our Paper I we present a set of Compton-$y$ maps with various deprojection choices to remove other components from the final $y$-map.  There are, in particular, (at least) two components that must be removed in order to make an unbiased detection of the $\left<y\kappa\right>$ signal: the integrated Sachs--Wolfe (ISW) signal~\cite{1967ApJ...147...73S} signal and the cosmic infrared background (CIB) signal.  These two components are present in the mm-wave intensity maps that we use to construct our $y$-maps and are correlated with large-scale structure (and thus with $\kappa$).  Thus, they can induce biases in our $C_\ell^{y\kappa}$ measurement if they are not sufficiently removed from the $y$-map prior to the cross-correlation measurement.  We describe these signals and the associated biases below, and our techniques to mitigate them.

\subsection{Mitigating the ISW biases}

The ISW signal~\cite{1967ApJ...147...73S} is sourced when primary CMB photons travel through a time-varying gravitational potential. In a constant potential, a photon will not gain or lose energy as it enters and leaves the potential. However, if the potential evolves while the photon passes through it, it can result in the photon requiring less or more energy to leave the potential than it gained entering the potential. This effect generates anisotropies in the CMB during the periods in the history of the Universe when potentials were time-varying. During matter domination, potentials are constant.  However, during the more recent dark energy domination, potentials have been decaying, and so there is an ISW signal imprinted on the primary CMB, which is most significant on large angular scales as these correspond to the projection of late-time linear modes. Note that the ISW effect preserves the blackbody spectrum of the primary CMB photons.

The ISW-$\kappa$ cross-correlation on large scales is dominant (or comparable) to the tSZ-$\kappa$ signal up to $\ell\sim100-200$ (see, e.g., Fig.~1 of~\cite{2018PhRvD..98h3542H}) at typical tSZ-sensitive frequencies (e.g., $\approx 100$ GHz), so it is important to deproject the primary CMB on these scales when building the $y$-map for use in our tSZ -- CMB lensing cross-correlation measurement.  Fortunately, such a deprojection is simple due to the well-known blackbody SED of the primary CMB anisotropies. Ideally, we would deproject the CMB everywhere in our NILC $y$-map, but we will find that we do not have enough frequency coverage to deproject the CMB and the CIB along with its first moments on small scales. As such, we only deproject the CMB in the first five needlet scales in our $y$-map; we will refer to this as CMB${}^5$-deprojection.

\subsection{Mitigating the CIB biases}

\subsubsection{CIB mitigation: constrained ILC}

The CIB is highly correlated with the CMB lensing signal (e.g.,~\cite{2014A&A...571A..18P}), and it is imperative to remove this signal from the $y$-map before measuring $C_\ell^{y\kappa}$. Thus, we need to deproject the CIB using an estimate of its frequency dependence.  However, unlike the tSZ and the CMB signals, which display no frequency decorrelation and which have SEDs that are well-understood from first principles and can thus be calculated theoretically, the CIB is not described perfectly by one SED. It is sourced by the line-of-sight-integrated thermal emission of different objects; in particular, different frequency channels are sensitive to slightly different objects, as source emission at different redshifts will be redshifted into different frequency bands. This leads to frequency decorrelation between the CIB channels.  However, as the correlation coefficients are $\gtrsim 90$\% at the frequencies of interest~\cite{2013ApJ...772...77V,2014A&A...571A..30P,2017MNRAS.466..286M,2019ApJ...883...75L}, it is still possible to clean the CIB using multifrequency measurements. Its SED does not need to be known or modeled in order for the CIB to be (partially) cleaned in a standard, unconstrained ILC.  However, if we wish to explicitly deproject it, we must model its SED.  We model the CIB SED as a modified blackbody:

\be
 \Theta_\nu^{CIB}=\nu^\beta B_\nu(T^{\mathrm{eff}}_{\mathrm{CIB}})\label{CIB_SED}
\ee
where $B_\nu(T)$ is the Planck function
\be
B_\nu(T) = \frac{2 h \nu^3}{c^2}\frac{1}{e^{\frac{h \nu}{k_BT}}-1} \,.
\label{mbb}
\ee
The SED $\Theta_\nu^{CIB}$ depends on two parameters: the spectral index $\beta$ and the dust temperature $T^{\mathrm{eff}}_{\mathrm{CIB}}$. In Paper I, we fit this SED to estimates of the CIB monopole, and find best-fit parameters of $\beta=1.77,\,T_{\mathrm{CIB}}^{\mathrm{eff}}=10.14\,\mathrm{K}$. However, there is significant uncertainty on these parameters; additionally, it is not a perfect approximation to the CIB emission to describe it as an exact modified blackbody. Thus, in addition to deprojecting the CIB, we also deproject the first moment(s) of the CIB with respect to these parameters.  We describe this approach in detail in Paper I; in short, it involves additionally deprojecting components whose SEDs are exactly defined by the derivative with respect to the appropriate parameter of the modified blackbody SED. We refer to these components as $\delta\beta$ (for the $\beta$-moment) and $\delta T^{\mathrm{eff}}_{\mathrm{CIB}}$ (for the $T^{\mathrm{eff}}_{\mathrm{CIB}}$ moment).

\subsubsection{CIB mitigation: decreased frequency coverage}

We note that, while using more frequency channels in the {NILC} allows for better characterization of the spatial structures of foregrounds and lower variance in the final map, it also requires better characterization of the CIB SED in order to deproject it. Thus, we explore whether using fewer frequencies than standard can lead to an improved CIB deprojection, in particular by dropping the 545 GHz information from the {NILC}. {We additionally always leave out the 857 GHz channel, both because it is not calibrated as precisely as the other channels, and because it requires extrapolation of the CIB modified blackbody SED to higher frequencies than we have been using to constrain it.}

\subsection{Analysis masks}\label{sec:fiducialmask}

After performing the NILC analysis to build the $y$-map, non-negligible residual foreground power remains in the dustiest areas of the sky (e.g., close to the Galactic plane), especially in the $\delta\beta$-deprojected tSZ map. This can add significant variance to our measurement. To avoid this, we mask the dustiest areas of the sky, which we define by thresholding the highest $(1-f_{\mathrm{sky}})\times 100 \%$ of pixels in the \textit{Planck} 857 GHz single-frequency map, where $f_{\mathrm{sky}}$ is the fraction of sky on which we wish to make our measurement. We use $f_{\mathrm{sky}}=0.8$ as our fiducial choice; we explore effects of varying the masked sky area in Appendix~\ref{app:skyarea}.

Additionally, the reconstructed CMB lensing maps from \textit{Planck} are provided with analysis masks applied. We multiply our $f_{\mathrm{sky}}$-thresholded mask by this analysis mask. {Finally, we also multiply this mask with the union of the \textit{Planck} HFI and LFI point source masks, which were used in the preprocessing of the single-frequency  maps before performing the NILC~\cite{Paper1}(as a result, the resulting maps contain no information at the locations of these point sources).} In principle, this slightly suppresses the Compton-$y$ signal, due to the tSZ--source correlation, but this effect is negligible at \emph{Planck} signal-to-noise~\cite{2022JCAP...08..029L,Omori2022}.  {We apodize the resulting mask with an apodization scale of 10 arcminutes, using the ``\texttt{C1}'' apodization procedure of \texttt{NaMaster}~\cite{2019MNRAS.484.4127A}\footnote{\url{https://github.com/LSSTDESC/NaMaster}}.} {Most of the sky area that is covered by the $f_{\mathrm{sky}}=80\%$ threshold mask is indeed covered by the $\kappa$ analysis masks, such that our overall masks are essentially limited by the size of these $\kappa$ masks.}

Note that, as is standard, the $\kappa$ analysis masks remove regions containing massive tSZ-selected clusters~\cite{2022JCAP...09..039C}, and so the use of these masks has the risk of biasing our tSZ -- CMB lensing cross-correlation signal low. To avoid this, it is desirable to use a $\kappa$ map which does not mask these clusters, and so we perform a measurement with a \textit{Planck} 2018 $\kappa$ reconstruction in which these clusters are not masked.  However, as discussed in Section~\ref{sec:cmblensing_data}, we also perform a measurement using the \textit{Planck} 2018 $\kappa$ reconstruction performed on tSZ-deprojected temperature maps, which avoids a potential $\left<yyy\right>$ intrinsic bispectrum bias in our measurement, as well as on the $\kappa$ reconstruction from the \texttt{NPIPE} maps, which has the lowest noise. These final two $\kappa$ reconstructions are provided with analysis masks that remove tSZ clusters, meaning that there is potential for the signal to be biased low.  However, we find that the measurements with the three $\kappa$ reconstructions are all statistically consistent, indicating that any low bias due to the masking of tSZ clusters is negligible.

{The relevant masks are shown in Figure~\ref{fig:analysis_masks}. The combination of the $f_{\mathrm{sky}}=0.8$ threshold mask and the lensing analysis mask does not remove a large sky area, with the remaining sky fraction decreasing from $80\%$ to $61.57\%$ for the tSZ-deprojected $\kappa$ analysis mask; $62.74\%$ for the tSZ-clusters-restored mask; and $62.24\%$ for the $\texttt{NPIPE}$ mask.}

\begin{figure*}
\includegraphics[width=0.32\textwidth]{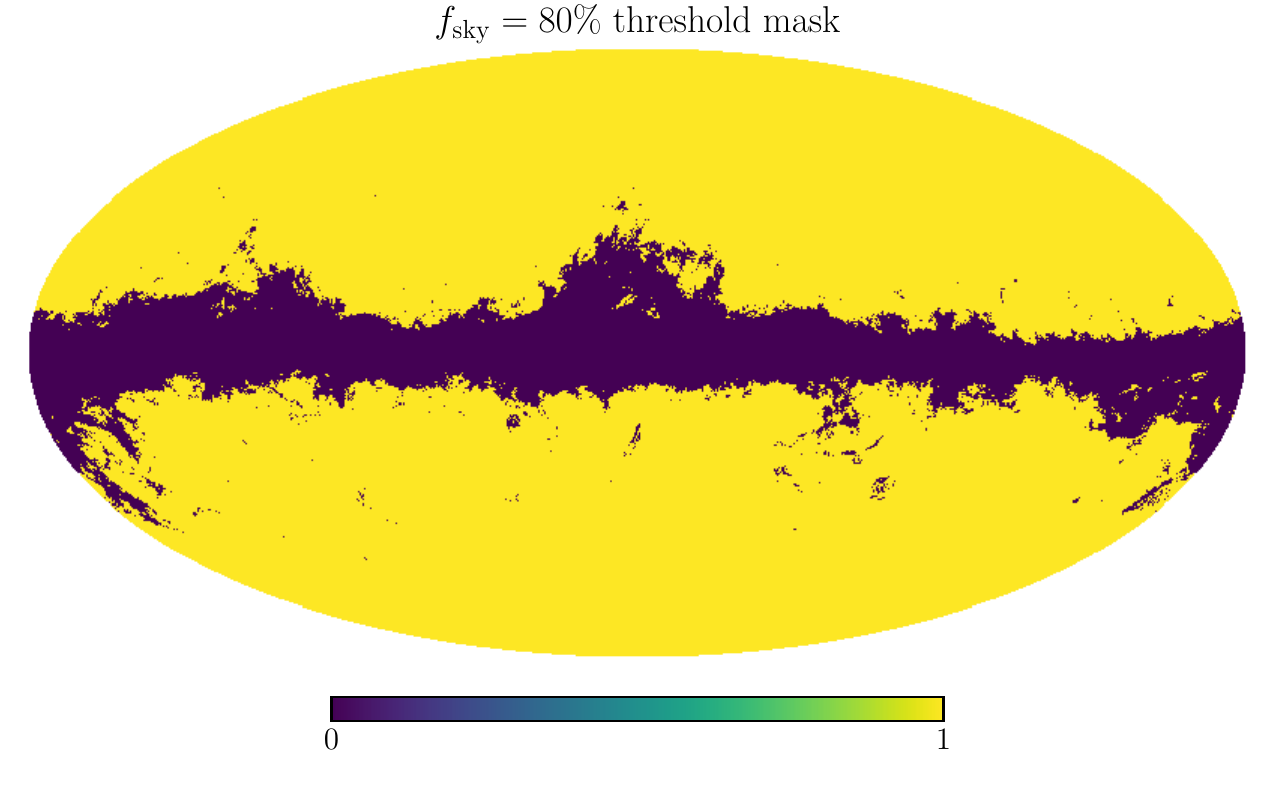}
\includegraphics[width=0.32\textwidth]{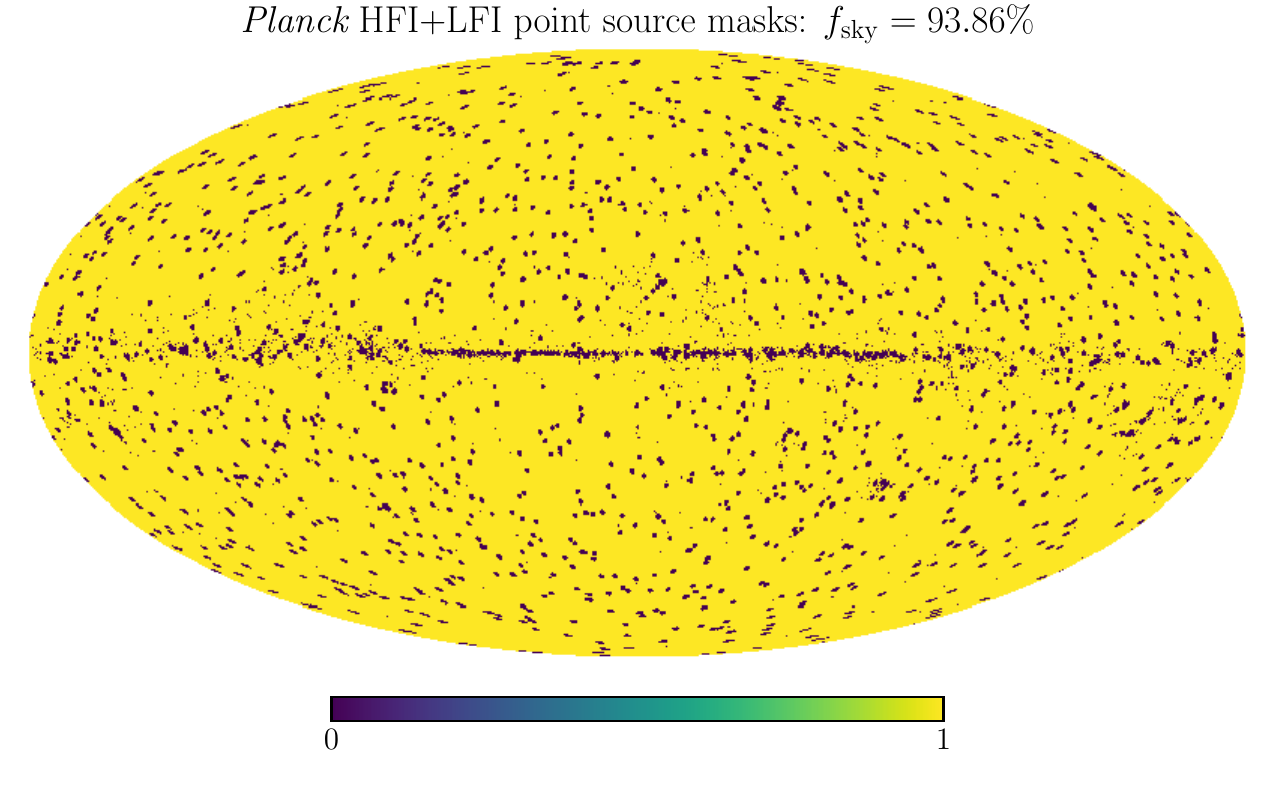}
\\
\includegraphics[width=0.32\textwidth]{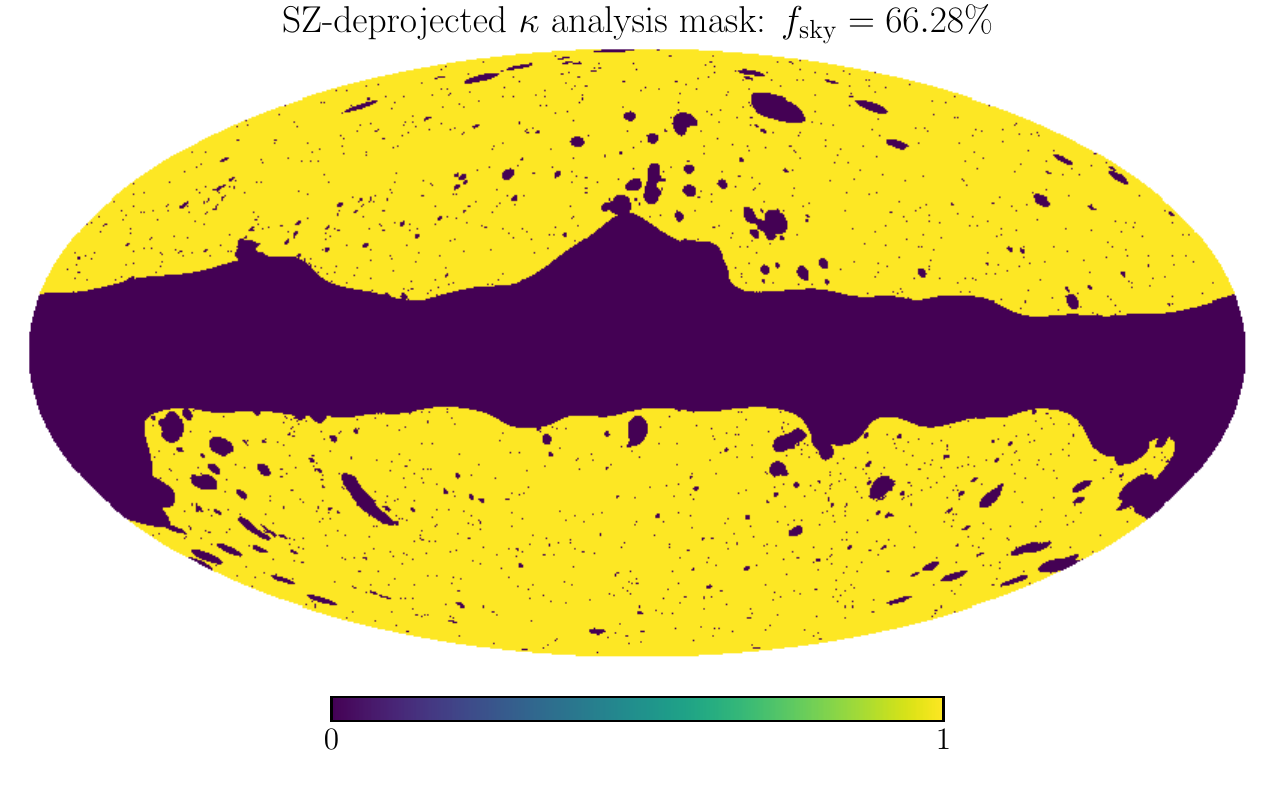}
\includegraphics[width=0.32\textwidth]{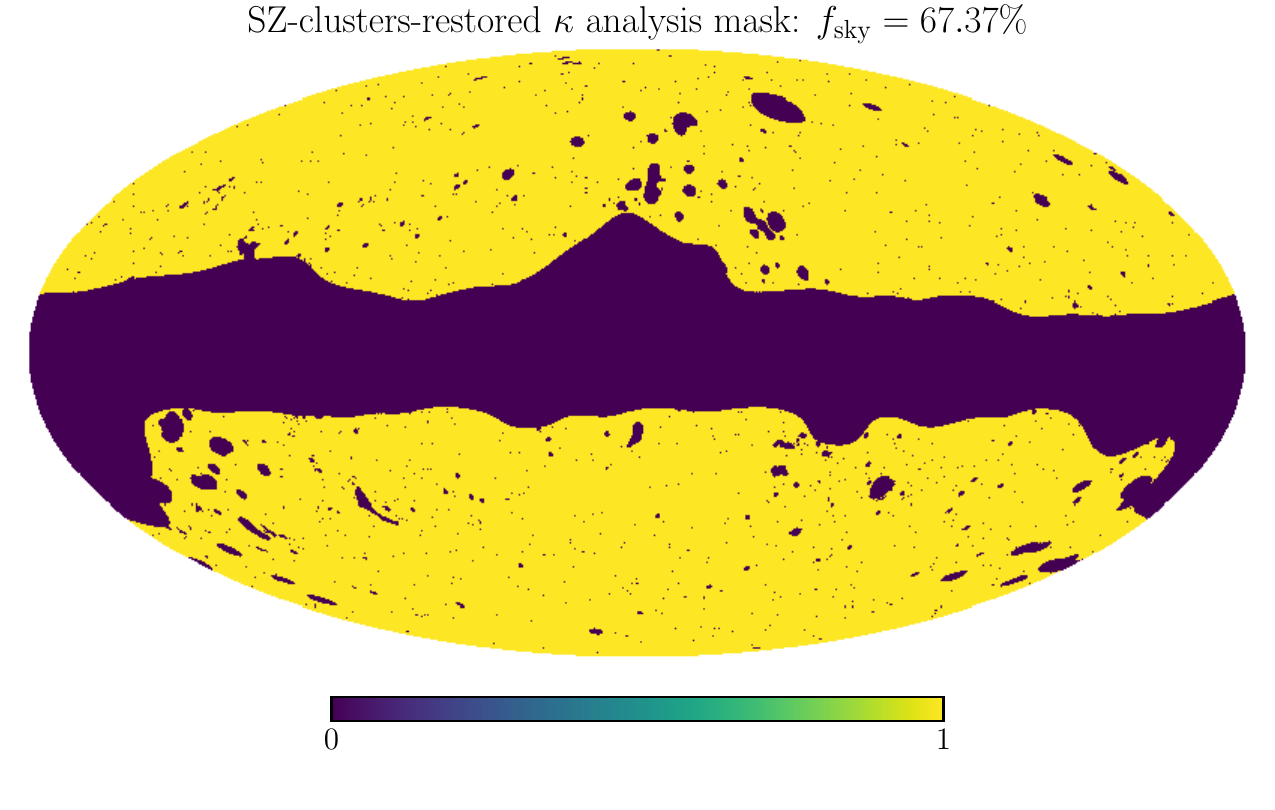}
\includegraphics[width=0.32\textwidth]{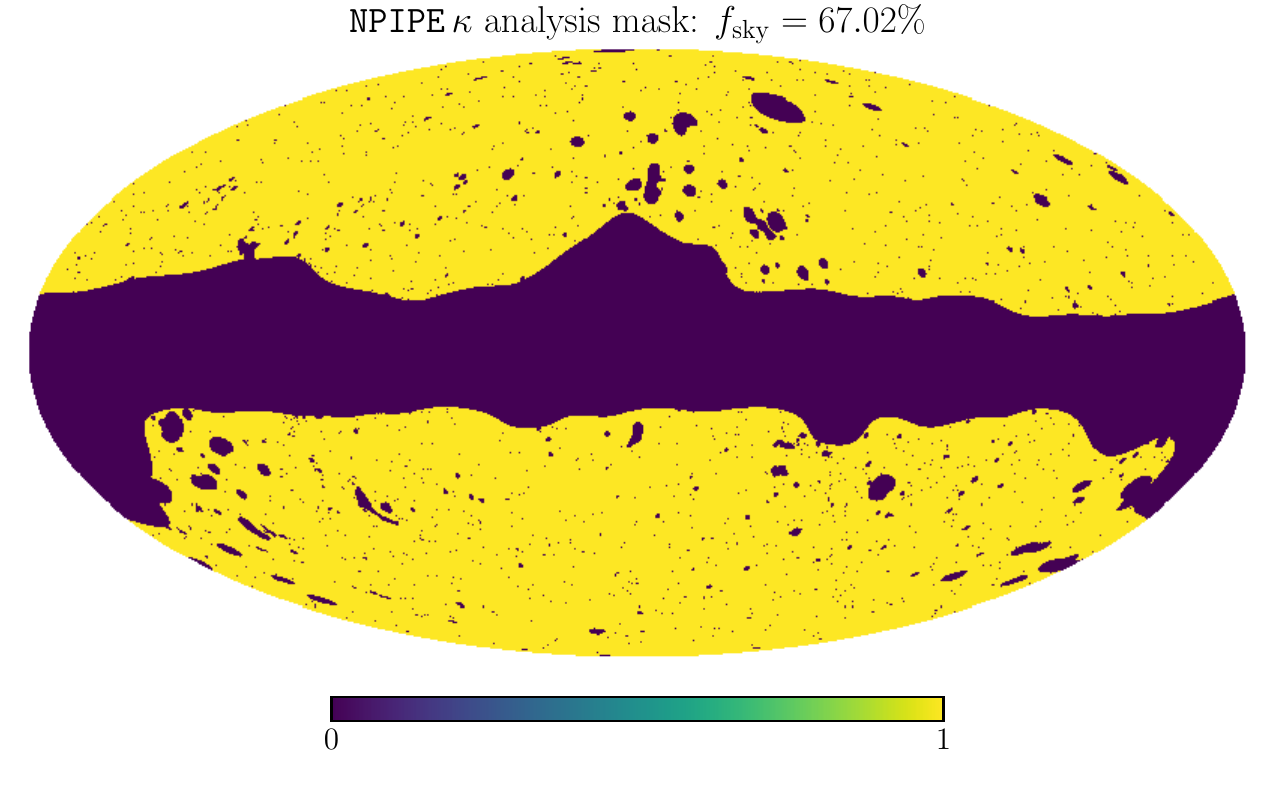}
\includegraphics[width=0.32\textwidth]{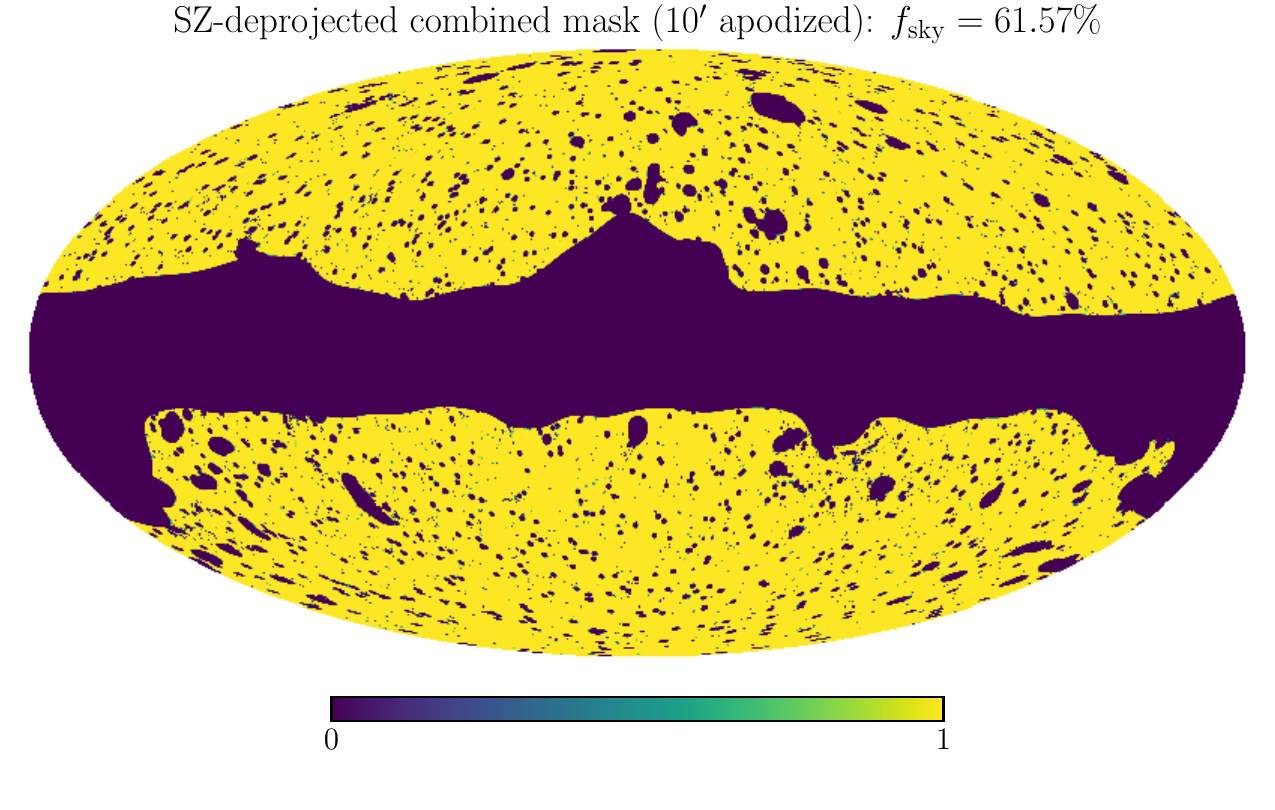}
\includegraphics[width=0.32\textwidth]{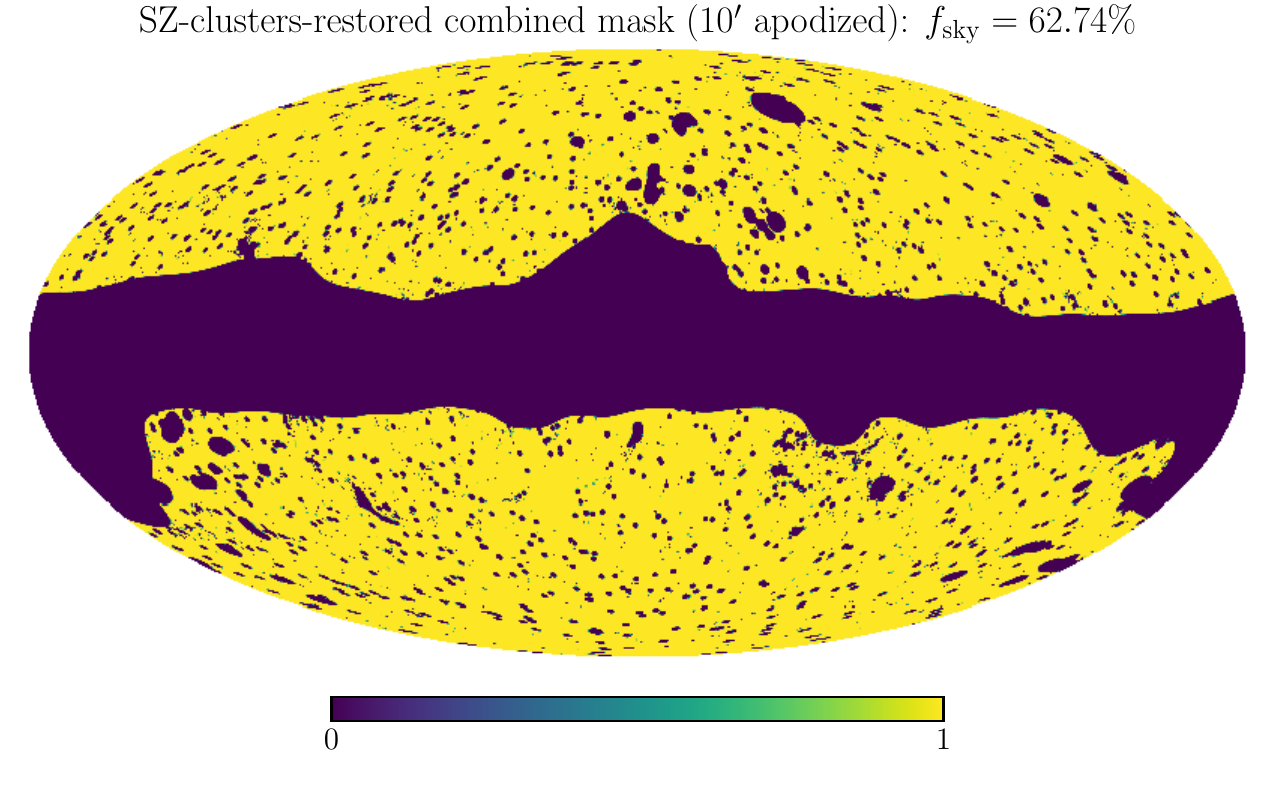}
\includegraphics[width=0.32\textwidth]{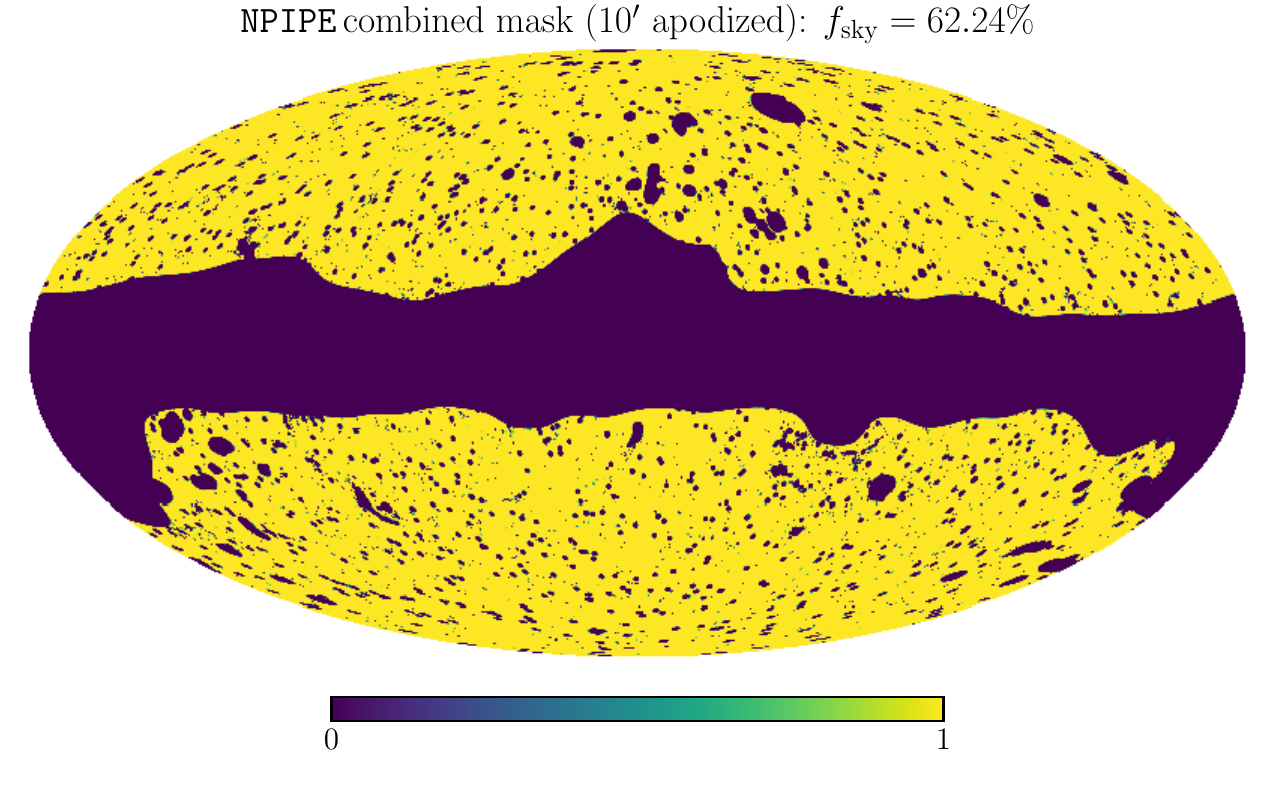}

\caption{The various masks used in our analysis; the top two rows include the unapodized masks that we multiply to make the ``combined'' masks in the bottom row. Every ``combined'' mask is made of the product of the $f_{\mathrm{sky}}=80\%$ threshold mask with the point source mask and one of the lensing masks; the final product has a $10^\prime$ apodization applied as described in the text.}\label{fig:analysis_masks}
\end{figure*}

\section{Thermal SZ -- CMB lensing cross-correlation measurement}\label{sec:clyk_meas}

In this section, we present our measurement of the $C_\ell^{y\kappa}$ signal. In particular, we describe in Section~\ref{sec:pow_error_yk} our estimate of the cross-correlation data points and associated covariance. In Section~\ref{sec:CIB_systematics} we discuss systematics induced by the CIB, in particular by comparing the data points as measured using different assumptions for the parameters of the CIB SED in the deprojection scheme used in the $y$-map construction {(this analysis provides an explicit example of how to use the CIB-cleaned $y$-maps of Paper I to remove CIB residual bias)}.  In Section~\ref{sec:different_kappas} we show our final results for the different $\kappa$ maps we use.

\subsection{Power spectrum measurement and covariance estimation}\label{sec:pow_error_yk}

We measure the cross-power spectrum $C_\ell^{y\kappa}$ between the tSZ maps and the CMB lensing convergence maps using \texttt{NaMaster}. We make the measurement in linearly spaced multipole bins with $\Delta\ell=125$, using a minimum $\ell$ of $\ell_{\mathrm{min}}=10$. The maximum multipole that we use is $\ell_{\mathrm{max}}=2010$.  {Thus we use 16 $\ell$ bins centered on $\ell=\{72,  197,  322,  447,  572,  697,  822,  947, 1072,
       1197, 1322, 1447, 1572, 1697, 1822, 1947\}$. }

Following from the standard expression for the covariance in the measurement of a cross-power spectrum of Gaussian fields,
\be
\mathcal C( \hat C^{\alpha\beta }_\ell,\hat C^{\gamma\delta }_{\ell^\prime})= \frac{\delta_{\ell \ell^\prime}}{(2\ell+1 ) f_{\mathrm{sky}}}\lb\lb C_\ell^{\alpha\gamma} + N_\ell^{\alpha\gamma}\rb\lb C_\ell^{\beta\delta} + N_\ell^{\beta\delta}\rb+\lb C_\ell^{\alpha\delta} + N_\ell^{\alpha\delta}\rb\lb C_\ell^{\beta\gamma} + N_\ell^{\beta\gamma}\rb\rb \,
\label{general_cov_matrix_cell}
\ee
{(where $\delta_{\ell \ell^\prime}$ is the Kronecker delta function)},
the error bars on the power spectrum $\hat C_\ell^{y\kappa}$ measured on a fraction of the sky $ f_{\mathrm{sky}}$ are given by
\begin{widetext}
\be
\sigma^2(\hat C_\ell^{y\kappa}) = \frac{\delta_{\ell \ell^\prime}}{ \lb2\ell+1\rb f_{\mathrm{sky}}}\left(\left(C_\ell^{yy}+N_\ell^{yy}\right)\left(C_\ell^{\kappa\kappa}+N_\ell^{\kappa\kappa}\right)+\left(C_\ell^{y\kappa}+N_\ell^{y\kappa}\right)^2\right) \,,
\label{eq.ykerror}
\ee
\end{widetext}
where $C_\ell^{AB}$ are the true underlying power spectra and $N_\ell^{AB}$ are the noise power spectra (including instrumental noise and residual foregrounds).  In our case, we use \textit{measured} auto-power spectra $\hat C_\ell^{yy}$ and $\hat C_\ell^{\kappa\kappa}$ to estimate $C_\ell^{yy}+N_\ell^{yy}$ and $C_\ell^{\kappa\kappa}+N_\ell^{\kappa\kappa}$ respectively,  we use our fiducial theoretical model (see Section~\ref{sec:theory}) to calculate the (sub-dominant) $C_\ell^{y\kappa}$ term, and we assume $N_\ell^{y\kappa}=0$, i.e., that all sources of noise bias and foreground power on this measurement are zero (this assumes that any Galactic foregrounds that remain in the tSZ map do not appear in the $\kappa$ reconstruction and that the \emph{Planck} instrument noise has vanishing skewness). {These assumptions are } robust due to the fact that the noise power spectra of the $y$-map and CMB lensing map highly dominate the error bars on $C_\ell^{y\kappa}$.  Again, we use \texttt{NaMaster} to decouple the mask in the final calculation of our error bars (using the function \texttt{gaussian\_covariance()}); note that this requires us to interpolate our measured values of $\hat C^{yy}_\ell$ and $\hat C^{\kappa\kappa}_\ell$, as this function requires a theoretical estimate of $C_\ell^{XY}$ at all multipoles.  
{Note that the decoupling of the mask with \texttt{Namaster} induces off-diagonal ($\ell\ne\ell^\prime$) terms in the covariance matrix}.\footnote{The non-Gaussian covariance (i.e., the connected trispectrum of two $y$ and two $\kappa$ fields) would also induce off-diagonal elements, but we neglect this contribution as it is much smaller than the Gaussian error bars in our measurement (e.g., HS14).}

We measure $C_\ell^{y\kappa}$ using $y$-maps with various deprojection choices and frequency coverage. In Figure~\ref{fig:different_freqs_and_deproj}, we plot the data points for various deprojection choices in the NILC.  We plot separately the points for the cases when we have different frequency coverage in our {NILC} (the ``standard frequency coverage'' case corresponds to when we include all frequencies except for 857 GHz). {In general, we see an extremely strong signal in the case when we do not deproject the CIB (labeled ``CMB deprojection'') in the figures. When we deproject the CIB, the signal is significantly lowered, indicating that the non-CIB-deprojected measurement is biased.  Deprojecting further moments results in points with larger error bars but much less scatter (with respect to the chosen parameters of the CIB SED), eventually converging on a stable, robust measurement. We discuss the different deprojections below.}

\begin{figure}
    \includegraphics[width=0.49\textwidth]{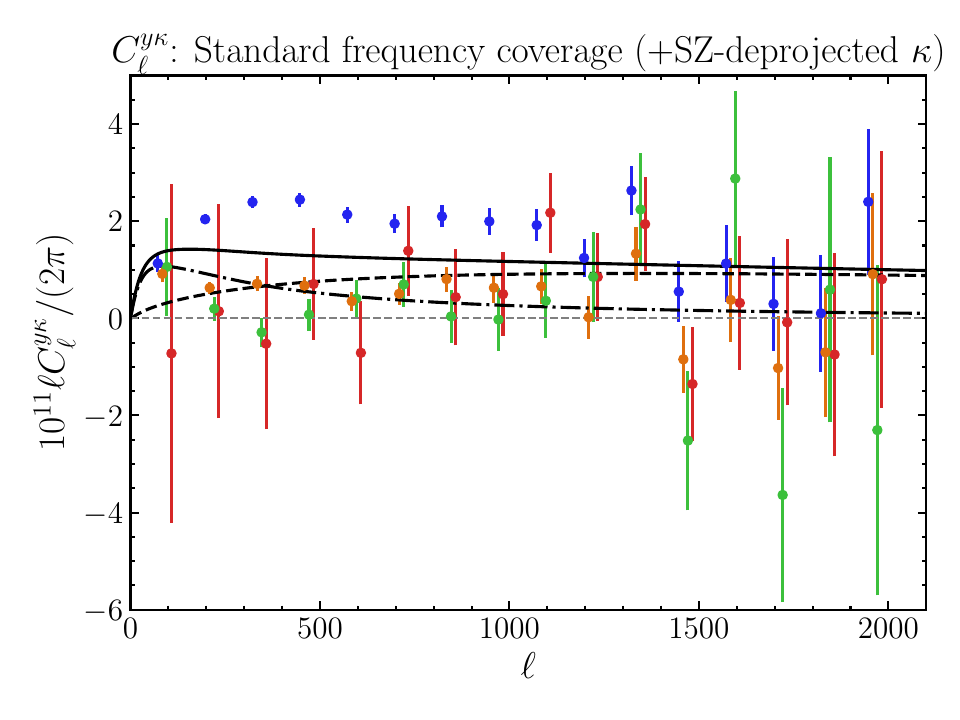}
    \includegraphics[width=0.49\textwidth]{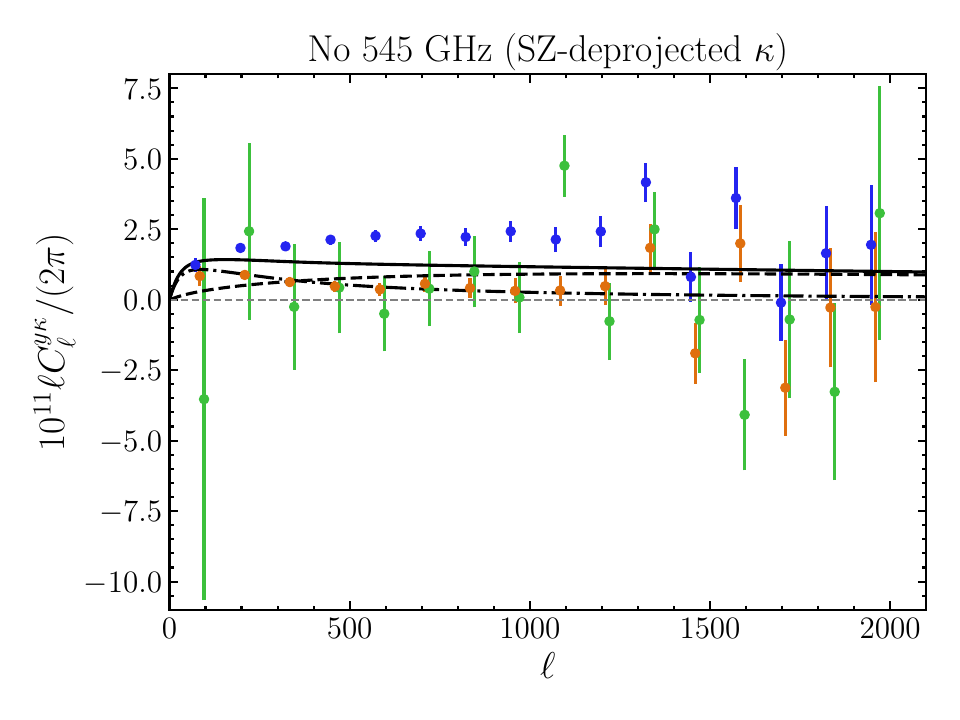}
    \includegraphics[width=0.2\textwidth]{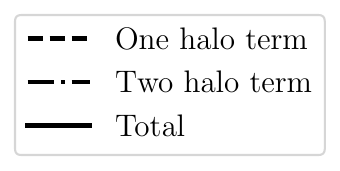}\hspace{10em}
    \includegraphics[width=0.4\textwidth]{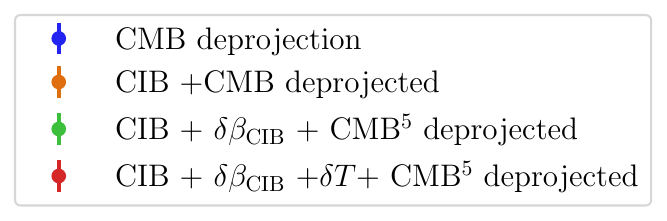}
    \caption{The tSZ -- CMB lensing cross-power spectrum, $C_\ell^{y\kappa}$, as measured for different frequency coverage (left panel including all frequencies from 30 - 545 GHz and right panel with 545 GHz removed) and different deprojection combinations (as labeled in the legend). Note that, for the case when we exclude 545 GHz, we do not have enough frequency channels to simultaneously deproject {the CIB and both} $\delta \beta_{\mathrm{CIB}}$ and $\delta T_{\mathrm{CIB}}^{\mathrm{eff}}$ on the smallest scales; it would be possible to do so on large scales, but we do not consider such a measurement.  The solid black curve shows our fiducial theoretical model, with the one-halo (two-halo) term shown in dashed (dash-dotted). {Note that, in the legend, CMB$^5$ refers to the CMB deprojected on the largest five needlet scales. {Also note that some points are slightly offset on the $x$ axis to allow for better visualization.} } {Both plots were made with the SZ-deprojected reconstruction of the $\kappa$ map.}}
    \label{fig:different_freqs_and_deproj}
\end{figure}

\subsection{Systematics due to residual CIB contamination}\label{sec:CIB_systematics}

Our first step to remove the CIB contribution to our measurement is a simple deprojection in the NILC of a component with the appropriate frequency scaling for the CIB. We discuss this deprojection in detail in Paper I, in which we directly fit a modified blackbody to the CIB monopole calculated in Ref.~\cite{2014A&A...571A..30P}, and find best-fit values for the spectral index $\beta$ and the dust temperature $T_{\mathrm{CIB}}^{\mathrm{eff}}$ of $(T_{\mathrm{CIB}}^{\mathrm{eff}},\beta)=(10.14\, \mathrm{K},1.77)$. These are the values that we use for our standard CIB deprojection, as presented in Figure~\ref{fig:different_freqs_and_deproj}.

However, there is significant uncertainty on the CIB SED, including significant parameter degeneracy between $\beta$ and $T_{\mathrm{CIB}}^{\mathrm{eff}}$, as well as uncertainties on the CIB monopole estimate to which the SED is fit, along with the fact that the CIB is not perfectly described by a modifed blackbody SED {(and that the CIB monopole is not necessarily the appropriate observable to consider in this case, as it is sourced at different redshifts and thus has different frequency dependence to the CIB anisotropies)}. Thus we also make our measurement for $y$-maps on which various values of ($\beta$,$T_{\mathrm{CIB}}^{\mathrm{eff}}$), {which are drawn from the posterior for these parameters~\cite{Paper1} (with values as indicated in the figure legend)}, have been used in the deprojection.   The resulting changes in the data points are shown in Figure~\ref{fig:deproj_CIBdifferentbeta}, where we note that we have included the primary CMB deprojection in the first five needlet scales in all cases. It is clear that there is a significant systematic uncertainty associated with the exact CIB SED used in the deprojection, as the spread of these data points is much larger than the statistical error bars.  However, we note that on small scales ($\ell\greaterthanapprox 1000$), when we drop 545 GHz from the NILC, this systematic is significantly reduced (see the right panel of Figure~\ref{fig:deproj_CIBdifferentbeta}).

\begin{figure}
\includegraphics[width=0.49\textwidth]{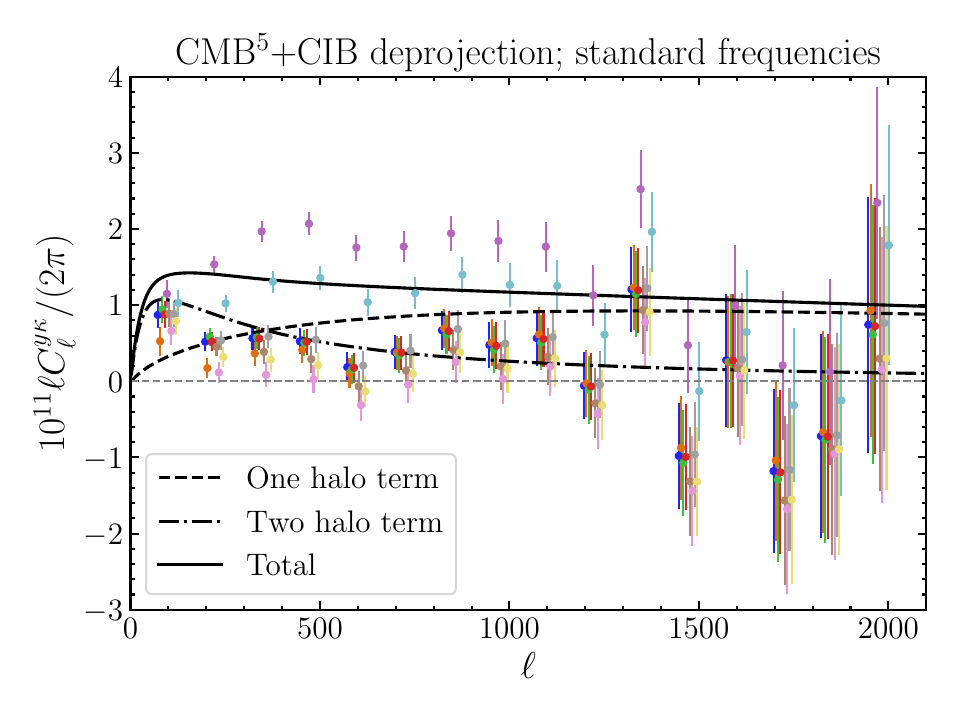}
\includegraphics[width=0.49\textwidth]{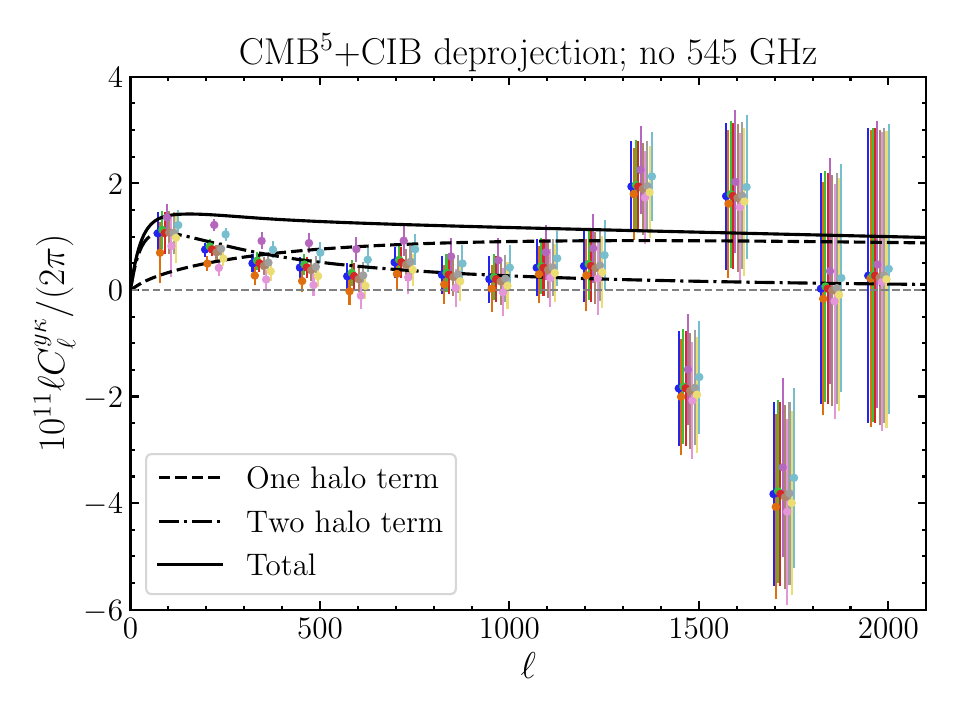}

\includegraphics[width=\textwidth]{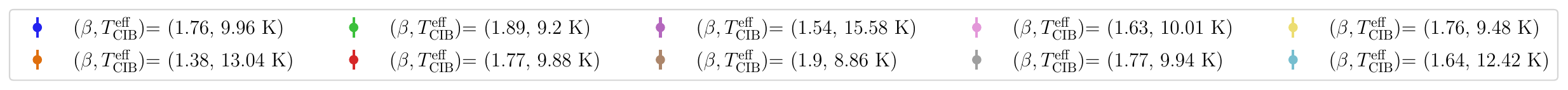}
\caption{The tSZ -- CMB lensing cross-power spectrum, $C_\ell^{y\kappa}$, measured on maps with the CIB deprojected, for various choices of the CIB SED parameters. The SED parameters used here are obtained by taking ten samples from the posterior of $(\beta,T_\mathrm{CIB}^{\mathrm{eff}})$ which we calculate in Paper I. On the left, we show the standard-frequency-coverage case (where we include 545 GHz but no 857 GHz), and on the right the case where we remove 545 GHz from the NILC. In all cases, the primary CMB is deprojected in the first five needlet scales of the NILC. {The points are systematically offset along the $x$ axis for better visibility of overlapping points.} {Both plots were made with the SZ-deprojected reconstruction of the $\kappa$ map.}}\label{fig:deproj_CIBdifferentbeta}
\end{figure}

Thus, in order to reduce the systematic uncertainty associated with the CIB SED deprojection, we proceed to measure the tSZ -- CMB lensing cross-correlation using $y$-maps from which the first moment of the CIB with respect to $\beta$ has been deprojected. We show the resulting measurements in Figure~\ref{fig:deproj_CIBdbetadifferentbeta}. We note that the scatter in the ``standard-frequency'' case is still significantly larger than the size of the statistical error bars for some multipole bins. However, in the ``no-545 GHz'' case, this behavior is no longer seen. Thus, for the ``no-545 GHz'' case, we use this deprojection configuration to make our measurement, as we have achieved stability. For the final data points that we use in our analysis, we use the deprojection done with the best-fit SED ($\beta=1.77, T_{\mathrm{CIB}}^{\mathrm{eff}}=10.14\,\mathrm{K}$.)

\begin{figure}

\includegraphics[width=0.49\textwidth]{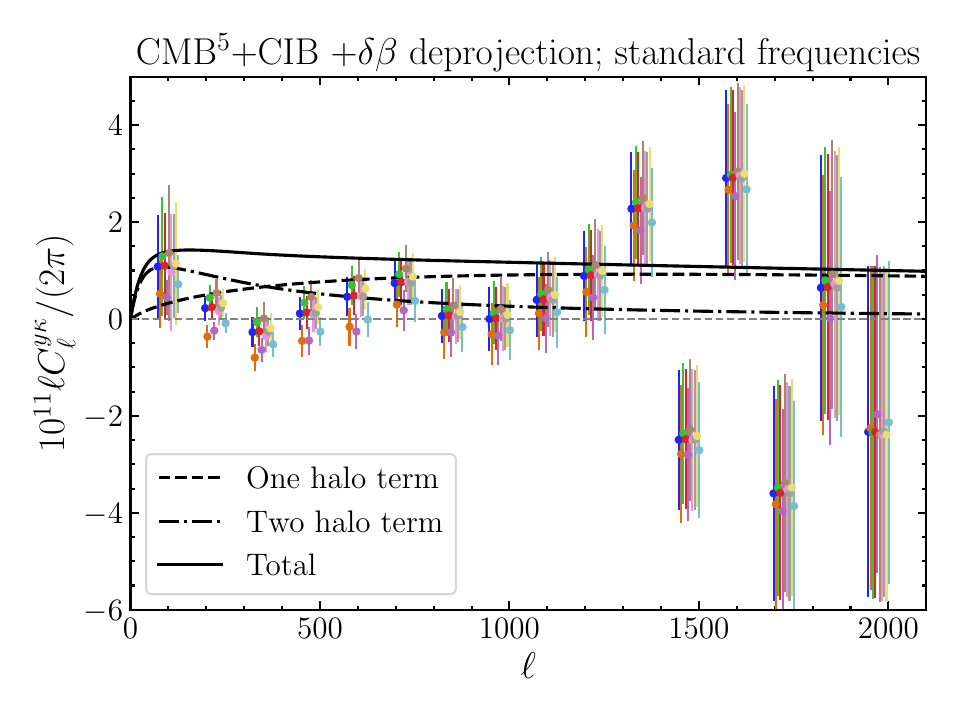}
\includegraphics[width=0.49\textwidth]{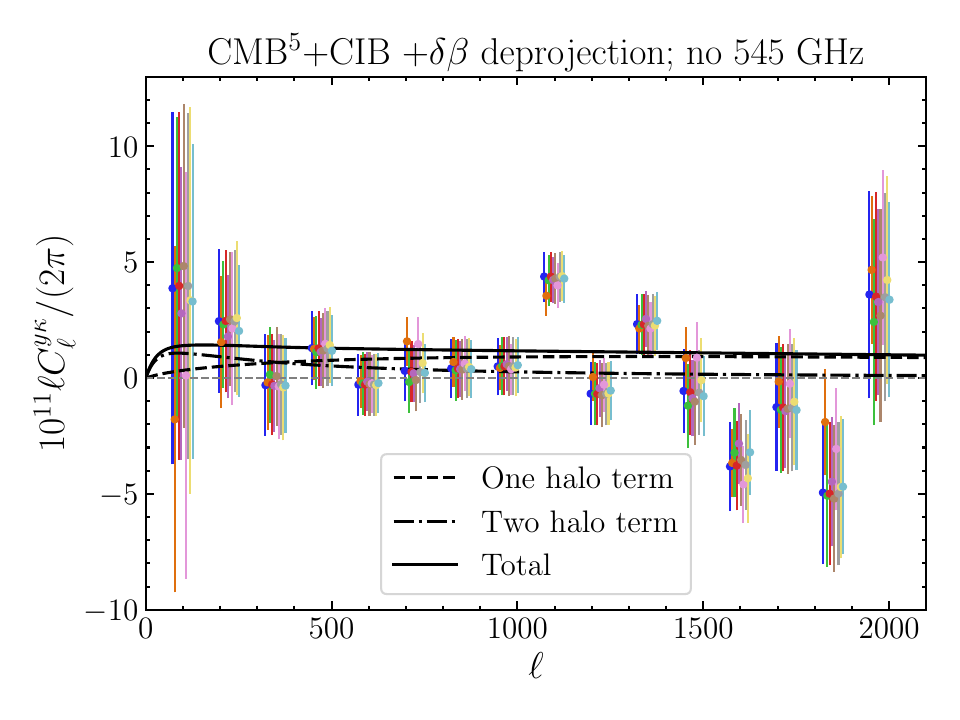}

\includegraphics[width=\textwidth]{legend_differentsamples_labelled.pdf}

\caption{The tSZ -- CMB lensing cross-power spectrum, $C_\ell^{y\kappa}$, measured on $y$-maps with the CIB and the first moment of the CIB SED with respect to $\beta$ deprojected, for various choices of the CIB SED parameters. The SED parameters are the same as those in Figure~\ref{fig:deproj_CIBdifferentbeta}.  On the left, we show the standard-frequency-coverage case (where we include 545 GHz but no 857 GHz), and on the right the case where we remove 545 GHz from the NILC.  In all cases, the primary CMB is deprojected in the first five needlet scales of the NILC. {The points are systematically offset along the $x$ axis for better visibility of overlapping points.} {Both plots were made with the SZ-deprojected reconstruction of the $\kappa$ map.}}\label{fig:deproj_CIBdbetadifferentbeta}
\end{figure}

However, for the standard-frequency case it still remains to find a stable configuration for which we can proceed. Thus, we additionally deproject the first moment of the CIB SED with respect to $T_{\mathrm {CIB}}^{\mathrm{eff}}$. The resulting points are shown in Figure~\ref{fig:deproj_CIBdbetadTdifferentbeta}. {These points are stable, i.e., the variation with the CIB SED used in the deprojection is much less than the statistical error bars for all multipole bins.  For the final analysis we use the best-fit CIB SED from Paper I to perform the deprojections.}

\begin{figure}
\includegraphics[width=0.49\textwidth]{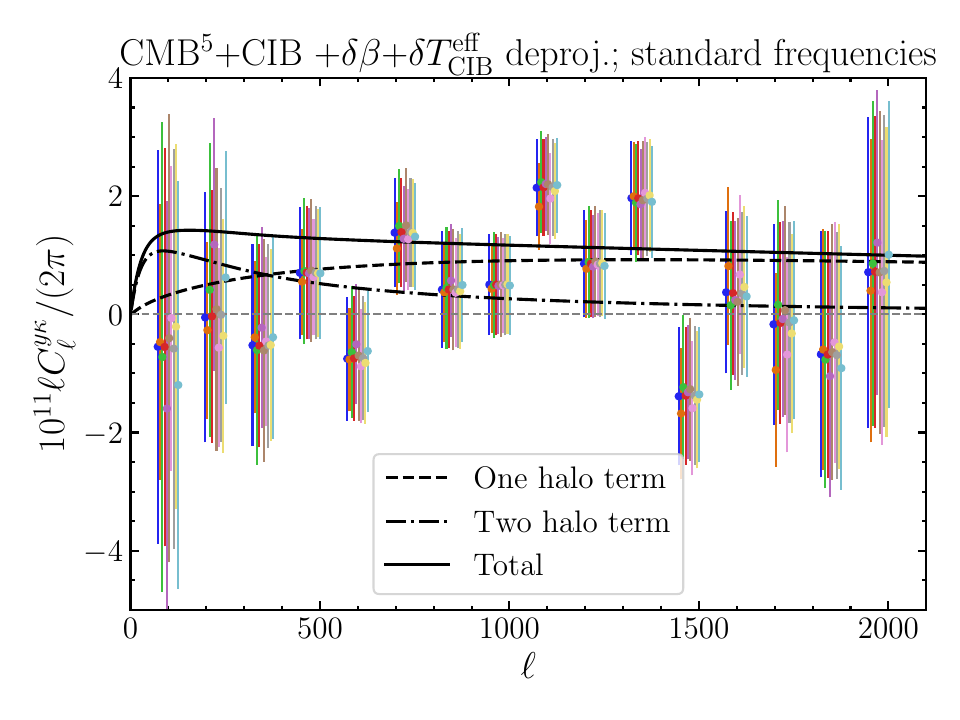}

\includegraphics[width=\textwidth]{legend_differentsamples_labelled.pdf}

\caption{The tSZ -- CMB lensing cross-power spectrum, $C_\ell^{y\kappa}$, measured on $y$-maps with the CIB deprojected as well as its first moments with respect to both $\beta$ and $T_{\rm CIB}^{\rm eff}$, for various choices of the CIB SED parameters. The SED parameters are the same as those in Figure~\ref{fig:deproj_CIBdifferentbeta}. In all cases, the primary CMB is deprojected in the first five needlet scales of the NILC. {The points are systematically offset along the $x$ axis for better visibility of overlapping points.} {This plot was made with the SZ-deprojected reconstruction of the $\kappa$ map.}
}\label{fig:deproj_CIBdbetadTdifferentbeta}
\end{figure}

\subsubsection*{Summary of the CIB removal and deprojection choices}

{To summarize, our method for removing the CIB residual bias is as follows:}

\begin{enumerate}
\item Make a standard, non-deprojected (or CMB-deprojected) NILC $y$-map with a given frequency coverage and measure the observable ($C_\ell^{y\kappa}$).

\item Make a CIB-deprojected (or CMB+CIB-deprojected) NILC map for various choices of the CIB SED and measure the observable. If the measurement is stable to the choice of CIB SED, use this measurement. If it is not, move on to Step 3.

\item Make a CIB$+\delta\beta$-deprojected (or CMB+CIB$+\delta\beta$-deprojected) NILC $y$-map for various choices of the CIB SED and measure the observable. If the measurement is stable to the choice of CIB SED, use this measurement. If it is not, move on to Step 4.

\item Make a CIB$+\delta\beta+\delta T^{\mathrm{eff}}_{\mathrm{CIB}}$-deprojected (or CMB+CIB$+\delta\beta+\delta T^{\mathrm{eff}}_{\mathrm{CIB}}$-deprojected) NILC $y$-map for various choices of the CIB SED and measure the observable. If the measurement is stable to the choice of CIB SED, use this measurement. If not, continue deprojecting higher moments of the CIB SED until stability is found.
\end{enumerate}

{For the standard-frequency-coverage case, we obtain stability in our measurement after Step 4 (see Figure~\ref{fig:deproj_CIBdbetadTdifferentbeta}). For the no-545-GHz case, we obtain stability after Step 3 (see Figure~\ref{fig:deproj_CIBdbetadifferentbeta}, right-hand side).

\subsection{Results for the different $\kappa$ maps }\label{sec:different_kappas}

We also explore the changes in our measurement when we use different $\kappa$ maps. Using the CMB$^5$+CIB+$\delta\beta$-deprojected no-545 GHz $y$-map and the CMB$^5$+CIB+$\delta\beta$+$\delta T_{\mathrm{CIB}}^{\mathrm{eff}}$-deprojected standard-frequency $y$-map  described above, we show in Figure~\ref{fig:different_kappas_measurement} the signal as measured with different $\kappa$ maps.  These include the $\kappa$ map released with an analysis mask that restores the signal at the location of tSZ clusters; the $\kappa$ map reconstructed from tSZ-deprojected temperature maps; and the $\kappa$ map reconstructed from the \textit{Planck} PR4 (\texttt{NPIPE}) map. It is encouraging that our measurements with different $\kappa$ reconstructions look broadly consistent, as they have different sensitivity to systematics (in particular, the tSZ-deprojected map should not suffer from a potential $\langle yyy\rangle$ bias from spurious tSZ signal in the $\kappa$ map).

{While it is encouraging that our measurements are broadly consistent, we
continue to use all three $
\kappa$ maps to present our results, although for ease of presentation in Figures~\ref{fig:different_freqs_and_deproj},~\ref{fig:deproj_CIBdifferentbeta},~\ref{fig:deproj_CIBdbetadifferentbeta}, and ~\ref{fig:deproj_CIBdbetadTdifferentbeta} we have chosen the tSZ-deprojected $
\kappa$ map for the plots. The corresponding results with the different $
\kappa$ maps are similar.}

\begin{figure}
\includegraphics[width=0.49\textwidth]{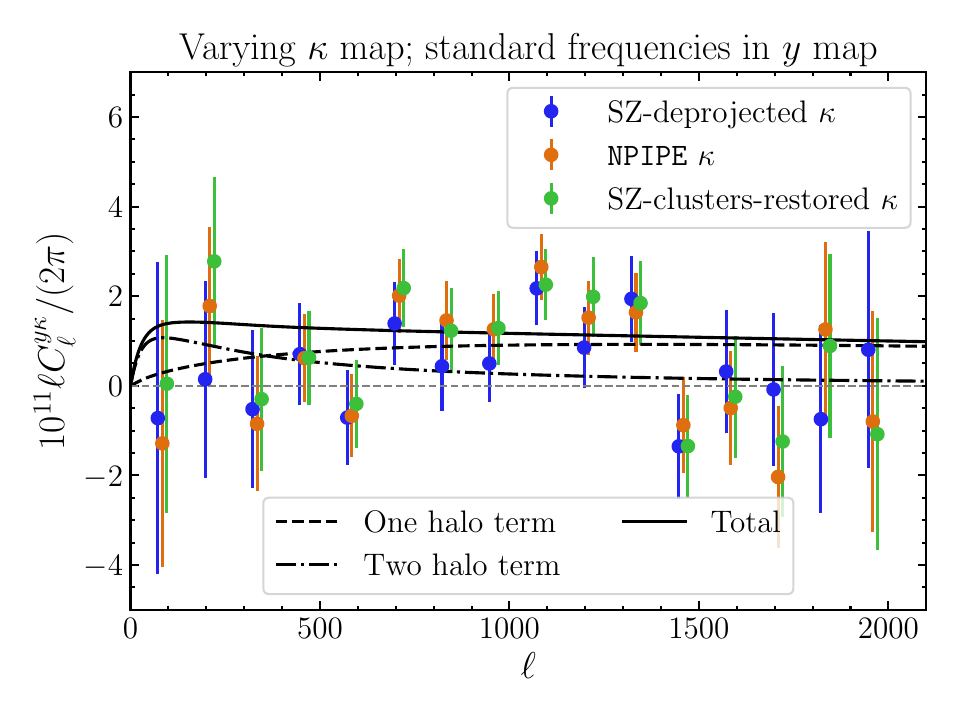}
\includegraphics[width=0.49\textwidth]{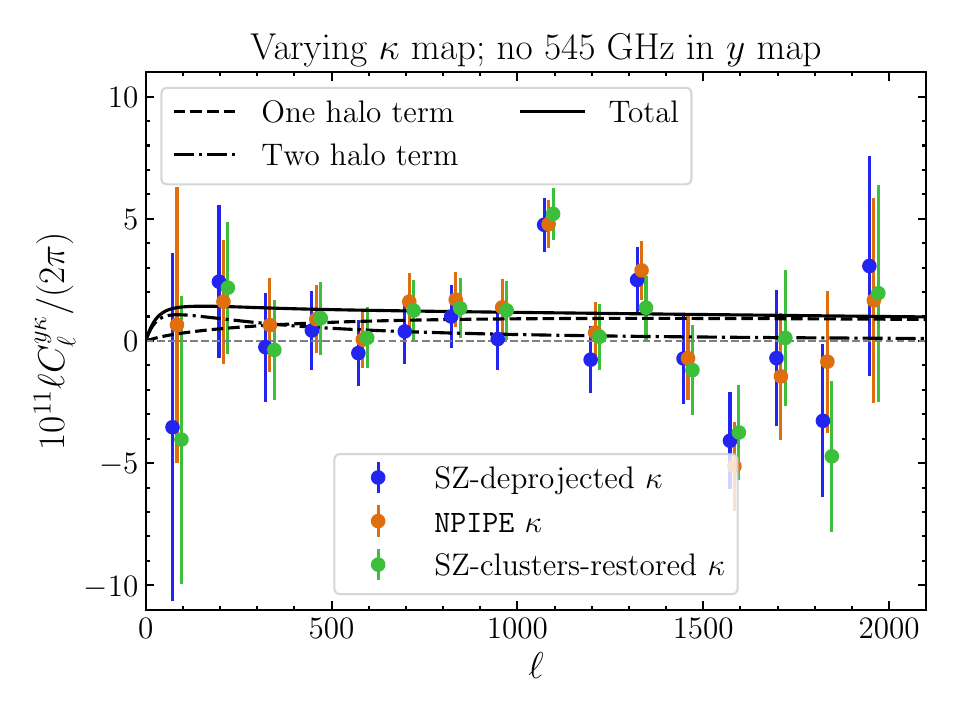}
\caption{The tSZ -- CMB lensing cross-correlation measured on different $\kappa$ maps, as labeled in the legend.  The left (right) panel uses a NILC $y$-map built with the CMB${}^5$+CIB+$\delta \beta$+$\delta T_{\rm CIB}^{\rm eff}$ deprojection applied to the standard frequency coverage (CMB${}^5$+CIB+$\delta \beta$ deprojection applied to the no-545-GHz frequency coverage).  The measurements are broadly consistent for all three $\kappa$ maps. {Some points are slightly offset along the $x$ axis for better visibility of overlapping points.} }\label{fig:different_kappas_measurement}
\end{figure}

\section{Pipeline validation}\label{sec:pipeline_validation}

\subsection{Application on Websky simulations}

In the previous section, we perform{ed} various null tests and stability tests on the data that inspire confidence in our results. The stability with respect to different CIB SED parameter choices encourages us that we have truly deprojected the CIB, and the stability of our data points with respect to different sky fractions (see Appendix~\ref{app:skyarea}) reassures us that we are not seeing signal from residual Galactic foregrounds in the CMB lensing reconstruction correlating with those in the $y$-map. However, we wish to also validate our pipeline on simulations, to ensure that performing the $\delta\beta$ and $\delta T_{\rm CIB}^{\rm eff}$ deprojections satisfactorily removes the CIB contamination and preserves our signal of interest.

As such, we build simulations of the microwave sky at the frequencies we use in our NILC, in particular $\{30,44,70,100,143,217,353,545\}$ GHz. We include simulated cosmological signals as well as the Galactic foregrounds, using the Websky~\cite{2020JCAP...10..012S} simulations for the cosmological signals and the Python Sky Model (PySM3\footnote{\url{https://pysm3.readthedocs.io/en/latest/}})~\cite{2017MNRAS.469.2821T} for the Galactic foregrounds. From Websky, we include the lensed primary CMB  (blackbody in its frequency dependence), the CIB, and the tSZ field.\footnote{We also check that the simulation analysis results obtained using our fully-deprojected $y$-maps are unchanged if we include the Websky radio galaxy component~\cite{2022JCAP...08..029L} in the simulated sky maps.} 
The Websky CIB model, which is based on the halo model of Refs.~\cite{2012MNRAS.421.2832S,2013ApJ...772...77V}, includes appropriate light-cone evolution and inter-frequency decorrelation at the frequencies at which it is simulated ($\nu>93$ GHz). At lower frequencies than this, we simply scale the lowest-frequency CIB map\footnote{In fact, we use the 100 GHz CIB map for this rescaling, as it is the lowest-frequency CIB map at frequencies corresponding to \textit{Planck}.} by the SED of Equation~\eqref{CIB_SED} (note that the CIB is extremely subdominant to other sky components at these frequencies).

The Galactic foregrounds that we include are those corresponding to the [$\texttt{d1}$,$\texttt{s1}$,$\texttt{a1}$,$\texttt{f1}$] model of PySM3.  Thus we include components due to Galactic dust ($\texttt{d1}$); synchrotron radiation ($\texttt{s1}$); anomalous microwave emission ($\texttt{a1}$); and free-free emission ($\texttt{f1}$). The dust is the  dominant foreground at high frequencies, and the synchrotron at low frequencies. We refer the reader to the PySM3 documentation for details of these models, which are built from analyses of \emph{Planck}, \emph{WMAP}, and other data sets. We find that these simulations reproduce appropriately well (for the purposes of this pipeline validation, which we stress is \textit{not} aiming to assess the level of Galactic foreground removal but instead the recovery of the true cosmological signal for our NILC deprojection choices) the measured power spectra of the \textit{Planck} data both on the full sky and on a patch of sky masked by the smallest-area Galactic mask released by the \textit{Planck} collaboration, which has a sky area of 20\%.

As we use the true, simulated Websky $\kappa$ map as our proxy for the CMB lensing map, it would be futile to build an overly-complicated Galactic foreground map, as our pipeline validation is not sensitive to the true source of systematics from Galactic foregrounds, i.e., the correlations between residual foregrounds in the $\kappa$ reconstruction and the NILC $y$-map. A true pipeline validation would propagate the Galactic biases through the lensing reconstruction pipeline by \textit{performing lensing reconstruction} on the simulated single-frequency maps (or, ideally, on an ILC combination or for various independent simulations of the noise). Such a thorough procedure is outside the scope of this work, as we do not perform lensing reconstruction ourselves but instead use publicly-available $\kappa$ maps.  As we find results that are stable to large changes in the sky area used in our measurements (see Appendix~\ref{app:skyarea}), we do not expect these Galactic foreground biases to be a significant issue. However, for future measurements of this signal (e.g., with forthcoming high-resolution CMB data), such a simulation validation would be interesting and useful.

After adding the Galactic foregrounds to the cosmological signals, we convolve our maps with Gaussian beam window functions with full-width-at-half maxima (FWHMs) corresponding to those of \textit{Planck}.  These FWHM values are listed in Table~\ref{tab:fwhm}. Finally, we add isotropic Gaussian white noise corresponding to the \textit{Planck} noise levels, which are also listed in Table~\ref{tab:fwhm}. In particular, we simulate fields with white-noise power spectra corresponding to
\be
N_\ell = \left(\frac{\pi}{180\times 60} T_n\right)^2\label{noise_conv}
\ee
where $T_n$ is the noise level in $\mathrm{\mu K}$ arcmin and $N_\ell$ is the noise power spectrum in $\mathrm{\mu K}^2$.
\begin{table*}
\begin{tabular}{|c||c|c|c|}\hline
Frequency (GHz) & Beam FWHM (arcmin) & Noise power spectrum amplitude ($\mathrm{\mu K}$ arcmin) & Noise ($\mathrm{\mu K}^2$)\\\hline\hline
30 & 32.239& 195& 0.00322\\\hline
44 & 27.005& 226&0.00433 \\\hline
70 & 13.252&199 & 0.00335\\\hline
100 & 9.69& 77.4&0.000507 \\\hline
143 & 7.30&33 & 9.21$\times10^{-5}$\\\hline
217 & 5.02&46.8& 0.000185\\\hline
353 & 4.94&154 &0.00200 \\\hline
545 & 4.83&818 &0.0566\\\hline
\end{tabular}
\caption{Characterizing features of the \textit{Planck} experiment, in particular the beam FWHM and white noise levels for each frequency. We quote the noise both in $\mathrm{\mu K}$ arcmin and in $\mathrm{\mu K}^2$; the conversion between the two is given by Equation~\eqref{noise_conv}. See Ref.~\cite{2016A&A...594A...6P} for the LFI characterization and Ref.~\cite{2016A&A...594A...8P} for the HFI characterization. Note that we have applied a $\mathrm{Jy/sr}$-to-$\mathrm{\mu K}$ conversion factor to the values quoted for 545 GHz, as we simulate the maps in $\mathrm{\mu K}$.  }
\label{tab:fwhm}
\end{table*}

 Thus our final sky model is
\begin{widetext}
\be
T(\nu, \hat n) =  W\left(T^{\mathrm{CMB}}(\hat n)+g_\nu y(\hat n) + T^{\mathrm{CIB}}(\nu,\hat n) + T^{\mathrm{Galaxy}}(\nu, \hat n)\right) + T_n(\nu, \hat n)
\ee
\end{widetext}
where $T^{\mathrm{CMB}}(\hat n)$ is the Websky lensed CMB temperature map,  $y(\hat n)$ is the Websky Compton-$y$ parameter map; $T^{\mathrm{CIB}}(\nu,\hat n) $ is the Websky CIB temperature map at frequency $\nu$; $T^{\mathrm{Galaxy}}(\nu, \hat n) $ is the simulated Galactic temperature map from PySM3; $g_\nu $ is the spectral response function of the tSZ effect; $W(\cdot)$ is the beam-convolving operation in real space (which corresponds to a filtering by the beam window function
\be
B_\ell=\exp\lb\frac{-\ell\lb\ell+1\rb }{2}\frac{\theta_{FWHM}^2}{8\ln 2}\rb
\ee
in harmonic space); and $T_n(\nu, \hat n)$ is the simulated instrumental noise. Note that all of our simulations are in units of $ \mathrm{\mu K}_{\mathrm{CMB}}$, i.e., units in which the CMB has unit SED; thus we need to convert the Websky CIB maps, which are supplied in $\mathrm{Jy/sr}$, to $ \mathrm{\mu K}_{\mathrm{CMB}}$.

After applying our NILC pipeline (see Paper I for details on the pipeline) to construct $y$-maps from the simulated intensity maps, we measure $C_\ell^{y\kappa}$ using the simulated $\kappa$ map provided by Websky. While we do not include any systematics due to the fact that the true $\kappa$ is reconstructed from temperature data, we do include a realization of Gaussian noise with power spectrum matching the noise power spectrum of the {2018} \textit{Planck} lensing reconstruction. 
We also measure the cross-correlation on a realization of the lensing map with no noise included, to demonstrate the effectiveness of the CIB moment deprojections in the limit of zero CMB lensing noise.

The resulting data points, measured with the CIB deprojections done using the default best-fit SED, are shown in Figure~\ref{fig:websky_validation}. Comparing the measurement on the left of Figure~\ref{fig:websky_validation}, which includes appropriate Gaussian lensing noise, with the measurement on the right, which is performed on the noiseless $\kappa$ map, it is clear that (especially for the cases when $>1$ components are deprojected) there is significant scatter due to the lensing noise.  Note that the deprojection that we use for our true data points corresponds to the red points, where both $\delta \beta$ and $\delta T^{\mathrm{eff}}_{\mathrm{CIB}}$ are deprojected. It is also clear, in both cases, that in the case when no CIB components are deprojected, there is a significant bias on the measurement.  The CIB deprojection removes some of this bias, but (as is especially clear on the right) a significant amount remains. The moment deprojections remove significantly more of this bias.  We investigate this further in Figure~\ref{fig:validation_moments} where we vary the SED used for the CIB and moment deprojections to illustrate the reduction in scatter they yield.

\begin{figure}
\includegraphics[width=0.49\textwidth]{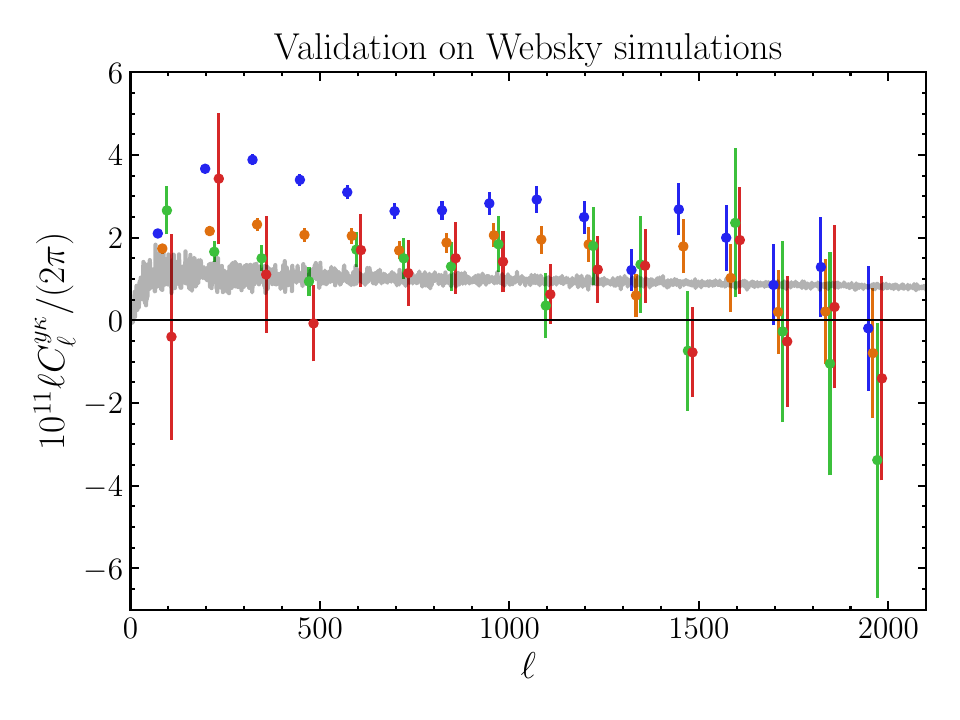}
\includegraphics[width=0.49\textwidth]{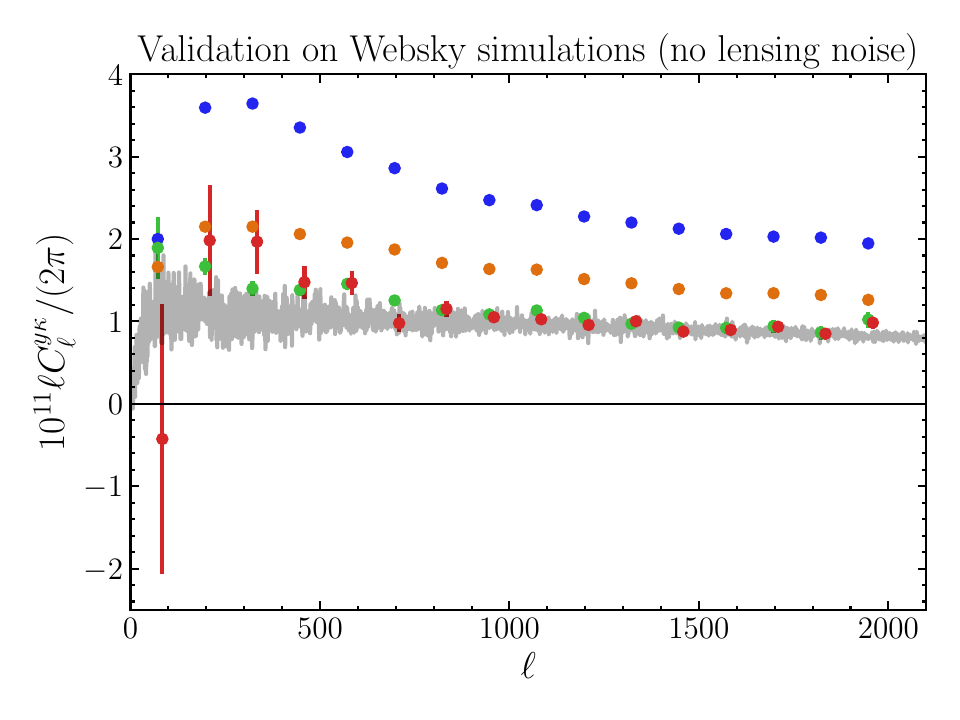}
\includegraphics[width=\textwidth]{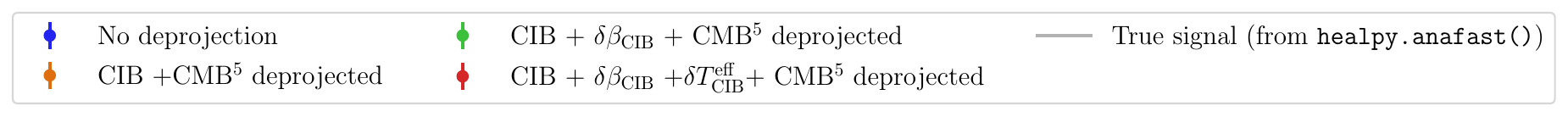}
\caption{Validation on the simulations. Note that the ``true signal'' that we compare to is estimated by using $\texttt{healpy.anafast()}$ on the full-sky maps. In both cases, we have applied our NILC pipeline with various deprojections (as labeled) to the simulated microwave sky maps to build $y$-maps, and measured $C_\ell^{y\kappa}$ using the true $\kappa$ map from Websky, with appropriate Gaussian noise added (\textit{left}) to approximate the scatter in our own measurement on data, and also with no lensing noise added (\textit{right}) to illustrate the biases associated with the different deprojections in the zero-lensing-noise case. {Some points are slightly offset along the $x$ axis for better visibility of overlapping points.}} \label{fig:websky_validation}
\end{figure}

In Figure~\ref{fig:validation_moments}, we show the signal measured on the simulations for the different moment deprojections as we vary the CIB SED. Again, we show the points measured with lensing noise on the left, and without lensing noise on the right; the biases and scatter are easier to see when there is no lensing noise. We see that deprojecting $\delta \beta$ indeed reduces some of the bias induced when the CIB alone is deprojected with different SEDs;  {the bias is removed by jointly deprojecting moments with respect to both parameters $\delta \beta$ and $\delta T^{\mathrm{eff}}_{\mathrm{CIB}}$, in the bottom plots, when we see scatter no greater than $\approx1\sigma$ and an unbiased (by eye) measurement.}

\begin{figure}
\includegraphics[width=0.49\textwidth]{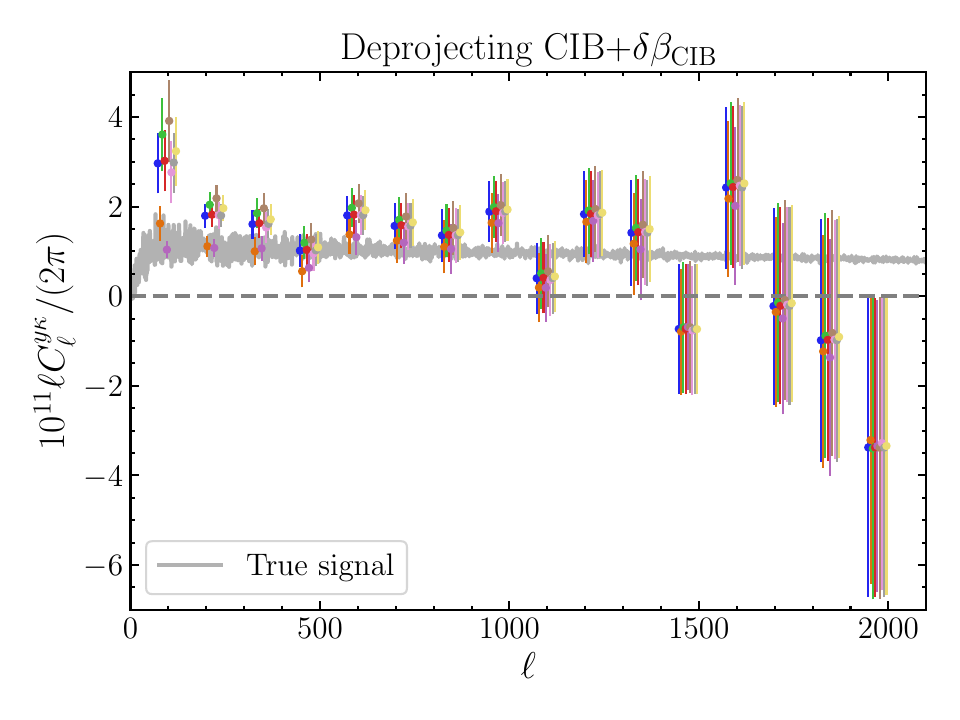}
\includegraphics[width=0.49\textwidth]{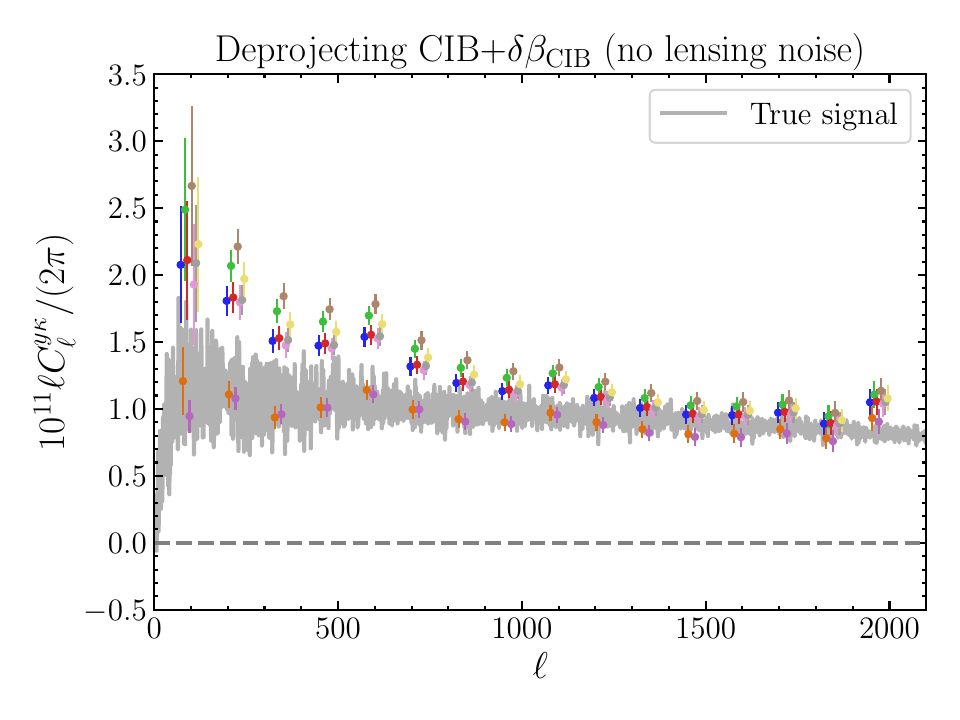}
\includegraphics[width=0.49\textwidth]{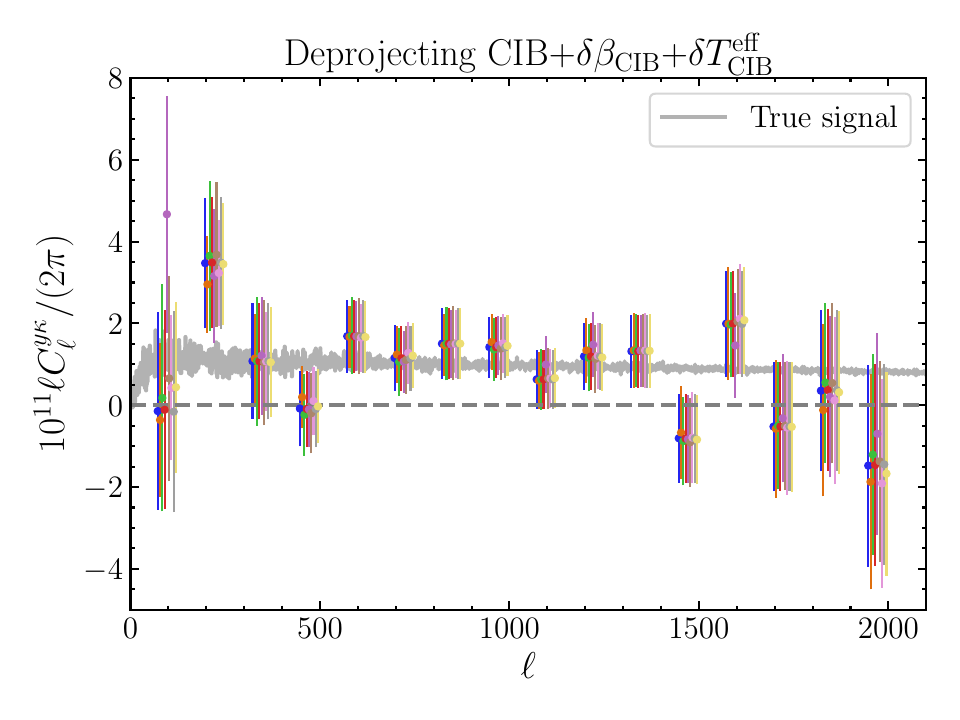}
\includegraphics[width=0.49\textwidth]{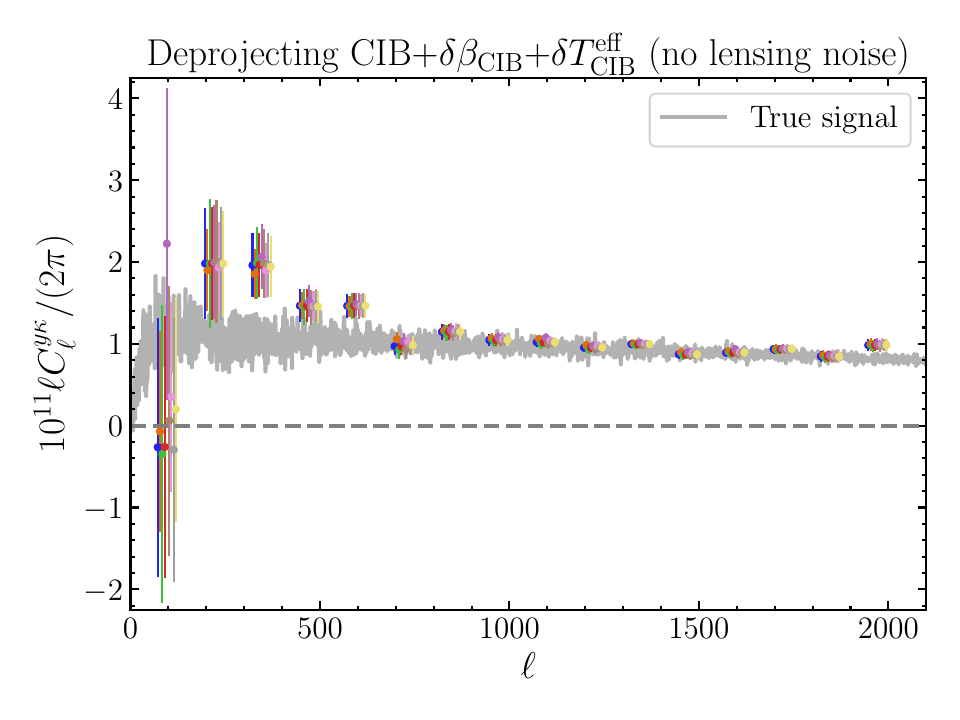}

\includegraphics[width=\textwidth]{legend_differentsamples_labelled.pdf}

    \caption{Deprojecting the CIB and various first moments thereof, measured on simulations with appropriate lensing noise (\textit{left}) and no lensing noise (\textit{right}). We use the same values for the CIB SED parameters as in the samples we use when we vary the SED parameters for our deprojections on the data (e.g., in Figure~\ref{fig:deproj_CIBdifferentbeta}). {The points are systematically offset along the $x$ axis for better visibility of overlapping points.} }
    \label{fig:validation_moments}
\end{figure}

Our overall conclusion from these simulation tests is that a CIB+$\delta\beta+\delta T^{\mathrm{eff}}_{\mathrm{CIB}}$-deprojected map can be used to make an unbiased estimate of $C_\ell^{y\kappa}$, for a wide range of SED parameters assumed for the CIB modified blackbody spectrum.  This reinforces our findings from the actual data in the previous section, and substantiates the overall robustness of our measurement.

 \section{analysis}\label{sec:analysis}
 
\begin{figure}
\includegraphics[width=0.65\textwidth]{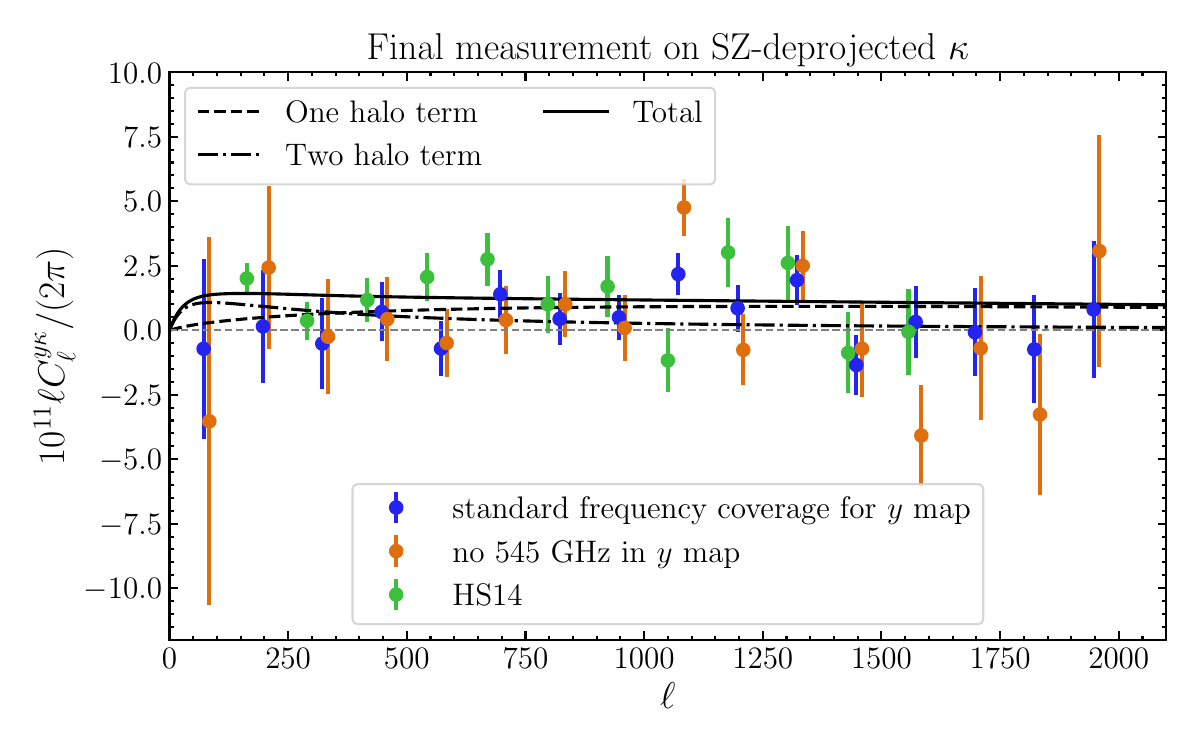}\\
\includegraphics[width=0.65\textwidth]{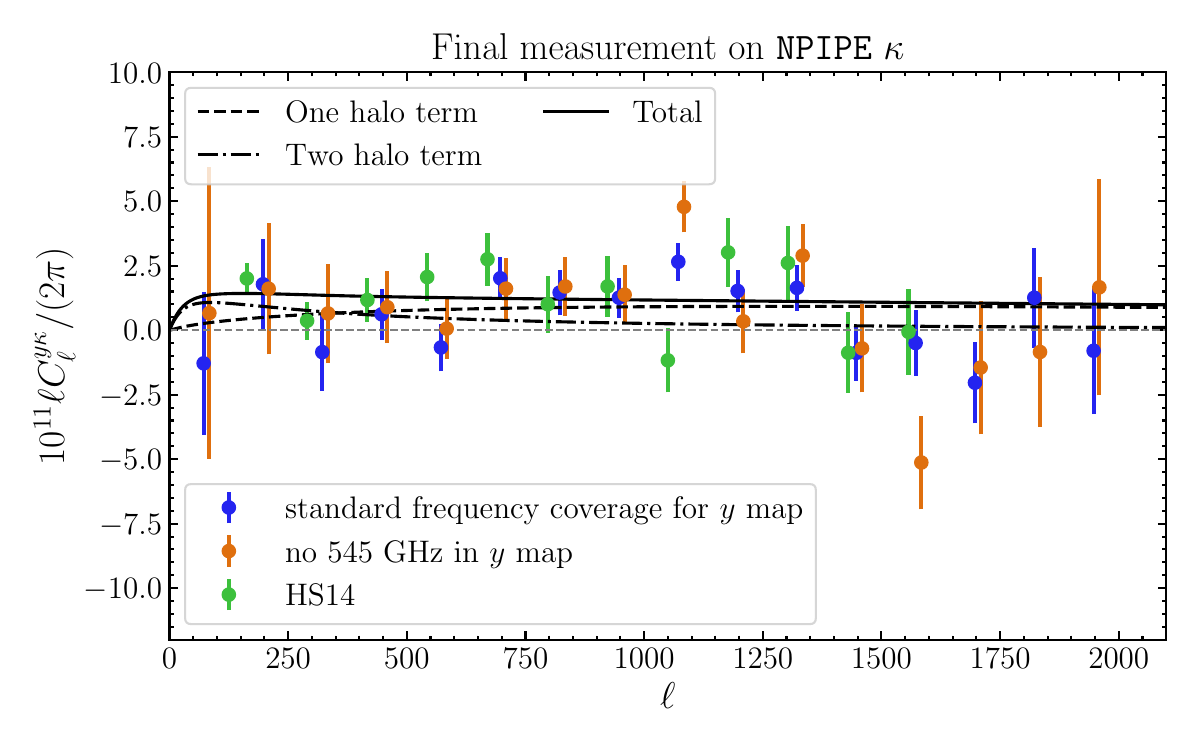}\\
\includegraphics[width=0.65\textwidth]{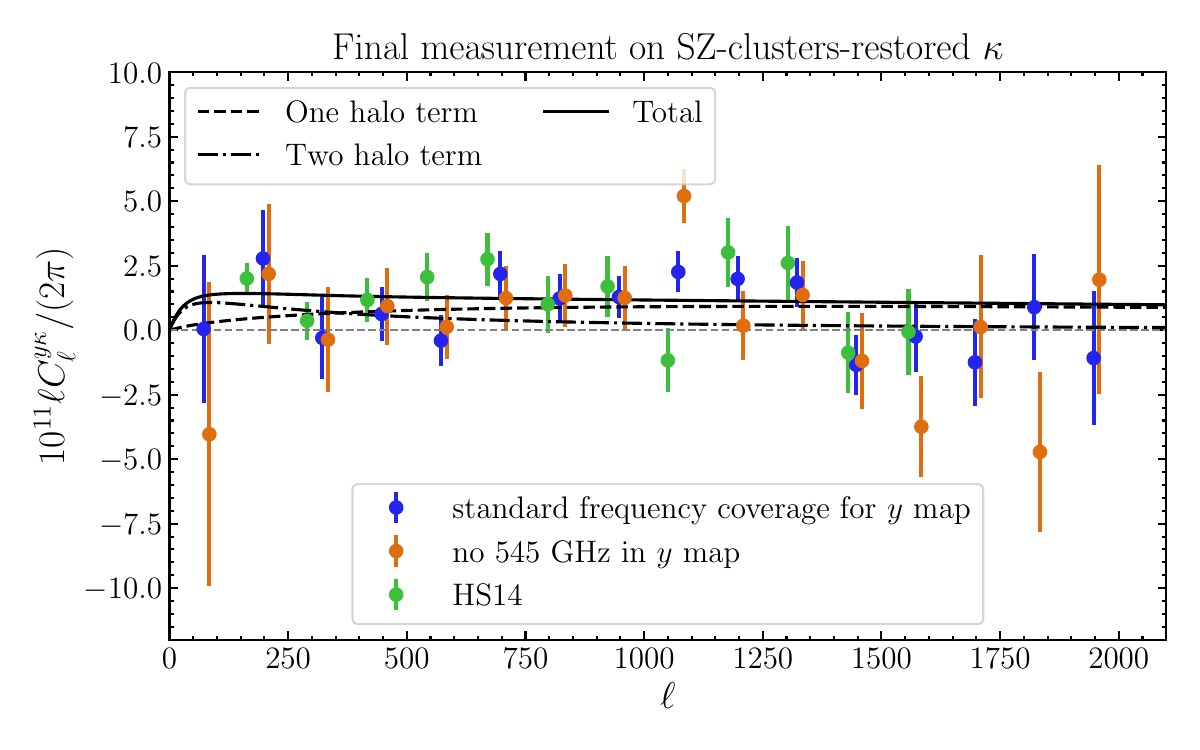}
\caption{Our final data points for the tSZ -- CMB lensing cross-power spectrum, measured using the  CMB$^5$+CIB+$\delta\beta$-deprojected map with 545 GHz removed (orange), and the CMB$^5$+CIB+$\delta\beta$+$\delta T^{\rm eff}_{\rm CIB}$-deprojected map with standard frequency coverage (blue). {We show the data points for all three $\kappa$ maps as indicated in the titles}, and deproject the primary CMB in the first five needlet scales. We also include the data points from HS14 (green) for comparison. We note that, on large scales, our error bars are significantly larger than those of HS14, as we pay a large SNR penalty due to our many deprojections needed to robustly clean CIB contamination (the width of our multipole bins is the same as in HS14).  The black curve is not a fit to the data, but rather shows our fiducial theoretical model (see Section~\ref{sec:theory}), with the one-halo (two-halo) term shown in dashed (dash-dotted).  {Some points are slightly offset along the $x$ axis for better visibility of overlapping points.}}\label{fig:final_data}
\end{figure}

Our final data points, measured with all three $\kappa$ maps, with the maximally-deprojected $y$ maps (CMB$^5$+CIB+$\delta\beta$-deprojection for the no-545 GHz case and CMB$^5$+CIB+$\delta\beta$+$\delta T_{\mathrm{CIB}}^{\mathrm{eff}}$-deprojection for the standard-frequency case) are shown in Figure~\ref{fig:final_data}. Also included on this plot, for comparison, are the data points from HS14; as the $\ell$-binning schemes are similar (in particular, the widths of the $\ell$-bins are identical), the error bars can be compared directly by eye. It is clear that on large scales, our measurement has lower SNR than HS14, although on small scales our error bars are slightly smaller. The data points measured without 545 GHz and with the standard frequency coverage are broadly consistent.   We proceed to fit theoretical models to the { points in Figure~\ref{fig:final_data}.}

The tSZ effect is a powerful probe both of cosmology and of the physics of the ICM. Most of the signal is sourced in massive clusters in the late Universe ($z\lessthanapprox$ 1-1.5), which are rare objects; thus, the tSZ signal probes the tail of the HMF. This signal can be accessed through various tSZ statistics, for example through the power spectrum~\cite{2002MNRAS.336.1256K,2016A&A...594A..22P,2017MNRAS.469..394H,2018MNRAS.477.4957B,2022MNRAS.509..300T}, cluster counts~\cite{2013JCAP...07..008H,2016A&A...594A..24P,2019ApJ...878...55B}, or the 1-point PDF~\cite{2014arXiv1411.8004H,2019PhRvD..99j3511T}. In every case, while there is high sensitivity to the cosmological parameters that govern the distribution of the halos ($\sigma_8$, which governs the amplitude of matter clustering on scales of $8 \,h^{-1} \mathrm{Mpc}$ and $\Omega_m$, which parametrizes the matter density of the Universe), there is also a degeneracy with the parameters that describe the tSZ flux-mass relation (i.e., pressure-mass relation), which is required to translate from the HMF to the flux distribution that is measured. Thus, these parameters, which are poorly constrained in regimes where ICM physics is not well understood, must be jointly constrained with the cosmological parameters. In our case, where we consider just one statistic of the tSZ effect, it is not possible to fully break the degeneracy between ICM physics and cosmology.  However, as different tSZ statistics depend on various combinations of these parameters in different ways, this degeneracy can in principle be broken by combining our data with different tSZ probes, such as the auto-power spectrum. We leave such an approach to future work and instead only constrain ICM physics while holding the cosmology fixed, or alternatively constrain cosmology while holding the pressure-mass relation fixed.

The model that we constrain is described in Section~\ref{sec:theory}. When we fix the cosmology, and unless otherwise stated, we use the cosmological parameters from the best-fit $\Lambda$CDM model to the \textit{Planck} TT+TE+EE+lowE+lensing \texttt{Plik} likelihood~\cite{2020A&A...641A...6P}: \{$\Omega_b h^2$=0.022383; $\Omega_c h^2$ = 0.12011; $H_0= 67.32$ {km/s/Mpc}; $A_s=2.101\times10^{-9}$; $n_s=0.96605$\}, where $\Omega_b h^2$ is the physical density of baryons today; $\Omega_c h^2$ is the physical density of cold dark matter today; $H_0\equiv 100 h$ km/s/Mpc is the Hubble parameter; and $A_s$ and $n_s$ are respectively the amplitude and spectral index of scalar fluctuations with a pivot scale of $0.05\, \mathrm{Mpc}^{-1}$. For this cosmology, $\sigma_8 = 0.8120$.

\subsection{Likelihood for $C_\ell^{y\kappa}$}

To constrain the models for $C_\ell^{y\kappa}$, we construct a simple Gaussian likelihood $\mathcal L$: 
\be
\log \mathcal L(\mathbf{p}) \propto -\frac{1}{2} \chi^2(\mathbf{p}) \,,
\label{loglike}
\ee
where
\be
\chi^2 ={ \left(\hat C_{\ell}^{y\kappa}-  C_{\ell}^{y\kappa}(\mathbf{p})\right)^T \mathcal C_{\ell\ell^{\prime}}^{-1} \left( \hat C_{\ell^\prime}^{y\kappa}-  C_{\ell^\prime}^{y\kappa}(\mathbf{p})\right)}
\ee
with $C_\ell^{y\kappa}(\mathbf{p})$ the theoretical model for the signal evaluated for parameters $\mathbf{p}$ (calculated with \texttt{class\_sz}), $\hat C_\ell^{y\kappa}$ the cross-correlation data points, and {$\mathcal C_{\ell \ell^\prime}$ the full covariance matrix}. 
The theoretical model for the signal is a function of cosmological parameters (e.g., $\sigma_8$ and $\Omega_m$) and pressure profile parameters (e.g., $A_0^{P_0}$ and $A_0^{\beta}$, cf.~Equations~\eqref{pressure_profile} and~\eqref{parameter_scaling_def}).

We explore the parameter posteriors with Markov chain Monte Carlo (MCMC) sampling, using the Metropolis--Hastings algorithm implementation of Refs.~\cite{2002PhRvD..66j3511L,2013PhRvD..87j3529L} in \texttt{Cobaya}\footnote{\url{https://cobaya.readthedocs.io/en/latest/}}~\cite{2019ascl.soft10019T,Torrado:2020dgo}.  Priors on the sampled parameters are discussed below.  We run all chains until they are converged with a Gelman-Rubin convergence criterion~\cite{1992StaSc...7..457G} of $|R-1|<0.001$ (for the $P_0$ fits) or $|R-1|<0.01$ (for the cosmology fits). We quote each parameter's best-fit value as the point that maximizes the posterior, which is determined using the BOBYQA~\cite{bobyqa,2018arXiv180400154C,2018arXiv181211343C} minimizer implemented in \texttt{Cobaya}.  We use \texttt{GetDist}\footnote{\url{https://getdist.readthedocs.io/en/latest/}}~\cite{Lewis:2019xzd} to analyze our chains and plot our parameter posteriors.

\subsection{Constraints on pressure profile parameters}\label{sec:pressure_constraints}

We constrain the normalization of the amplitude of the pressure-mass relation, $A_0^{P_0}$ (see Equations~\eqref{pressure_profile} and~\eqref{parameter_scaling_def}).  For ease of notation, we redefine this amplitude parameter as simply $P_0$, i.e., $P_0 \equiv A_0^{P_0}$.  This procedure is equivalent to simply fitting a one-parameter amplitude that rescales the entire theoretical model. In our MCMC sampling, we use a wide, uninformative, flat prior on $P_0$; we use no other priors, so the posterior is given by the log-likelihood in Equation~\eqref{loglike}.  For convenience, we further rescale $P_0$ by its fiducial value given in Table~\ref{tab:pressure_profile_parameters}, i.e., we plot $P_0/P_{\rm fid}$, where $P_{\rm fid} = 18.1$ is the fiducial amplitude of the pressure-mass relation.

{The constraints on $\frac{P_0}{P_{\mathrm{fid}}}$ for different data combinations are summarized in Table~\ref{tab:P0_values}, and }the posteriors for $P_0/P_{\mathrm{fid}}$ are shown in Figure~\ref{fig:posteriors_P0}. For comparison, we also constrain the model using the measurements from HS14.  {Using the HS14 data,} we find a very similar result to that quoted in HS14, with $\frac{P_0}{P_{\mathrm{fid}}}=1.03\pm 0.21$  (cf.~the quoted value of $\frac{P_0}{P_{\mathrm{fid}}}=1.10\pm 0.22$ in that work). {The small discrepancy is due to slightly different choices in the halo model calculation for the template signal. First, HS14 used a \textit{WMAP9} cosmology~\cite{2013ApJS..208...19H}, which had a slightly lower value of $\Omega_m$ than the \textit{Planck} cosmology we use ($\Omega_m=0.282$ in \textit{WMAP9} versus $\Omega_m=0.3144$ in  \textit{Planck}). Second, our halo model calculation treats the low-mass halos differently to HS14, with the contribution from halos below $M_{200m} = 10^{10} \,h^{-1} M_\odot$ calculated in this work with the counter-term prescription, which is different to the extrapolation of the HMF used in HS14, which integrated directly the contribution of all halos with $10^5 \, h^{-1} M_\odot < M_{200c} < 5 \times 10^{15} \, h^{-1} M_\odot$. Third, we cut our halo profiles at $2 r_{200m}$, while HS14 do so at $1.5 r_{\mathrm{vir}}$.  Nevertheless, these differences all lead to relatively small changes in the template signal, and hence the central value of $\frac{P_0}{P_{\mathrm{fid}}}$ in our reanalysis of HS14 is quite close to theirs.}

{All of our constraints are consistent with each other to within $1\sigma$. We get much better fits from the standard frequency case than the no-545 GHz case, as well as tighter constraints. The tightest constraint comes from the standard-frequency-coverage map in combination with the $\texttt{NPIPE}$ $\kappa$ map, resulting in~\fiducialmeasurementbestP. The SZ-deprojected results are slightly  ($\approx 1\sigma$) slower, resulting in~\fiducialmeasurementP, perhaps hinting at a $\left<yyy\right>$ bias; the SZ-clusters-restored measurement is very slightly higher than the \texttt{NPIPE} measurement. It would be interesting to perform the measurement on a tSZ-deprojected + cluster-signal-restored \texttt{NPIPE} map, to see if these conclusions hold.}

It is notable that our final constraints are \text{not} tighter than those from HS14, with all of our error bars comparable to or larger than theirs, despite the higher-quality data we use. However, our results are significantly more robust to potential CIB contamination due to the battery of deprojection operations implemented in our analysis, which are also responsible for inflating our error bars significantly.  We show posteriors in Figure~\ref{fig:posteriors_P0} for all combinations of our fully-deprojected $y$-maps (i.e., CIB+$\delta\beta$+CMB$^5$-deprojected for the no-545 GHz case and CIB+$\delta\beta$+$\delta T_{\mathrm{CIB}}^{\mathrm{eff}}$+CMB$^5$-deprojected for the standard-frequency case) with the three $\kappa$ maps we consider. We also list the best-fit values in Table~\ref{tab:P0_values}.  It is generally true that the tSZ-deprojected $\kappa$ map yields a lower  measurement (by $\approx 1\sigma$) of $P_0$ than the other two maps. However, they are all statistically consistent. It is also generally true that including 545 GHz and deprojecting $\delta T^{\mathrm{eff}}_{\mathrm{CIB}}$ improves the quality of the fit (and also decreases the error bars), with the reduced $\chi^2$ values decreasing in every case.

\begin{figure}
\includegraphics[width=0.49\textwidth]{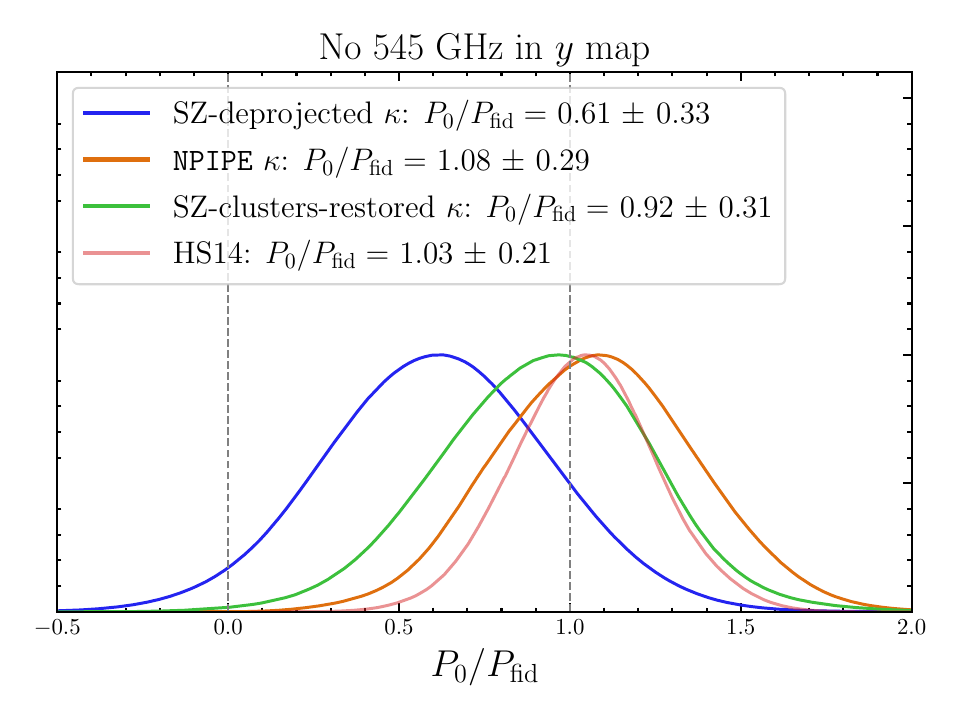}
\includegraphics[width=0.49\textwidth]{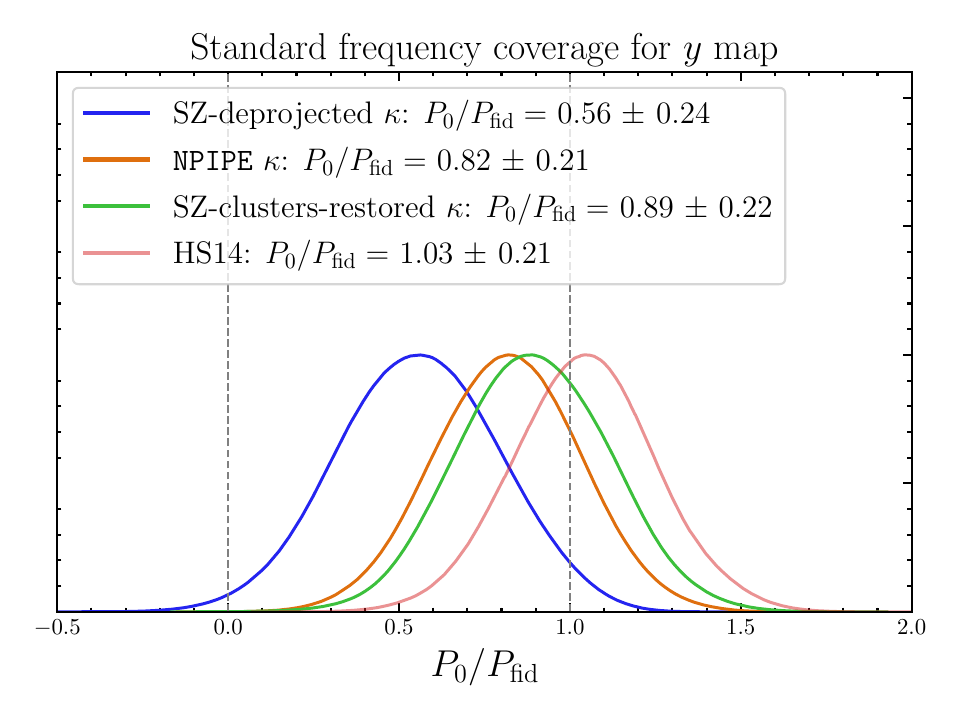}
\caption{Posteriors for the parameter $P_0$, which normalizes the amplitude of the pressure-mass relation. $P_0 / P_{\rm fid} = 1$ corresponds to the fiducial model.  A value of $P_0$ consistent with zero would imply a non-detection.  Results are shown for two different $y$-maps (left and right panels) and various $\kappa$ reconstructions (as labeled), as well as for the HS14 data. {In both cases we use the maximally-deprojected $y$ maps: CMB$^{5}$+CIB+$\delta\beta$-deprojection for the no-545 GHz case, and CMB$^{5}$+CIB+$\delta\beta$+$\delta T_{\mathrm{CIB}}^{\mathrm{eff}}$-deprojection for the standard-frequency case.}  For all of the analyses shown here, the signal is consistent with the fiducial prediction.  We find $P_0 / P_{\rm fid}$ consistent with unity at the $1$-$1.5\sigma$ level in all cases except for the tSZ-deprojected $\kappa$ map analysis in the right panel, which is consistent with unity at $\approx 2\sigma$. }\label{fig:posteriors_P0}
\end{figure}

\begin{table}
\begin{tabular}{|c|c||c|c|c|c|}\hline\hline
\multicolumn{2}{|c|}{Configuration} & Best-fit $P_0/P_{\mathrm{fid}}$ & Mean $P_0/P_{\mathrm{fid}}$ &$\chi^2_r$&PTE\\\hline\hline
SZ-deprojected $\kappa$    &    Standard frequency coverage in $y$ map    &    0.55    &    0.56 $\pm$ 0.24     &    1.03 &0.42\\ \hline
\texttt{NPIPE} $\kappa$    &    Standard frequency coverage in $y$ map    &    0.82    &    0.82 $\pm$ 0.21     &    1.52 &0.088\\ \hline
SZ-clusters-restored $\kappa$    &    Standard frequency coverage in $y$ map    &    0.89    &    0.89 $\pm$ 0.22     &    1.18&0.29 \\ \hline\hline
SZ-deprojected $\kappa$    &    No 545 GHz in $y$ map    &    0.61    &    0.61 $\pm$ 0.33     &    1.88&0.020 \\ \hline
\texttt{NPIPE} $\kappa$    &    No 545 GHz in $y$ map    &    1.08    &    1.08 $\pm$ 0.29     &    2.11&0.0071 \\ \hline
SZ-clusters-restored $\kappa$    &    No 545 GHz in $y$ map    &    0.93    &    0.92 $\pm$ 0.31     &    1.89&0.019 \\ \hline\hline
\end{tabular}
\caption{The best-fit and posterior mean values of the amplitude $P_0/P_{\mathrm{fid}}$ (quoted with the $1\sigma$ errors from the MCMC chains). We also list the best-fit reduced $\chi_r^2\equiv\chi^2/N_\mathrm{dof}$, where $N_\mathrm{dof}=15$ is the number of degrees of freedom, given by the number of data points (16) minus the number of free parameters (1).  { We also indicate the probability-to-exceed (PTE).}}\label{tab:P0_values}
\end{table}

\subsubsection*{Including a correction from hydrodynamical simulations in the theory}\label{sec:simulation_correction_constraints}

{In Appendix~\ref{app:cutoff}, we compare our halo model to a measurement of $C_\ell^{y\kappa}$ from the hydrodynamical simulations from which our $P-M$ relations were calibrated (see Figure~\ref{fig:compare_sims_calculation}). It is clear that some portion of the signal is unaccounted for by the halo model in the regime where we are measuring $C_\ell^{y\kappa}$. We can incorporate this by measuring the ratio of the halo model prediction to the simulation measurement (as calculated for cosmological parameters corresponding to those in the simulation, as we do in Appendix~\ref{app:cutoff}). Under the assumption that this ratio is cosmology-independent, we can then rescale the halo model prediction at the \textit{Planck} cosmology we are using in this work, in order to include this contribution. When we do this, and constrain an amplitude for the model, the posteriors are shown in Figure~\ref{fig:simratio_posteriors}. As expected, the posterior means are shifted to lower amplitudes, since the fiducial model prediction is now higher. Nevertheless, the results remain consistent with the fiducial model at $\lesssim 2\sigma$, apart from those obtained with the SZ-deprojected $\kappa$ map, which lie $\approx 3\sigma$ below unity.}

\begin{figure}
\includegraphics[width=0.49\textwidth]{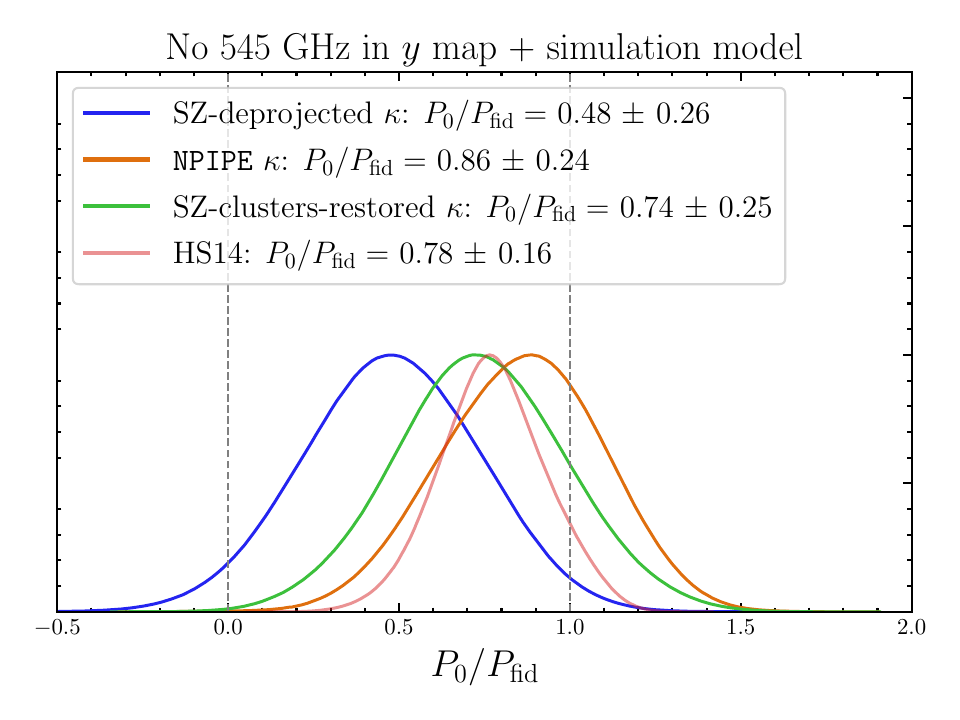}
\includegraphics[width=0.49\textwidth]{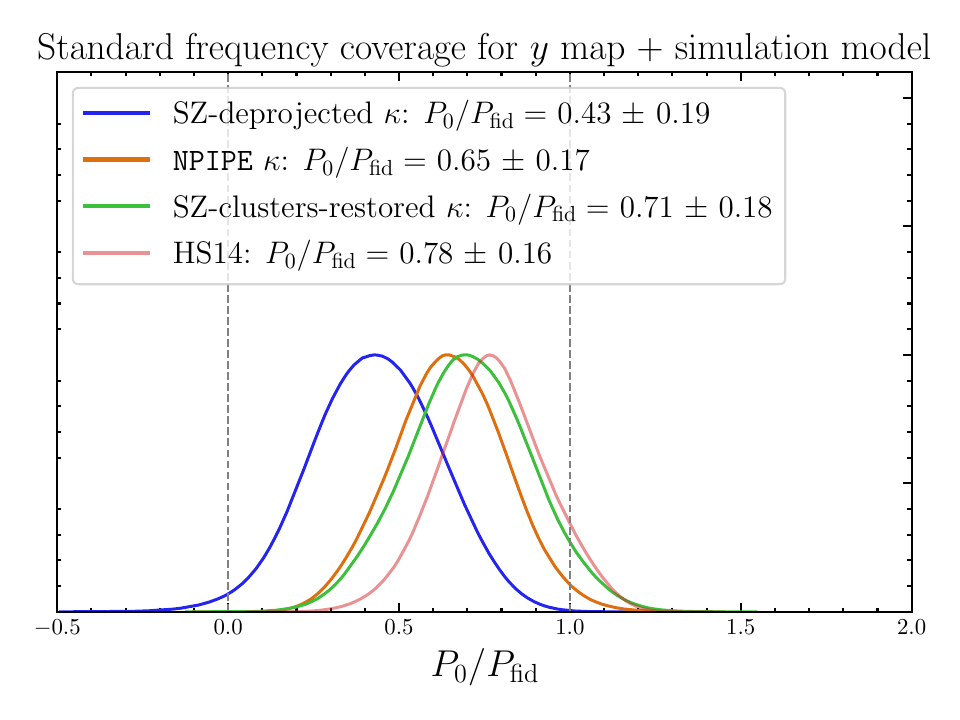}
\caption{Posteriors for the parameter $P_0$, which normalizes the amplitude of the pressure-mass relation. Here, $P_0 / P_{\rm fid} = 1$ corresponds to the simulation-corrected theory model (see Appendix~\ref{app:cutoff}), i.e., a model that includes contributions from unbound gas and low-mass halos that our halo model approach may miss.  Results are again shown for two different $y$-maps (left and right panels) and various $\kappa$ reconstructions (as labeled), as well as for the HS14 data. {In both cases we use the maximally-deprojected $y$ maps: CMB$^{5}$+CIB+$\delta\beta$-deprojection for the no-545 GHz case, and CMB$^{5}$+CIB+$\delta\beta$+$\delta T_{\mathrm{CIB}}^{\mathrm{eff}}$-deprojection for the standard-frequency case.}  {We find a shift of $\approx 1\sigma$ downwards compared to the posteriors in Section~\ref{fig:posteriors_P0}, which is expected due to the additional contributions in the theory model used here.}}\label{fig:simratio_posteriors}
\end{figure}

\subsection{Constraints on cosmological parameters}\label{sec:cosmology}

The tSZ effect is a powerful probe of cosmology, in particular of the matter density and the amplitude of clustering (parameterized by $\Omega_m$ and $\sigma_8$, respectively). However, the effects of cosmological parameter variations on tSZ observables are highly degenerate with changes in the ICM astrophysics. Nonetheless, given a fixed model for the pressure profile, we can constrain cosmology with our measurement of the tSZ -- CMB lensing cross-power spectrum. We find that, for our model, the cross-power spectrum depends on a combination of $\Omega_m$ and $\sigma_8$ in the following way:
\be
C_\ell^{y\kappa} \propto P_0 \lb\sigma_8\rb^{\approx 5.7} \lb\Omega_m\rb^{\approx 1.3} \,,\label{cross_dep}
\ee
with the exact values of the power-law indices depending on the multipole $\ell$. Importantly, this is a different cosmological dependence to that of the auto-power spectrum, which displays the following dependence\footnote{{The $y$ signal depends stronly on the baryon fraction $f_b\equiv\frac{\Omega_b}{\Omega_c}$. Thus the scaling with $\Omega_m$ depends on what parameters are held fixed. Here, we have kept $\Omega_b$ fixed and changed $\Omega_m$ by changing $\Omega_c$. The scaling with $\Omega_m$ is much stronger when the baryon fraction  is kept constant, and both $\Omega_b$ and $\Omega_m$ changed consistently. }}:
\be
C_\ell^{yy} \propto P_0^2 \lb\sigma_8\rb^{\approx 7.5} \lb\Omega_m\rb^{\approx 1} \,.\label{auto_dep}
\ee
As the relative dependences between cosmology and astrophysics (represented by $P_0$ in Equations~\eqref{cross_dep} and~\eqref{auto_dep}) are different, this means that while both probes separately cannot separate cosmology and astrophysics, a combined analysis can break this degeneracy (see, e.g., HS14). However, in this work we focus on constraining cosmology only from the cross-power spectrum, leaving a joint analysis to future work.

Due to the degeneracy between $\sigma_8$ and $\Omega_m$, we can only constrain simultaneously the product of these parameters.  We find the tightest constraints on the parameter combination $\sigma_8 \left(\Omega_m\right)^{\approx0.25}$, which is very close to the best-determined parameter combination $\sigma_8 \left(\Omega_m\right)^{0.26}$ found in HS14. 

When we vary the cosmological parameters, we only vary $\Omega_\mathrm{cdm}$ and $\sigma_8$, with $H_0$ held fixed to $H_0=67.32 \,\,\mathrm{km/s/Mpc}$ and $\Omega_b h^2$ fixed to 0.02383. We impose a flat prior on $\sigma_8$ ($0<\sigma_8<2$) and a flat prior on $\Omega_{\mathrm{cdm}}$ with an upper limit that corresponds to $0<\Omega_m<1$. We thus obtain $\Omega_m$ as a derived parameter in the chains.

The posteriors for the best-determined parameter combination are shown in Figure~\ref{fig:posteriors_Cosmology}. As expected from the astrophysical constraints, these posteriors are $\approx 1\sigma$ lower than the fiducial value of $\sigma_8 (\Omega_m/0.3144)^{0.25} = 0.812$. {For the standard-frequency-coverage case with $\texttt{NPIPE}$ $\kappa$, we find (at 68\% confidence) $0.73<\sigma_8\left(\frac{\Omega_m}{0.3144}\right)^{0.25}<0.83$.  For the SZ-deprojected measurement, we find $0.53<\sigma_8\left(\frac{\Omega_m}{0.3144}\right)^{0.25}<0.79$ (also at $68\%$ confidence).  We find similar results for the no-545-GHz case, as shown in Figure~\ref{fig:posteriors_Cosmology}.}

\begin{figure}
\includegraphics[width=0.49\textwidth]{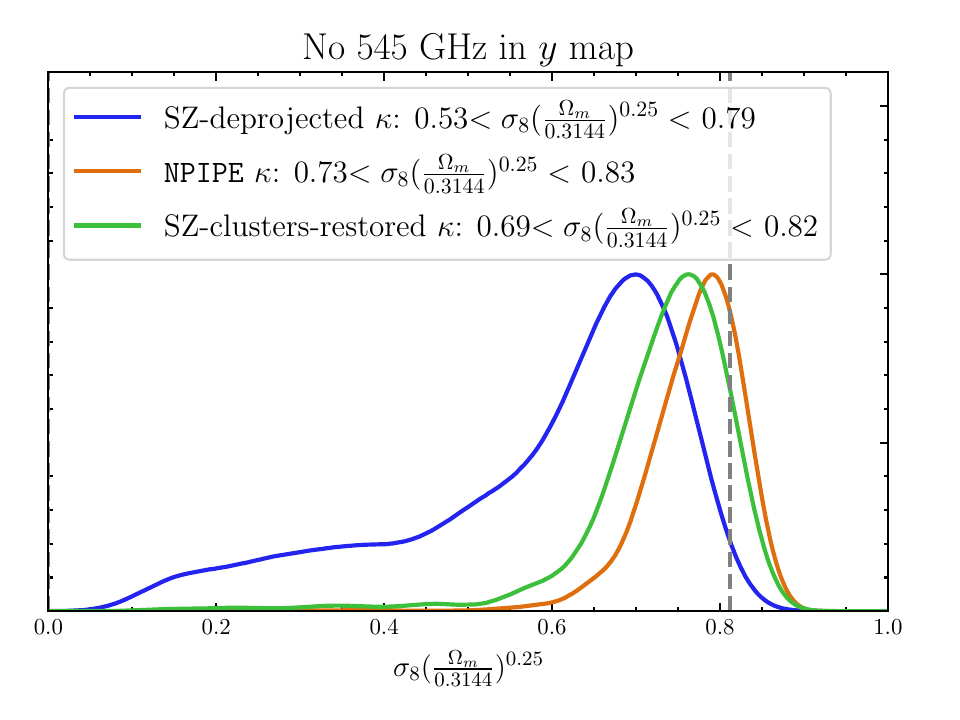}
\includegraphics[width=0.49\textwidth]{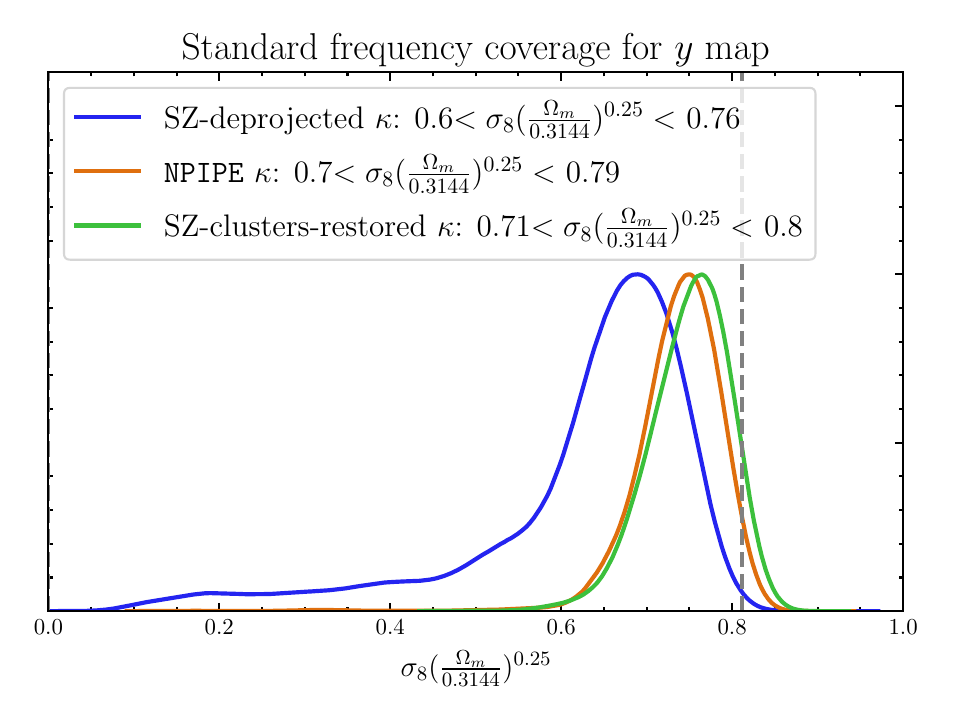}
\caption{Constraints on the cosmological parameter combination that is most tightly constrained by our $C_\ell^{y\kappa}$ measurement, namely, $\sigma_8 (\Omega_m/0.3144)^{0.2}$.  The dashed grey line shows the value of this parameter in our fiducial model, which matches the \emph{Planck} CMB result.  Our data are broadly consistent with this fiducial value.   } \label{fig:posteriors_Cosmology}
\end{figure}

\section{Discussion}\label{sec:conclusion}

In this work, we have used our CIB-cleaned $y$-maps from Ref.~\cite{Paper1} to measure and constrain the cross-correlation of the tSZ effect and CMB lensing in our Universe, with \textit{Planck} PR4 data. The tSZ effect has been cross-correlated with low-$z$ tracers of LSS (in particular galaxy catalogs and weak lensing maps) many times (e.g.~\cite{2014PhRvD..89b3508V,2020MNRAS.492.4780O,2021MNRAS.500.1806M,2022PhRvD.105l3525G,2022PhRvD.105l3526P}); however, the cross-correlation with CMB lensing has only been measured once before, in HS14. The difficulty in this measurement lies in the fact that at the high redshifts probed by CMB lensing, the tSZ signal is subdominant in our mm-wave sky maps to the CIB. Much of our work has thus focused on removing the CIB from this measurement in a model-independent way. The result is a less statistically significant, but much more robust, measurement than that of HS14.

We have chosen to remove the CIB by making constrained ILC maps from the \textit{Planck} single-frequency maps. It has been standard for previous $y$-LSS cross-correlations to {neglect the CIB (for low-$z$ tracers), or to} remove only a component corresponding to the CIB SED, by modelling the CIB as a modified blackbody with fixed parameters $\beta$ and $T_{\mathrm{CIB}}^{\mathrm{eff}}$ describing its SED (e.g., in~\cite{2022PhRvD.105l3525G,2022PhRvD.105l3526P}).  However, due to the uncertainty on the values of these parameters for the CIB, we have extended this approach by measuring our cross-correlation on moment-deprojected maps~\cite{2017MNRAS.472.1195C}. This approach allows us to account for errors incurred due to choosing incorrect parameter values for the SED, as well as for the spatial variation and line-of-sight variation of the CIB SED.

We have verified extensively that such an approach should measure an unbiased signal, by applying our NILC pipeline to detailed simulations of the microwave sky, incorporating appropriate inter-frequency decorrelation of the CIB. We have checked that this approach works on the simulations regardless of the SED parameters chosen to deproject the CIB. We have also tested that our measurement on the actual data is stable to the choice of parameters used for the CIB SED. We report our measurements with two different $y$-maps: one with all frequency channels up to 545 GHz used in the NILC and with the CIB and its two first-order moments $\delta\beta$ and $\delta T^{\mathrm{eff}}_{\mathrm{CIB}}$ deprojected, and one without 545 GHz used, which has the CIB and only $\delta\beta$ deprojected. Both maps have the CMB deprojected in the first five needlet scales to avoid ISW bias in the measurement.

As such, we have made the most robust CIB-free, unbiased measurement of the tSZ-$\kappa$ cross-correlation to date. We have made the measurement with three different choices for the $\kappa$ map used in the cross-correlation, each chosen either to minimize a certain systematic (a $\kappa$ map made with tSZ-deprojected $T$ maps to minimize residual $\left<yyy\right>$ bias; a $\kappa$ map which restores signal in the regions of the sky where the tSZ clusters are originally masked in the reconstruction; and the PR4 \texttt{NPIPE} $\kappa$ map, which has the largest signal-to-noise). Our measurements are broadly consistent, both on a point-by-point basis and in the overall constraint we make on the amplitude of the theory model, although the tSZ-deprojected amplitude is lower than the other amplitudes by $\approx 1\sigma$. Overall, our highest-significance detection of the cross-correlation is at the level of $\approx 3.6\sigma$, with the PR4 $\kappa$ map and the no-545-GHz CMB$^{5}$+CIB+$\delta\beta$-deprojected $y$-map. Note that if the data points perfectly matched the fiducial theory template, the SNR in this case would reach $\approx 5 \sigma$.

While our detection is not currently at extremely high significance, we note that the noise in the $\kappa$ map is a limiting factor that will be lowered very soon. Indeed, it will be extremely interesting to make this measurement with the much lower-noise-per-mode ACT DR6 CMB lensing map~\cite{2023arXiv230405203M,2023arXiv230405202Q}. Additionally, improvements in the small-scale measurement of the $y$-map~\cite{2023arXiv230701258C} using the ACT DR6 data will allow us to probe this signal on smaller scales than we have in this work. The methods to remove the CIB developed here will be of great use for such a measurement, as well as other future LSS-$y$ cross-correlation measurements, which will need to remove CIB bias (see, e.g.,~\cite{2023arXiv230801856F}).  Further exploration of the trade-off between ILC deprojections (to mitigate CIB biases) and the measurement SNR will also be of interest, e.g., using the methods of~\cite{2021PhRvD.103j3510A}.  Such a measurement will allow us to learn more about currently unprobed astrophysics in the ICM of high-$z$ groups and clusters. 

Finally, we note that it would be of great interest to combine an analysis of $C_\ell^{y\kappa}$ with a measurement of the auto-power spectrum (or indeed of any other statistic of the $y$-map), to break the degeneracies between cosmology and astrophysics and to access the large potential that the $y$ signal contains to constrain cosmology.

\begin{acknowledgements}
We are very grateful to Will Coulton for many useful discussions, including comparisons of NILC pipelines. We are also very grateful to Boris Bolliet for discussions about the halo model and \texttt{class\_sz}. We also thank Shivam Pandey for useful consistency checks of our $y$-maps. We additionally thank James Bartlett, Nick Battaglia, Aleksandra Kusiak, Jean-Baptiste Melin, Blake Sherwin, and David Spergel for useful conversations. We thank the Scientific Computing Core staff at the Flatiron Institute for computational support. The Flatiron Institute is supported by the Simons Foundation. JCH acknowledges support from NSF grant AST-2108536, NASA grants 21-ATP21-0129 and 22-ADAP22-0145, the Sloan Foundation, and the Simons Foundation. This work was completed at the Aspen Center for Physics, which is supported by National Science Foundation grant PHY-1607611.
\end{acknowledgements}

\bibliography{references}

\appendix

\section{Truncation radii for the halos}\label{app:cutoff}

{In this appendix, we discuss the halo-centric radius at which we cut off our halo model integrals, as well as compare our signal to a direct measurement from simulations~\cite{2015ApJ...812..154B}. As these simulations were performed with a \textit{WMAP}-like cosmology, we use the appropriate cosmological parameters throughout in all our plots: $\{h=0.72, \Omega_m = 0.25, \Omega_b = 0.043, n_s = 0.96, \sigma_8 = 0.8$\}.}

\subsection{Choice of truncation radius for calculation of $C_\ell^{y\kappa}$}

A key ingredient for many halo model observables is the Fourier transform of the relevant profiles of the halos. For lensing, the relevant profile is the lensing convergence profile $\kappa(r,M,z)$, which is a weighted integral of the density profile $\rho(r,M,z)$. For the tSZ effect, the relevant profile is that of the electron pressure $P_e(r,M,z)$. 

Formally, the Fourier transform of a spherically symmetric profile $A(r,M,z)$ is calculated by performing the following integral:
\begin{equation}
\tilde A(k,M,z) = 4\pi\int_0^\infty dr \, r^2 \frac{\sin(k r)}{kr} A(r,M,z) \,.
\end{equation}
In particular, the integral on the right-hand side should be taken to $r\rightarrow\infty$. For profiles that decay sufficiently fast as $r\rightarrow\infty$, such that the integral converges, this is not a problem.  In practice, this is the case for the generalized NFW expressions used to fit the pressure profiles in~\cite{2012ApJ...758...75B}, which we use in this work. \textit{However}, the NFW model used for the density profiles does not decay fast enough for the integral to converge; this has the related consequence that the mass of a given halo within an arbitrary radius $R$, which is given by
\be
M(R) = 4 \pi \int_0^R dr \, r^2 \rho(r) \,,\label{mofr}
\ee
diverges as $R\rightarrow\infty$. 

This means that, in practice, a cut-off (or ``threshold'') radius must be chosen when calculating such integrals. It is common to choose the virial radius of the halos $r_{\mathrm{vir}}$ for this cut-off. However, the corresponding integral when calculating the Fourier transform of the pressure profile $\tilde y_\ell$, which converges as $r\rightarrow\infty$, is not converged by $r=r_{\mathrm{vir}}$. Indeed, calculations of $C_\ell^{y\kappa}$ are much closer to convergence when we use a cut-off radius of $2 r_{200m}$.

Thus, we decide to use a cut-off radius of $2 r_{200m}$ when we calculate $\tilde y_\ell(M,z)$ for use in modelling $C_\ell^{y\kappa}$. We must also decide what cut-off to use when calculating the lensing profile $\tilde \kappa_{\ell}(M,z)$. One option is to use the standard cut-off for density of $r_{{\mathrm{vir}}}$ (or $r_{200m}$, which is close to $r_{{\mathrm{vir}}}$). The effects of this choice for the density profile cut-off on $C_\ell^{\kappa\kappa}$ are shown in Figure Figure~\ref{fig:clkk_rcut}.  We find that using a cut-off of $r_{{\mathrm{200}m}}$ ($2r_{{\mathrm{200}m}}$) reproduces the nonlinear Halofit~\cite{2012ApJ...761..152T} prediction to within 10\% (1\%) in the range $\ell \lesssim 1000$.  While using $r_{\rm cut}=2r_{200 m}$ is slightly further away from the Halofit prediction at low $\ell$ than $r_{\rm cut}=r_{200 m}$, it reproduces the Halofit calculation to within $3\%$ at $\ell \lesssim 5000$, a much larger range than the $r_{\rm cut}=r_{200 m}$ case.

\begin{figure}
\includegraphics[width=0.32\textwidth]{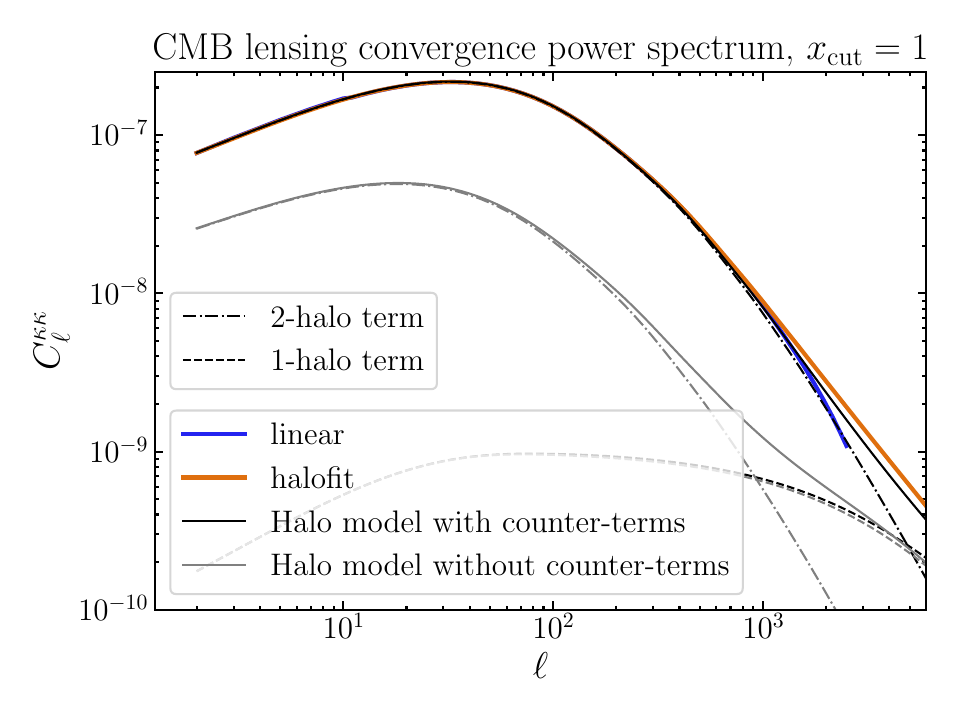}
\includegraphics[width=0.32\textwidth]{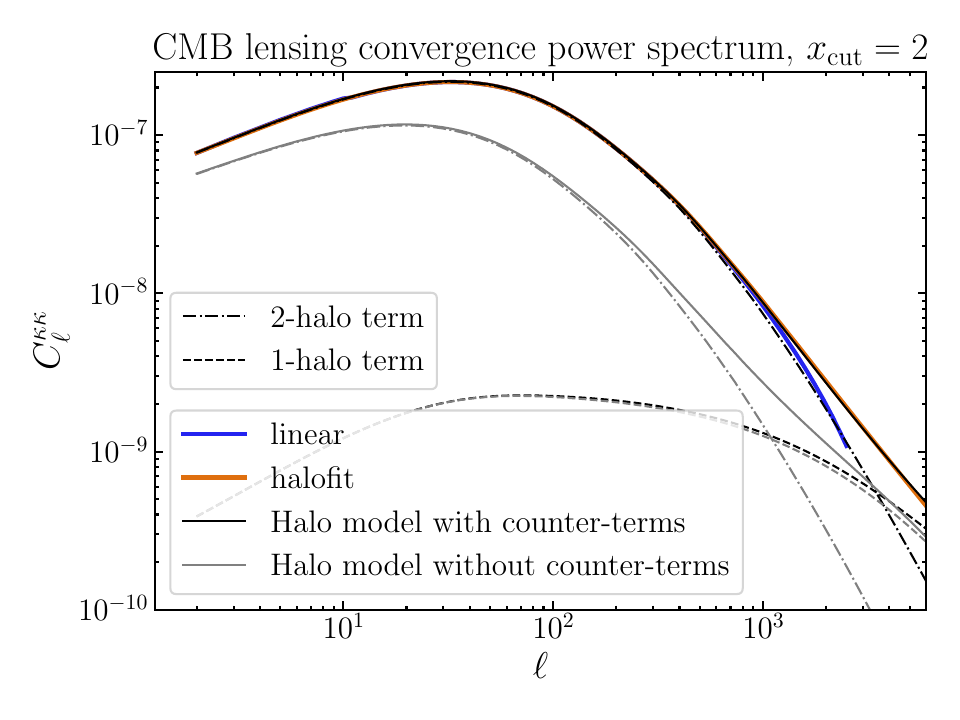}
\includegraphics[width=0.32
\textwidth]{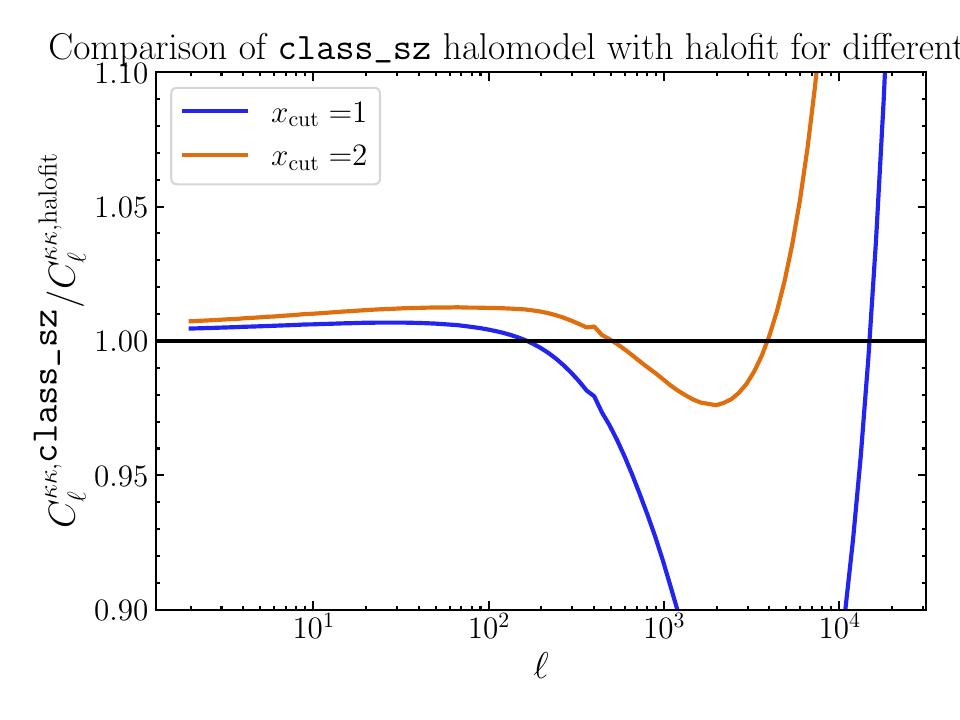}
\caption{{In the left two plots, we show the CMB lensing power spectrum computed with the halo model described in the text (implemented in \texttt{class\_sz}), compared with the linear calculation of \texttt{class} and the non-linear Halofit calculation (also calculated using \texttt{class}). We use $x_{\mathrm{cut}}=1,2$ as indicated in the titles of the plots. On the right, we show the ratio of the \texttt{class\_sz} calculation with the Halofit calculation for these values of $x_{\rm cut}$.} Note that $x_{\rm cut}$ is defined as $r_{\rm cut}\equiv x_{\rm cut} r_{200 m}$.}\label{fig:clkk_rcut}
\end{figure}

However, setting a sharp cut-off at a smaller threshold than we use for calculating $\tilde y_{\ell }(M,z)$ results in our model allowing non-zero gas pressure in a region where there is zero matter density. We thus choose to also extend our integrals to $2 r_{200m}$ in $\tilde \kappa_\ell(M,z)$. This results in the assignment of more mass to each halo than for the standard $r_{200m}$ threshold (i.e., $M(2r_{200m})$ as opposed to $M(r_{200m})\equiv M_{200m}$, where $M(r)$ is defined by Equation~\eqref{mofr}).  This mass must be accounted for correctly and consistently when we calculate the amount of mass in an interval of halo masses via the HMF.  In particular, the total mass $M$ between two halo masses of mass $M_1$ and $M_2$ is 
\be
M = \int _{M_1}^{M_2}\frac{dN}{dM^\prime}dM^\prime  M(M^\prime)\label{totalmass}
\ee
where $M^\prime$ in Equation~\eqref{totalmass} is a dummy variable that refers to some mass definition, which may or not be the same mass definition that we use to count the mass assigned to each halo. In practice, we use the HMF of~\cite{2008ApJ...688..709T}, which is defined for the mass definition $M_{200m}$. Thus, it counts halos within a given range of $M_{200m}$, which is the variable that we integrate over in our HMF expressions. However, we must calculate $M(M^\prime)$, the mass within $2 r_{200m}$ for a halo of given $M_{200m}=M^\prime$, and count \textit{this} mass when we perform the integral. As this quantity does not appear in $\tilde \kappa_\ell(M,z)$, the only thing we change in $\tilde \kappa_\ell(M,z)$ is the cutoff for the Fourier transform; however, when we calculate the \textit{counter-terms}, it is important to modify these as the consistency condition they preserve is defined in terms of the total mass of the Universe, which we now calculate using the modified Equation~\eqref{totalmass}.

\subsubsection*{Validity of extending the NFW profile beyond $r_{\mathrm{vir}}$}

It is clear that it is not valid to extend the NFW profile to arbitrarily large radii. For example, the NFW model is not a good fit for the density profiles of halos at extremely large radii; see, e.g.,~\cite{2014ApJ...789....1D,2017ApJS..231....5D,2016ApJ...825...39M}. Ideally we would use a modified density profile, not NFW, at such large radii, ensuring that the appropriate integrals converge. We leave such modelling to future work.

\subsection{Comparison to direct measurement from simulations}

{In order to validate our choice of $r_{\mathrm{cut}}$, we compute  $C_\ell^{y\kappa}$ in the halo model for various choices of $r_{\rm cut}$ and compare to a direct measurement of this signal from cosmological hydrodynamics simulations~\cite{2015ApJ...812..154B}. Importantly, these are the same simulations from which our pressure profile model was extracted~\cite{2010ApJ...725...91B}, and thus such a comparison is valid as a test of the halo model approximation for our signal.  Note that the direct simulation measurement is an integral over the full Compton-$y$ and $\kappa$ fields in the simulation box, including contributions from outside the halos, and so does not depend on the halo model prescription. We show our comparisons in Figure~\ref{fig:compare_sims_calculation}. Note that, as the simulations were performed with a \textit{WMAP}-like cosmology ($h=0.72, \Omega_m = 0.25, \Omega_b = 0.043, n_s = 0.96, \sigma_8 = 0.8$), these are the cosmological parameters we have used for the halo model calculations in this plot. We find that for $x_{\mathrm{cut}}=2$, we capture 75\% of the power on large scales, with the power captured to within 10\% on small scales. It is clear that there is some missing power in our halo model calculation in the $\ell<2000$ regime where we have measured the signal in data, due to the unbound gas outside of halos (the same conclusion was reached by comparing halo model and simulation results in Ref.~\cite{2015ApJ...812..154B}).  This indicates that our constraints on $P_0$ are biased slightly high by neglecting the unbound-gas contributions; to avoid this bias we can incorporate this mismatch by rescaling our theory curves by the ratio of the simulation and halo model calculation in Figure~\ref{fig:compare_sims_calculation}.}

{Note that, while the mismatch between theory and simulation in Figure~\ref{fig:compare_sims_calculation} could be decreased by increasing $r_{\mathrm{cut}}$, it is unphysical to extended the $\kappa$ profile to arbitrary $r_{\mathrm{cut}}$ as the total enclosed mass in halos becomes larger than the total amount of matter in the Universe (i.e., the counter-terms required to account for lower-mass halos become negative to compensate).}

\begin{figure}
\includegraphics[width=0.5\textwidth]{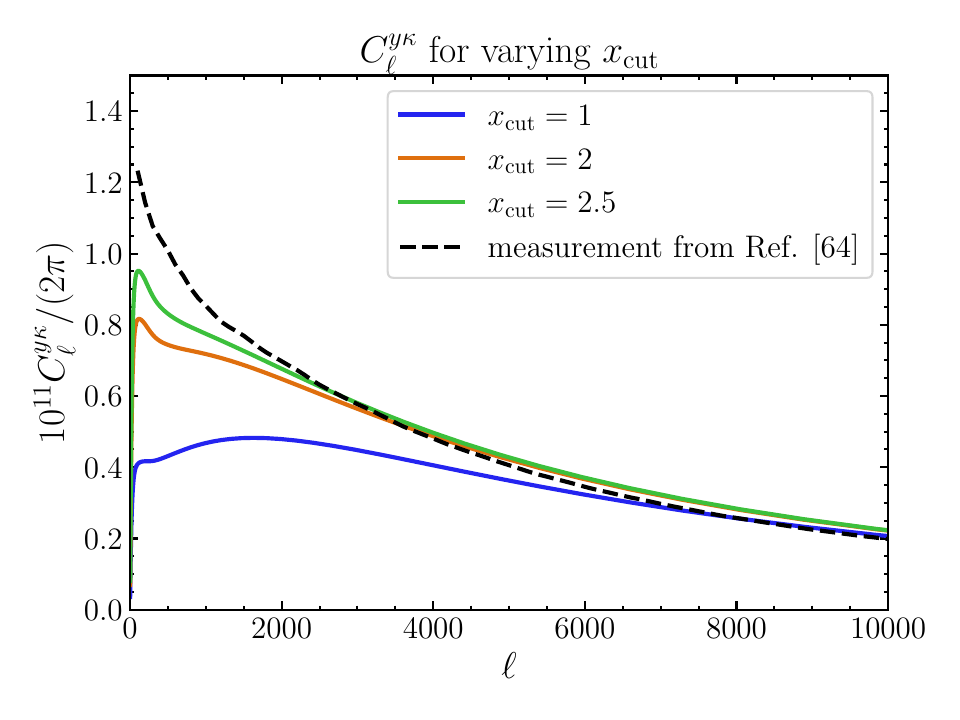}
\caption{A comparison between our halo-model calculation of the signal and the direct measurement from simulations in Ref.~\cite{2015ApJ...812..154B}. The halo model calculation for our fiducial cutoff choice ($x_{\rm cut} = 2$) is shown in orange.  It is clear that we miss some power on large scales, which arises from diffuse, unbound gas outside of halos and/or contributions from low-mass halos.}\label{fig:compare_sims_calculation}
\end{figure}

\section{Sky area}\label{app:skyarea}

We report our measurements on 61.57\% of the sky. In the absence of Galactic foregrounds, changing the sky area scales the total SNR by $\sqrt{f_{\mathrm{sky}}}$. However, it is important to avoid large-scale correlations between residual Galactic foregrounds in the $\kappa$ estimate correlating with those in the $y$-map. Thus, we wish to make a measurement on the largest sky area possible that is robust to this leakage of Galactic power.

To ensure that our measurement is indeed unbiased by the Galaxy, we perform stability tests on our data points over a range of values of $f_{\mathrm{sky}}$. We show in Figure~\ref{fig:fsky_stability} how the data points change when we change the sky fraction.  In general, we do not see any systematic trend with $f_{\rm sky}$.  Our standard analysis mask is created by thresholding the 857 GHz map --- using this as a tracer of Galactic dust, we mask the $20\%$ brightest pixels in the 857 GHz map (we refer to this as the $f_{\mathrm{sky}}=80\%$ threshold mask). We then multiply this mask by the point source mask and the lensing analysis mask. To explore different $f_{\mathrm{sky}}$ values, we create similar masks, removing the $40\%$, $60\%$, and $80\%$ brightest pixels (creating $f_{\mathrm{sky}}=$ 60\%, 40\%, and 80\% threshold masks respectively). We then multiply by the lensing analysis mask and the point source masks and apodize with a $10^\prime$ apodization scale; this results in total sky fractions of $51.44\%$, $32.92\%$,and $15.35\%$ respectively (our $f_{\mathrm{sky}}=80\%$ threshold mask results in a $61.57\%$ final mask).

Due to the stability observed with varying $f_{\mathrm{sky}}$ in  Figure~\ref{fig:fsky_stability}, we report our final results in the main text with the largest sky fraction (ie, with the 80\% threshold mask).

\begin{figure}
\includegraphics[width=0.49\textwidth]{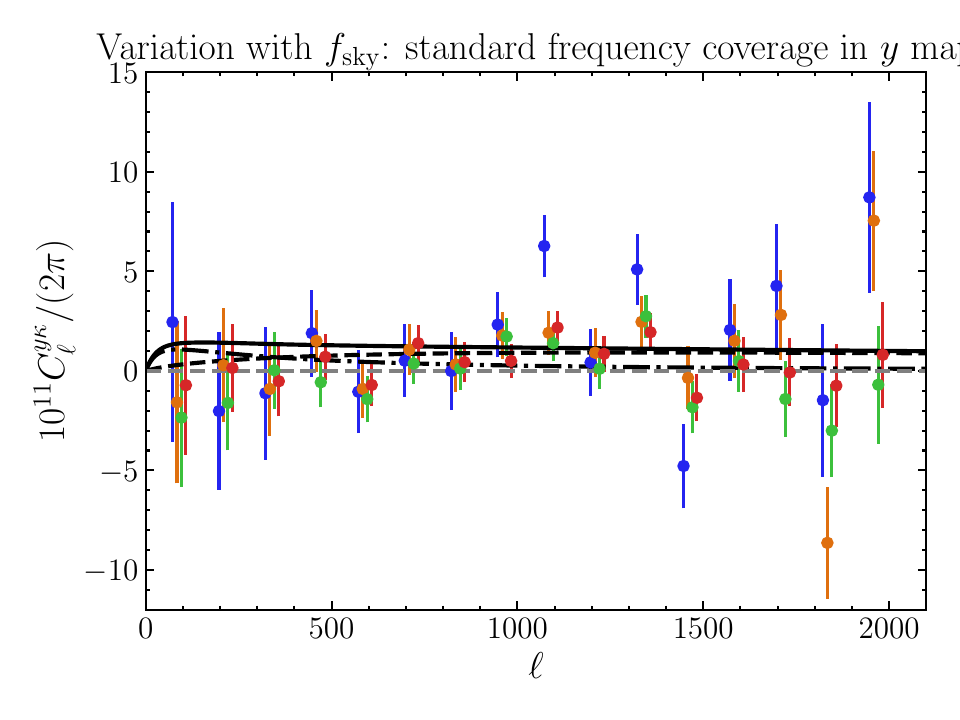}
\includegraphics[width=0.49\textwidth]{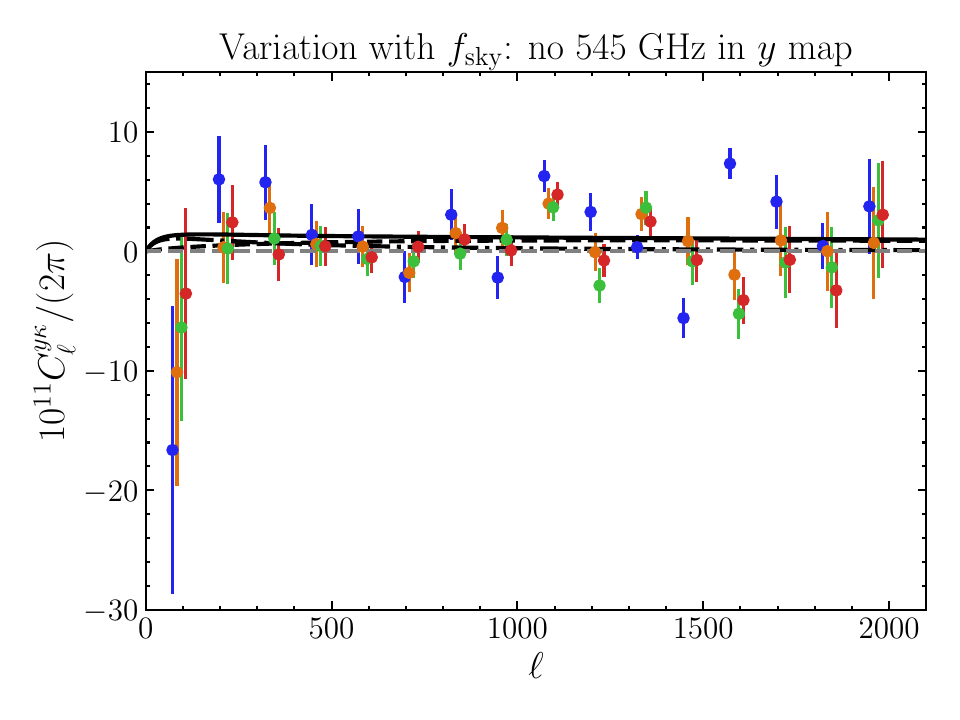}
\includegraphics[width=0.7\textwidth]{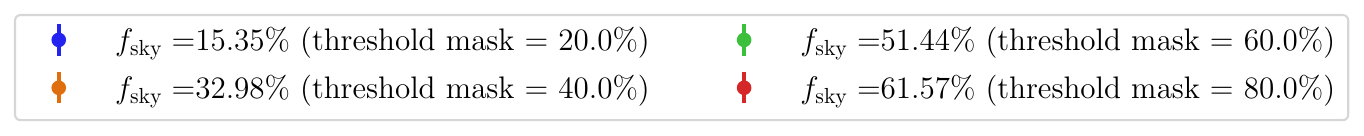}
\includegraphics[width=0.29\textwidth]{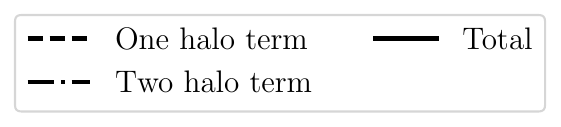}
\caption{The measured data points on different sky areas. In all cases, we use the maximally-deprojected $y$-maps (CMB$^5$+CIB+$\delta\beta$+$\delta T_{\mathrm{CIB}}^{\mathrm{eff}}$ for standard frequencies and CMB$^5$+CIB+$\delta\beta$-deprojected for the no-545 GHz case), and the tSZ-deprojected $\kappa$ map. Overall, we do not see a systematic variation in the data points as we decrease the sky area (and thus measure on regions of the sky that are less impacted by Galactic foregrounds). On the left is the standard frequency coverage case, and on the right is the no-545-GHz case.}\label{fig:fsky_stability}
\end{figure}

\end{document}